\documentclass[reprint,amsmath,amssymb,aps,floatfix]{revtex4-2}

\usepackage{hyperref}
\hypersetup{
    colorlinks=true,
    linkcolor=blue,
    citecolor=blue,
    urlcolor=cyan
}

\usepackage{graphicx}
\usepackage{mathdots}
\usepackage[normalem]{ulem}
\usepackage[svgnames]{xcolor}
\usepackage{bm}
\usepackage{multirow}

\newcommand{\Eq}[1]{Eq.~(\ref{#1})}
\newcommand{\Fig}[1]{Fig.~\ref{#1}}

\newcommand{\Sec}[1]{Sec.~\ref{#1}}
\newcommand{\cRef}[1]{Ref.\cc{#1}}
\newcommand{\cRefs}[1]{Refs.~\cc{#1}}
\newcommand{\cc}[1]{~\mbox{\cite{#1}}}

\newcommand{\norm}[1]{{\| #1 \|}}  
\newcommand{\ket}[1]{{ |{#1} \rangle }}

\newcommand{\ketbra}[2]{{ |{#1} \rangle\!\,\langle {#2} | }}

\newcommand{\EqDef}{\stackrel{\mathrm{def}}{=}}
\newcommand{\Id}{\mathbb{I}}

\DeclareMathOperator{\Tr}{Tr}

\newcommand{\mcU}{\mathcal{U}}
\newcommand{\mcV}{\mathcal{V}}

\newcommand{\mcN}{\mathcal{N}}
\newcommand{\mcR}{\mathcal{R}}

\newcommand{\vtheta}{{\bm{\theta}}}

\newcommand{\vsigma}{{\bm{\sigma}}}

\newcommand{\vPhi}{{\bm{\phi}}}
\newcommand{\vp}{{\bm{p}}}
\newcommand{\vn}{{\bm{n}}}

\newcommand{\vvr}{{\bm{r}}}
\DeclareMathOperator{\CX}{CX}

\DeclareMathOperator{\RESET}{RESET}

\newcommand{\ignore}[1]{}

\begin{document}

\title{Dissipative variational quantum algorithms for Gibbs state preparation}
  
\author{Yigal Ilin$^{1}$}
\email{yigal.ilin@gmail.com}

\author{Itai Arad$^{2}$}

\affiliation{$^1$Andrew and Erna Viterbi Department of Electrical
and Computer Engineering, Technion – Israel Institute of Technology,
3200003, Haifa, Israel}

\affiliation{$^2$Centre for Quantum Technologies, National
University of Singapore, 117543 Singapore, Singapore}

\begin{abstract} 
  In recent years, variational quantum algorithms
  (VQAs) have gained significant attention due to their adaptability
  and efficiency on near-term quantum hardware.  They have shown
  potential in a variety of tasks, including linear algebra, search
  problems, Gibbs and ground state preparation.  Nevertheless, the
  presence of noise in current day quantum hardware, severely limits
  their performance. In this work, we introduce \emph{dissipative
  variational quantum algorithms} (D-VQAs) by incorporating
  dissipative operations, such as qubit RESET and stochastic gates,
  as an intrinsic part of a variational quantum circuit.  We argue
  that such dissipative variational algorithms posses some natural
  resilience to dissipative noise. We demonstrate how such
  algorithms can prepare Gibbs states over a wide range of quantum
  many-body Hamiltonians and temperatures, while significantly
  reducing errors due to both coherent and non-coherent noise. An
  additional advantage of our approach is that no ancilla qubits are
  need.  Our results highlight the potential of D-VQAs to enhance
  the robustness and accuracy of variational quantum computations on
  NISQ devices.
\end{abstract}

{\hypersetup{hidelinks}
\maketitle
}

\section{Introduction}
\label{sec:gibbs-introduction}

In recent years, variational quantum algorithms (VQAs) have gathered
significant research interest due to their adaptability and
efficiency on near-term quantum hardware\cc{ref:Peruzzo2014,
ref:Farhi2014, ref:Troyer2015, ref:Endo2020, ref:Bittel2021,
ref:Benedetti2021, ref:Bonet-Monroig2023, ref:Yuan2019,
ref:Hastings2014, ref:Kandala2017}. A typical VQA involves preparing
a trial state on a quantum computer and calculating a cost function
based on measurements taken from that state\cc{ref:Ying2017,
ref:Hempel2018, ref:Romero2018, ref:Kandala2017, ref:Peruzzo2014}.
As of today, VQAs are one of our best candidates for a useful early
quantum advantage. Firstly, the framework is very rich and adaptive,
and can be applied to many practically important problems. VQAs have
been designed for various linear algebra problems\cc{ref:Xu2021,
ref:BravoPrieto2023, ref:Huang2019}, search
problems\cc{ref:Nakanishi2019}, determining the ground state of a
given Hamiltonian\cc{ref:Cerezo2022}, singular value
decomposition\cc{ref:Wang2021-SVD}, fidelity
estimation\cc{ref:Chen2021} and many other tasks\cc{ref:Romero2018,
ref:Yuan2019, ref:McArdle2019, ref:Ying2017}. Secondly, the VQA
framework leverages a significant portion of the computation to a
classical side, allowing for much shallower quantum circuits that
are crucial for our current day noisy devices\cc{ref:McClean2016,
ref:Grimsley2019}. Finally, the variational nature of these
algorithms provides some resilience against coherent errors, which
are prevalent in today's quantum hardware\cc{ref:Khodjasteh2009,
ref:McClean2016, ref:Sharma2020, ref:Movassagh2021,
ref:Berberich2024}.

Despite the above strengths, VQAs have not yet demonstrated a
convincing quantum advantage.  Several challenges remain, even on
noiseless devices. These include barren plateaus\cc{ref:Uvarov2021,
ref:Larocca2024}, high measurement costs for computing cost
functions involving numerous non-commuting
operators\cc{ref:Gonthier2022}, and additional
obstacles\cc{ref:Anschuetz2022, ref:Lee2023-noexpVQE}. However, the
primary limitation on today’s quantum hardware remains the
high noise levels. The accumulation of noise in a VQA
circuit degrades its performance. Specifically, non-coherent noise,
such as decoherence or amplitude damping, poses the greatest
challenge, as coherent noise can largely be mitigated by the
variational ansatz. In such cases, one must resort to error
mitigation techniques (see \cRefs{ref:Temme-PEC2023, ref:Seif2023,
ref:Strikis2021, ref:Cai2021, ref:Huggins2021, ref:Ravi2022}) to
rectify the effects of noise, which usually come at an exponential
cost, thereby again limiting the performance of the VQA, or wait for
error correction to become viable.

In this study, we expand the capabilities of variational quantum
algorithms by incorporating \emph{dissipative} operations in the
variational circuits. We call the resulting framework
\emph{dissipative VQA} (D-VQA). Our main claim is that the
introduction of these new elements can alleviate the effect of
non-coherent noise, much like variational unitaries can fight
coherent errors. A simple example of a dissipative operation within
a quantum circuit is a mid-circuit measurement, potentially followed
by a qubit reset. The inclusion of such operations in quantum
circuits has been investigated in various contexts, including local
quantum channel learning\cc{ref:Ilin2024}, measurement-induced
entanglement phase transitions\cc{ref:Roeland2023}, error mitigation
through postselection\cc{ref:Botelho2022}, quantum steering for
state preparation\cc{ref:Volya2024}, non-equilibrium phase
transitions\cc{ref:Chertkov2023}, and others\cc{ref:DeCross2023,
ref:Hua2022, ref:Brandhofer2023}. Additionaly, as shown in
\cRef{ref:Roeland2023}, introducing dissipation through random
measurements can help mitigate barren plateaus by significantly
increasing the variance of the gradients with respect to the
variational parameters of the VQA.

Here, we investigate the efficacy of dissipative VQA, where we
introduce a variational dissipative operation, denoted as
$\mathcal{R}_i$, alongside unitary gates. Our dissipative operation
includes both a stochastic element (a probabilistic gate), together
with a RESET operation. Specifically, it is a parametrized gate,
denoted by $\mathcal{R}_i(p,\vPhi)$, which resets qubit $i$ to the
pure state $\ketbra{\vPhi}{\vPhi}$ with probability $p$. We
demonstrate that incorporating such gates offers two significant
advantages: 1) The inclusion of dissipative operations allows for
the creation of mixed states, eliminating the need for ancilla
qubits when the trial state of the VQA is mixed. 2) D-VQA circuits
can mitigate some of the non-coherent errors, similar to how unitary
VQAs mitigate coherent errors, resulting in better performance in
the presence of noise. This resilience is particularly good when
targeting states with low purity.

We demonstrate the advantages of our D-VQA ansatz through a simple
toy model and a set of classical simulations of variational Gibbs
state preparation with high fidelity in both noisy and noiseless
scenarios across a wide range of quantum many-body Hamiltonians.

The structure of this paper is as follows: In
\Sec{sec:gibbs-background}, we provide a comprehensive background on
Gibbs state preparation on quantum devices. In
\Sec{sec:gibbs-dissipative-ansatz}, we introduce a single-qubit toy
model of a quantum circuit used in a D-VQA scheme, demonstrating its
noise resilience against non-coherent errors. We also present our
global ansatz, which employs the $\mcR$ gate and general 2-local
unitaries. \Sec{sec:gibbs-model-optimization} outlines the numerical
optimization scheme and defines the noise model used.
\Sec{sec:gibbs-numerical-results} presents the numerical results.
Finally, in \Sec{sec:gibbs-discussion-outlook} we present our
conclusions and discuss possible future research directions.

\section{Background}
\label{sec:gibbs-background}

Throughout this work we consider the problem of preparing a Gibbs
state on a quantum computer. Gibbs state preparation is a central
problem in quantum computation and information of both theoretical
and practical importance. Specifically, Gibbs states are used in quantum machine
learning\cc{ref:Biamonte2017, ref:Lifshitz2021-GibbsLearn,
ref:Zoufal2021-1, ref:Zoufal2021-2, ref:Torlai2020}, quantum
simulation\cc{ref:Childs2018, ref:Vernier2023}, and quantum
optimization\cc{ref:Somma2008, ref:Mohseni2022}, and for training
quantum Boltzmann machines by sampling from a well-prepared Gibbs
state\cc{ref:Wiebe2014, ref:Kieferova2017, ref:Amin2018}.

Given a Hamiltonian $H$, its Gibbs state $\rho_G$ at inverse
temperature $\beta=1/k_BT$ is given by
\begin{align}
\label{eq:gibbs-density-mat}
  \rho_{G} \EqDef \frac{1}{Z}e^{-\beta H},
\end{align}
where $Z\EqDef\Tr(\exp{(-\beta H)})$ is the partition function.

There are various approaches for preparing a Gibbs state on a
quantum computer, which can be broadly categorized into two types.
The first type includes non-variational algorithms that steer the
system toward the target Gibbs state. These algorithms may simulate
the physical process of thermalization, as seen in
\cRefs{ref:Poulin2009, ref:Boixo2010, ref:Riera-Eisert2012,
ref:Su2020, ref:Holmes2022-theoretical, ref:Temme2011, ref:Yung2012,
ref:Metcalf2020}, utilize imaginary time evolution like in
\cRefs{ref:Cirac2004, ref:Motta2019}, or use Monte Carlo style Gibbs
samplers as in \cRefs{ref:Gilyen2023, ref:Layden2023}. Other methods
include thermal shadow tomography\cc{ref:Benedetti2023}, sampling
from the Gibbs state using cluster expansions\cc{ref:Cohn2023},
sampling from the steady-state of a local Kraus
map\cc{ref:Gurevich2024}, dissipative sampling\cc{ref:Cubitt2023},
or various other approaches\cc{ref:Brando2018, ref:Chowdhury2016,
ref:Gilyen2022, ref:Lu2021, ref:Alhambra2023}. 

The second type consists of variational algorithms, where a
classical optimization routine iteratively updates the parameters of
the variational quantum circuit based on a cost function that
measures the proximity of the system state to the desired Gibbs
state\cc{ref:Endo2020, ref:Tilly2022, ref:Cerezo2021,
ref:Farhi2014}. These algorithms may leverage approaches similar to
non-variational algorithms, such as
thermalization\cc{ref:Sagastizabal2021, ref:Movassagh2021},
imaginary time evolution\cc{ref:Movassagh2021, ref:Wang2023}, or
Monte Carlo style methods\cc{ref:Patti2022}, among
others\cc{ref:Troyer2015, ref:Cohn2020}. Additionally, variational
algorithms have been successfully combined with classical neural
network architectures to enhance the classical component of Gibbs
sampling algorithms, as shown in \cRefs{ref:Verdon2017, ref:Liu2021,
ref:Guveina2024}. 

Some variational algorithms are designed specifically for NISQ-type
computers. These algorithms aim to prepare a Gibbs state using
minimal quantum resources, such as the number of gates and
measurements, often relying on heuristic approaches
\cc{ref:Verdon2017, ref:Liu2021, ref:Swingle2019, ref:Hidary2019,
ref:Zhu2020, ref:Wang2021, ref:Foldager2022,
ref:Consiglio-XY-Ising2024, ref:Selisko2023}. Most of these NISQ-type algorithms require ancillary qubits coupled to the system where the thermal equilibrium state is prepared.

The natural cost function for the variational algorithm is the
Helmholtz free-energy function $\mathcal{F}(\rho) \EqDef
\Tr(H\rho)-\beta^{-1}S(\rho)$, as it is minimized by the Gibbs state
$\rho_G$\cc{ref:Reif1965-StatMech}. While the energy term
$\Tr(H\rho)$ is a local observable, which can be estimated
efficiently on a quantum computer, the von Neumann entropy term
$S(\rho)\EqDef -\Tr(\rho\log\rho)$ is highly non-local, and cannot
be estimated efficiently. Therefore, in variational algorithms for
the Gibbs states one has to make additional assumptions and/or
approximations when using the free energy cost
function\cc{ref:Hsieh2019, ref:Swingle2022, ref:Swingle2019,
ref:Hidary2019, ref:Chowdhury2020, ref:Zhu2020, ref:Wang2021,
ref:Foldager2022, ref:Consiglio-XY-Ising2024,
ref:Selisko2023, ref:Araz2024}, or resort to a different cost
function altogether\cc{ref:Warren2022adaptive}.

The main purpose of this work is to investigate which Gibbs
states can be prepared with dissipative variational circuits and how
the accuracy changes in the presence of noise. Specifically, we aim
to study the \emph{expressibility} of the proposed ansatz. We shall
therefore ignore the question of efficiency in the cost function and
use the \emph{infidelity} of the resultant state with the ideal
Gibbs state as a cost function. Nevertheless, our ansatz
can readily accommodate other loss function choices, including those
with gradients calculated on quantum hardware, such as in
\cRefs{ref:Warren2022adaptive, ref:Wang2021}.

\section{Dissipative circuits and noise resilience}
\label{sec:gibbs-dissipative-ansatz}

In this section we describe the central building block of our
circuit, which is the dissipative single-qubit $\mcR$ gate. We study
its noise resilience properties through a simple one qubit 
example, and then describe our global
ansatz that uses the $\mcR$ gate and general 2-local
unitaries.

\subsection{Dissipative gates and noise resilience}
\label{sec:dissipative-gates}

Our variational circuit is based on a specific dissipative gate,
which is a single-qubit, probabilistic gate that combines a RESET
gate and a general 1-qubit rotation. For brevity, we shall refer to
it as the $\mcR$ gate. When acting on qubit $i$, its action is
described by the non-unital quantum channel
\begin{align}
\label{eq:gibbs-non-det-reset}
    \mcR_i(p, \vPhi)[\rho]  &\EqDef (1-p)\rho 
      + p\ketbra{\vPhi}{\vPhi} \cdot\Tr_i(\rho).
\end{align}
$\mcR_i(p, \vPhi)$ resets qubit $i$ to the state
$\ketbra{\vPhi}{\vPhi}$ with probability $p\in [0,1]$, and with
probability $1-p$ leaves the state unchanged. Since
$\ketbra{\vPhi}{\vPhi}$ is a pure state on the Bloch sphere, it is
described by two angles. Together with $p$, it uses three
parameters.

One advantage of variational dissipative quantum circuits over
unitary circuits is their partial resilience to dissipative noise.
To illustrate this, we consider a toy model system with one qubit,
assuming a simple depolarizing noise model.

Let $\vvr$ be a vector in the Bloch ball, and $\vsigma\EqDef
(X,Y,Z)$ be the vector of Pauli matrices. Consider a situation in
which the ideal circuit applies a single qubit unitary gate $U$ to a
mixed input state $\rho_0= \frac{1}{2}(\Id + |\vvr|\cdot Z)$, and
takes it to the ideal output state $\rho_1$:
\begin{align*}
  \rho_1 = \mcU[\rho_0] = U\rho_0 U^\dagger 
    = \frac{1}{2}(\Id + \vvr\cdot\vsigma),
\end{align*}
where $\mcU$ is the channel corresponding
to the unitary $U$, i.e., $\mcU[\rho] = U\rho U^\dagger$. For
simplicity, we assume that a depolarizing noise channel, given by
\begin{align}
  \mcN[\rho] \EqDef (1-\lambda)\rho + \lambda\Tr(\rho)
    \cdot \frac{\Id}{2} ,
\end{align}
acts on the qubit after each gate. In such case, we can look for a
unitary $V$ with a corresponding channel $\mcV$ such that
$\mcN\circ\mcV(\rho_0)$ is as close as possible to the ideal state
$\rho_1$. For any one-qubit state $\tau = \frac{1}{2}(\Id + \vn\cdot
\vsigma)$, the action of the depolarizing channel $\mcN$ is
to shrink its Bloch radius, i.e., $\tau \to
\mcN[\tau]=\frac{1}{2}\big(\Id + (1-\lambda)\vn\cdot
\vsigma\big)$. Therefore, it is easy to see
that the optimal unitary $V$ will be $V=U$, and the optimal noisy
output $\rho'_1$ is
\begin{align*}
  \rho'_1 = \mcN\circ\mcU[\rho_0] = \mcN[\rho_1] 
    = \frac{1}{2}\big(\Id + (1-\lambda)\vvr\cdot\vsigma\big) .
\end{align*}
In such case, the optimized noisy state suffers an error of 
\begin{align*}
  \norm{\rho_1 - \rho'_1}_1 = \big\|\frac{1}{2}
    \lambda\vvr\cdot\vsigma\big\|_1 = \lambda\cdot|\vvr| . 
\end{align*}

Now assume that we can use the dissipative $\mcR$ gate, and apply
a circuit of the form $\mcR\circ\mcU$. Together with
the noise, our circuit will apply the channel
$\mcN\circ\mcR\circ\mcN\circ \mcU$ so that
\begin{align*}
  \tilde{\rho}_1 = \mcN\circ\mcR\circ\mcN\circ \mcU(\rho_0)
   = \mcN\circ\mcR\big[ \frac{1}{2}\big(\Id 
     + (1-\lambda)\vvr\cdot\vsigma\big)\big] .
\end{align*}

To optimize the circuit, we first fix the $\vPhi$
parameter of the $\mcR$ gate to be such that $\ketbra{\vPhi}{\vPhi}
= \frac{1}{2}(\Id + \hat{\vvr}\cdot\vsigma)$, where $\hat{\vvr}
\EqDef \vvr/|\vvr|$, and leave $p$ unfixed. A simple calculation
shows that
\begin{align*}
  &\mcR\big[ \frac{1}{2}\big(\Id 
     + (1-\lambda)\vvr\cdot\vsigma\big)\big] \\
  &= \frac{1}{2}\big[\Id + \big((1-p)(1-\lambda)+p/|\vvr|\big)
         \vvr\cdot\vsigma\big],
\end{align*}
and therefore,
\begin{align*}
  \tilde{\rho}_1 &= \frac{1}{2}\big[\Id 
    + (1-\lambda)\big((1-p)(1-\lambda)+p/|\vvr|\big)
         \vvr\cdot\vsigma\big].
\end{align*}

It follows that as long as we can choose $p\in [0,1]$ such that
\begin{align*}
  (1-\lambda)\big((1-p)(1-\lambda)+p/|\vvr|\big) = 1 ,
\end{align*}
we can fully cancel the effect of the dissipative noise. A
straightforward calculation shows that this is possible iff
$\lambda\le 1-|\vvr|$.

\begin{figure}
  \includegraphics[width=\linewidth]{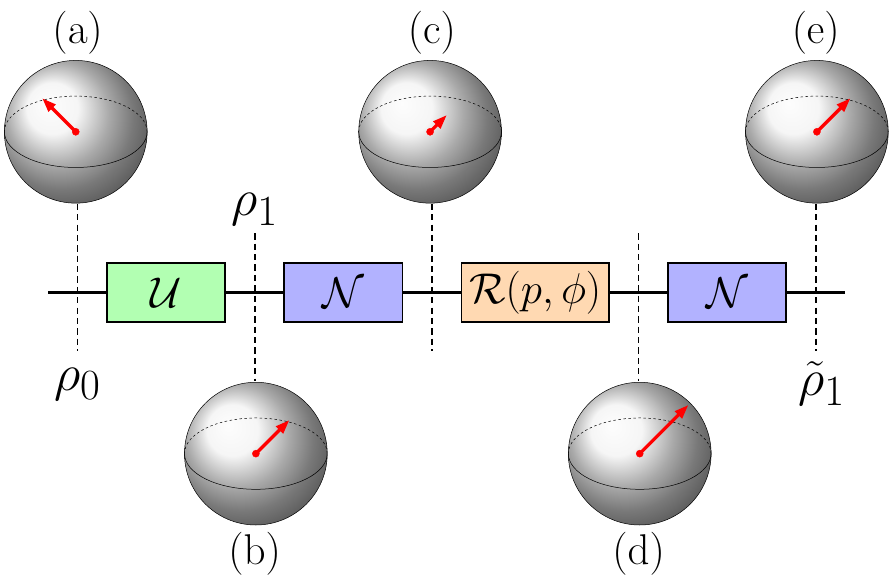}
    \caption{Toy model illustrating a noisy single qubit circuit
    with the input state $\rho_0= \frac{1}{2}(\Id + |\vvr|\cdot Z)$,
    and the output state $\tilde{\rho}_1 =
    \mcN\circ\mcR\circ\mcN\circ \mcU(\rho_0)$. The red arrows in the
    Bloch balls (a-e) represent the mixed state after each
    operation. From left to right: (a) The input state $\rho_0$. (b)
    Unitary channel $\mcU$ rotates the input state to the ideal
    output state $\rho_1$. (c) The depolarizing noise channel $\mcN$
    ``shrinks'' the vector $\vvr$. (d) The dissipative $\mcR$ gate
    ``expands'' the vector $\vvr$ beyond what is required to
    mitigate only the noise channel in (c). (e) The action of the
    noise channel $\mcN$ applied after the $\mcR$ gate reverts the
    state back to the ideal output of (b), such that in the best
    case $\tilde{\rho}_1=\rho_1$.} 
\label{fig:dissipative-toy-model}
\end{figure}

The above conclusion is further illustrated in
\Fig{fig:dissipative-toy-model}, and can be understood intuitively
as follows. The application of the noise increases the entropy of
the state, or, equivalently, shrinks its Bloch radius $|\vvr|$.
Therefore, to counter it, we first inflate the Bloch
radius using $\mcR$ so that the subsequent decrease by $\mcN$ will
take us back to the ideal output. However, if the initial state is
close to being pure (i.e., $|\vvr|$ is close to $1$), we cannot
increase its radius by much, and we cannot fully cancel the
shrinkage due to $\mcN$. Therefore, for $\mcR$ to cancel the noise,
it must be that the distance of $|\vvr|$ from $1$ is comparable to
$\lambda$. 

On a high level, the above analysis shows that the dissipative
variational circuit behaves like the probabilistic error mitigation
(PEC)\cc{ref:Temme-PEC2017, ref:Temme-PEC2023, ref:Temme-PEC2023-2},
as it effectively inverts the noise channel. Unlike, PEC, however,
it does this automatically without a prior characterization of the
noise model. On the other hand, it can only apply inverse noise
channels that are physical, and as such, it cannot fully correct
high noise levels.

Another potential advantage of using the dissipative $\mcR$ gate in
variational quantum circuits is that it can be seen as a stochastic,
parameterized version of a mid-circuit measurement. As shown in
\cRef{ref:Roeland2023}, adding dissipation through random
measurements can significantly increase the variance of gradients
with respect to the circuit's parameters. This may help the
classical optimization routine in VQA algorithms to avoid local
minima and prevent rapidly decaying gradients, which may lead to
barren plateaus \cc{ref:McClean2018, ref:Cerezo2021}.

\subsection{The dissipative circuit ansatz}
\label{sec:gibbs-dissipative-global-ansatz}

Having defined the dissipative $\mcR$ gate, our dissipative circuit
anstaz follows a simple brick-wall architecture that is common in
VQA circuits. 

\begin{figure}
  \includegraphics[width=\linewidth]{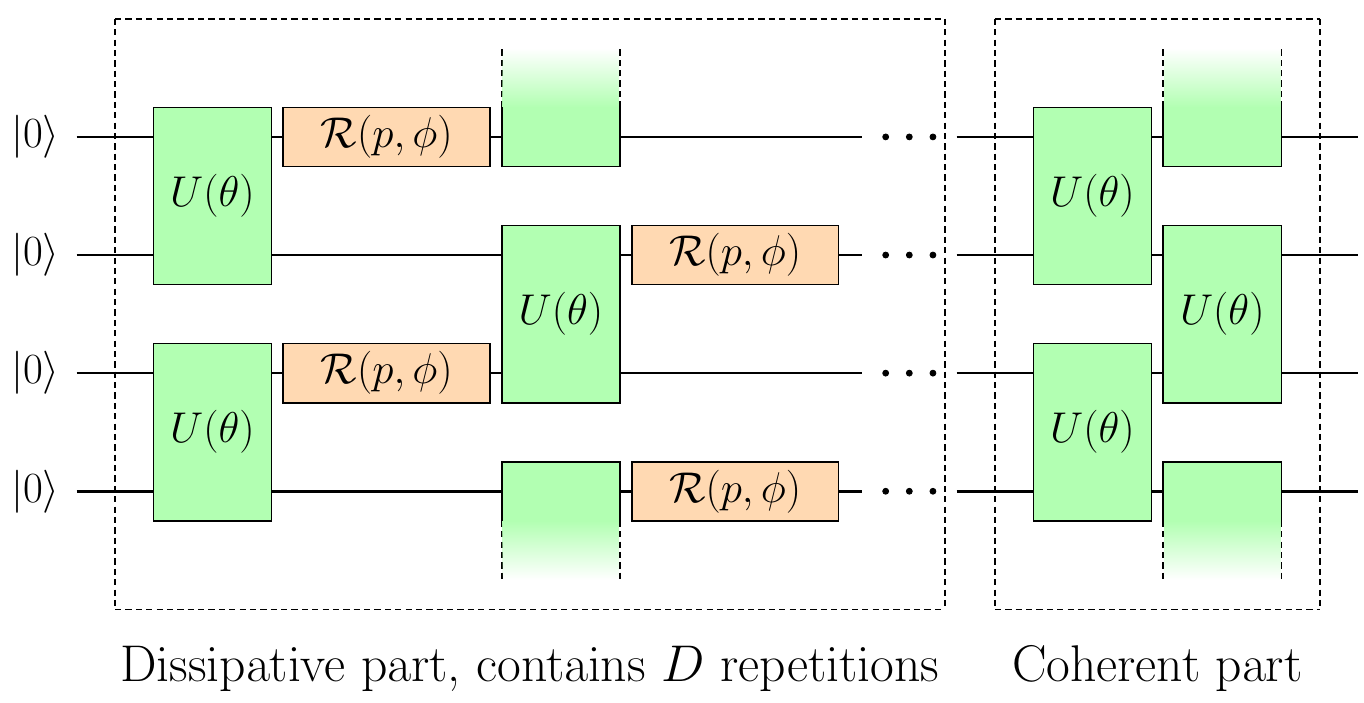}
    \caption{Dissipative circuit model for system size $n=4$ qubits
    arranged on a ring. The input state is initialized to
    $|0\rangle\langle 0|^{\otimes n}$. The dissipative parametrized
    layer is repeated $D$ times, and followed by a single unitary
    (coherent) layer. The unitary gates $U(\theta)$ are general
    two-qubit gates. The parametrized probabilistic $\RESET$ gates
    are represented as non-unital channels $\mcR$ given in
    \Eq{eq:gibbs-non-det-reset}. The parameters $\theta, p, \phi$
    are different from gate to gate.}
    \label{fig:dissipative-gibbs-ansatz}
\end{figure}

Alongside the dissipative $\mcR$ gates, we use general two-qubit
gates $U$, parameterized by a vector of rotation angles $\vtheta$,
applied on neighboring qubits. For a general unitary $U\in SU(4)$,
there exists a standard KAK decomposition into three $\CX$ gates and
$15$ elementary single qubit rotations between
them\cc{ref:Farrokh2004}. We used this decomposition and let each
two-qubit unitary to have its own set of $15$ rotation angles.

Using the dissipative $\mcR$ gates and general $SU(4)$ gates, our
ansatz uses a brick-wall structure, described in
\Fig{fig:dissipative-gibbs-ansatz}. While we focus on the periodic
1D case, our results can be readily generalized to other types of
lattices. From a high-level perspective, the circuit consists of two
parts. The first part is the dissipative part, which consists of $D$
even-odd brick-wall layers, where each layer is made of entangling
$SU(4)$ gates and dissipative gates (see
\Fig{fig:dissipative-gibbs-ansatz}). The second part is purely
coherent, consisting of a single even-odd brick-wall layer of
$SU(4)$ gates. For a system size $n$, with $D$ dissipative layers,
our numerical experiments were performed on variational circuits
with $(18D+15)n$ parameters.

The division of the circuit into two parts, loosely follows the
approach taken in \cRefs{ref:Swingle2019, ref:Hidary2019,
ref:Consiglio-XY-Ising2024}. There, the first
part prepares a mixed state of orthogonal states (usually in the
computational basis) with Boltzmann weights that approximate the
weights of the target Gibbs state. This mixed state is then rotated
to the actual Gibbs state using several layers of unitary gates.
Similarly, the dissipative part of our circuit can be viewed as
preparing a mixed state with the target Boltzmann weights, which is
then rotated to the target Gibbs state by the coherent part. We
note, however, that in our approach the dissipative part also
contains many unitary rotations, which help align the eigenbasis of
the initial mixed state with the target Gibbs state. This allows the
coherent part to be relatively shallow at the expense of increasing
the depth of the dissipative part, which, in view of
\Sec{sec:dissipative-gates}, is important for achieving dissipative
noise resilience.  

We conclude this section by discussing how our anstaz
can be used on current day quantum computers. Our
circuit is equivalent to an ensemble of circuit instances, each
instance representing a different branch of the various $\mcR_i(p,
\vPhi)$ probabilistic gates --- either the identity operator or the
reset operator $\ketbra{\vPhi}{\vPhi}\cdot \Tr_i(\rho)$. The latter
can be implemented by a standard RESET to $\ketbra{0}{0}$ gate,
followed by a local rotation $\ket{0}\to \ket{\vPhi}$. To run the circuit, we first sample a sufficient
number of instances, and then execute each of these instances on a
quantum computer. The final output state generated by
the ansatz is obtained by averaging the outputs of the sampled
circuit instances. Using $M$ samples will result in the usual
statistical error of $O(1/\sqrt{M})$.

Implementing our ansatz on quantum hardware also involves
calculating gradients of the cost function with respect to the
circuit's variational parameters. The ansatz has two gradient types:
(i) rotation angles $\vtheta, \vPhi$, computed using the well-known
parameter shift rule\cc{ref:Schuld2019, ref:crooks2019gradients},
and (ii) activation probabilities $\vp$ of the $\mcR$ gates. Since
the $\mcR$ gates use mid-circuit measurements (implicitly, as
part of the RESET gate), computing gradients with respect to the
activation probabilities $\vp$ typically requires additional circuit
executions, as shown in \cRef{ref:Roeland2023}. This overhead arises
if the measurement outcome (of the partially collapsed wave
function) impacts later stages of the quantum circuit. However, in
our ansatz, the final output state is independent of the mid-circuit
measurement outcome. Instead, we use the mid-circuit measurement
simply to reset the qubit to a variational pure state defined by a
single-qubit rotation $\ket{0} \to \ket{\vPhi}$. As a result, the
dependence of the final output state on $\vp$ appears only in the
relative weights of different circuit instances within the ensemble.
A straightforward calculation (based on \cRef{ref:Roeland2023})
shows that in this case, the gradient with respect to $\vp$ can be
computed without additional circuit runs.

\section{Numerical Simulations}
\label{sec:gibbs-model-optimization}

To evaluate the effectiveness of our dissipative variational circuit
in preparing Gibbs states, we conducted several numerical
simulations of it, together with simulations of other coherent
anstazes. These simulations were performed both under ideal,
noiseless conditions and using a simple noise model typically
assumed for gate-based quantum computers like those offered by
IBMQ\cc{ref:IBMQ}. In this section we describe the systems we
simulated and their noise model, together with technical details on
the optimization algorithm, and finally the numerical results.

\subsection{Simulated systems}
\label{sec:gibbs-simulated-systems}

We simulated system sizes between $n=2$ to $n=6$ qubits arranged on
a ring, with the initial state $\rho_0 = |0\rangle\langle
0|^{\otimes n}$. In all our simulations, we considered the Gibbs
states of $2$-local translation invariant Hamiltonians defined on a
ring of $n$ qubits as:
\begin{align}
\label{eq:gibbs-generic-hamiltonian}
  H = \sum_{i=1}^{n} h_{i,i+1} ,
\end{align}
where $h_{i,i+1}$ is the same operator, with
$\norm{h_{i,i+1}}\leq1$, acting on qubits $i,i+1$, with the $n+1$
qubit identified as qubit $1$. We note that while our target Gibbs
Hamiltonians were translationally invaraint, we
did not impose translation invariance on the underlying parameters
of the circuit, and let the optimizer pick the optimal parameters
independently.

\subsection{Noise model}
\label{sec:gibbs-noise-model}

For the noisy simulations of our ansatz, we assumed a local
Markovian noise model without crosstalks. Since we have two distinct
types of gates, there are two types of noises to consider: the noise
in the two-qubit unitary gates $U$ (using the KAK formula, which is
a combination of $\CX$ and rotation gates\cc{ref:Farrokh2004}) and
the noise in the non-unital single-qubit channel $\mcR$.

For the two-qubit unitary gates, we used a simplified noise model
similar to those in quantum noise simulators like IBM's
Qiskit\cc{ref:IBMQ, ref:Qiskit}, which is also often used for performing
noisy simulations of VQAs\cc{ref:Sharma2020, ref:Zeng2021,
ref:Fontana2021}. While being relatively simple, it captures two important aspects of noise:
decoherence, which is characterized by the $T_2$ time, and amplitude damping,
characterized by the $T_1$ time, governing thermal relaxation. In
this model, each qubit $j$ has its own noise channel $\mcN_j$, so
the global noise channel on $n$ qubits factors into a tensor product
of single-qubit noise channels $\bigotimes^{n}_{j=1} \mcN^{(j)}$.
Each $\mcN^{(j)}$ combines two noise sources: dephasing (phase
damping) and amplitude damping\cc{ref:nielsen2010quantum}. The
dephasing noise model with parameter $\lambda_j$ acting on a qubit
$j$ is defined as a Pauli noise channel:
\begin{align}
\label{eq:gibbs-dephasing-noise}
  \mcN_{dep}^{(j)}[\rho] 
    \EqDef (1-\lambda_j)\rho + \lambda_j Z \rho Z .
\end{align}
The amplitude damping noise model with parameter $\omega_j$ acting
on qubit $j$ is defined by:
\begin{align}
\label{eq:gibbs-amplitude-noise}
  \mcN^{(j)}_{amp}[\rho] 
    \EqDef K_0 (\omega_j) \rho K^{\dagger}_0 (\omega_j) 
      + K_1 (\omega_j) \rho_j K^{\dagger}_1 (\omega_j),
\end{align}
where
\begin{align}
\label{eq:gibbs-amplitude-noise-ops}
  K_0(\omega_j) = \begin{pmatrix}
                    1 & 0 \\
                    0 & \sqrt{1-\omega_j}
                \end{pmatrix}
  \,\,\,K_1(\omega_j) = \begin{pmatrix}
                    0 & \sqrt{\omega_j} \\
                    0 & 0
                \end{pmatrix}.
\end{align}

The dephasing parameters $\{\lambda_j\}_{j=1}^{n}$ and
amplitude damping parameters $\{\omega_j\}_{j=1}^{n}$ were chosen to
correspond to a low noise regime with values
$\sim 10^{-3}$ and were randomly generated before each optimization
run: each $\lambda_j$ and
$\omega_j$ were picked uniformly at random inside the range
$[1,2]\times10^{-3}$.

Given that the $\CX$ gates, which have the longest duration and
highest error rates, dominate the noise in two-qubit
unitaries\cc{ref:Kjaergaard2020, ref:Jurcevic_2021}, we assumed that
most noise originates from these gates. Consequently, each
application of a $\CX$ gate in the general two-qubit unitaries $U$
on qubits $j$ and $j+1$ was followed by the noise channel
$\mcN_{j}\otimes \mcN_{j+1}$, and each 1-qubit rotation was assumed
ideal.

As the non-unital channel $\mcR$ from \Eq{eq:gibbs-non-det-reset}
is accomplished by a regular RESET gate
\begin{align*}
    \mcR_i(p, \vPhi)[\rho] = (1-p)\rho 
      + pU(\vPhi)\ketbra{0}{0}U^\dagger(\vPhi)\cdot\Tr_i(\rho),
\end{align*}
where $U(\vPhi)$ is a single-qubit rotation taking $\ket{0}\to
\ket{\vPhi}$, we modeled its noise by a noise model of the
standard RESET gate. Following \cRef{ref:Rost2021}, we assumed a
phenomenological noise model where the reinitialization of the qubit
to the $\ketbra{0}{0}$ state might fail. Specifically, we
upperbounded the activation probability $p$ by a global value
$p^*=0.99$, i.e., $p \le p^* < 1$, so that a prefect RESET is never
possible.

\subsection{Simulation method}
\label{sec:gibbs-numerical-sim}

Our simulations were done by evolving the full density matrix
according to the dissipative variational circuit
from \Fig{fig:dissipative-gibbs-ansatz}, starting from
the initial state $\rho_0=\ketbra{0}{0}^{\otimes
n}$. This produced an output variational state $\rho(\vtheta,\vp)$,
where $\vtheta$ represents the unitary rotation angles for both the
$\mcR$ gates and the two-qubit unitary gates, and $\vp$ are the
activation probabilities of the $\mcR$ gates. We then optimized the
variational parameters to minimize the infidelity cost function
\begin{align}
\label{eq:gibbs-fidelity-loss}
  \Phi(\vtheta,\vp) \EqDef 
    1 - F\big(\rho(\vtheta,\vp), \rho_{G}\big),
\end{align}
where 
\begin{align}
\label{eq:gibbs-fidelity}
  F\big(\rho(\vtheta,\vp), \rho_{G}\big) 
    \EqDef \left[\Tr\sqrt{\sqrt{\rho(\vtheta,\vp)}
      \rho_{G}\sqrt{\rho(\vtheta,\vp)}}\right]^2
\end{align}
is the fidelity between the variational state $\rho(\vtheta,\vp)$
and the target Gibbs state $\rho_{G}$\cc{ref:Uhlmann2010}.

As said by the end of \Sec{sec:gibbs-background}, the main focus of
this work is to check the \emph{expressibility} of our dissipative
circuit and its resilience to noise. Consequently, we allowed
ourselves to choose the infidelity as the loss function, even though
it is hard to estimate it efficiently on a quantum computer.
Similarly, other measures of distinguishability, such as trace
distance or relative entropy between the target and variational
Gibbs states, can be equally employed to define the loss
function\cc{ref:nielsen2010quantum}.

After each simulation, gradients with respect to the variational
parameters were computed using PyTorch's automatic differentiation
engine\cc{ref:NEURIPS2019_9015}. The parameters were then updated
using the Adam\cc{ref:Kingma2015} stochastic optimization algorithm
based on the infidelity cost function $\Phi(\vtheta,\vp)$.

The initial values of the variational parameters $\vtheta$ were
randomly drawn at uniform in range $[-\pi, \pi]$, and the initial
activation probabilities $\vp$ were drawn at uniform in range $[0,
1]$. 

Heuristically, we fixed the maximum number of optimization steps to
be 2000. We added another heuristic termination criterion for the
optimization process, wherein the optimization ceased if at step
$i$, the value of the loss function $\Phi<10^{-3}$.

\section{Results}
\label{sec:gibbs-numerical-results}

In this section we present the results of our numerical simulations
for Gibbs state preparation using our dissipative ansatz for the
ideal and noisy cases.

\subsection{Representation capabilities of the ansatz}
\label{sec:gibbs-random-hamiltonians}

We first investigated the expressibility of our ansatz, by checking
how well it prepares the Gibbs state of a generic $2$-local
translation-invariant Hamiltonian, as defined by
\Eq{eq:gibbs-generic-hamiltonian}, across a broad range of inverse
temperatures $\beta$. To this end, we fixed the system size to $n=4$
qubits and set the number of repetitions of the dissipative
parametrized layer to $D=n$ (see
\Fig{fig:dissipative-gibbs-ansatz}).

To evaluate the expressibility of our ansatz, we sampled $N=1000$ $2$-local translation-invariant Hamiltonians defined by \Eq{eq:gibbs-generic-hamiltonian} across five inverse temperatures: $\beta=0.5,1,2,4,8$. This range allowed us to examine both high and low temperature regimes. For each $\beta$ value, 200 $2$-local translation-invariant Hamiltonians were sampled.

For each sample, we ran the optimization procedure, described in \Sec{sec:gibbs-numerical-sim}, to minimize the
cost function $\Phi(\vtheta,\vp)$ five times for the
noiseless case and five times for the noisy case, using the noise
model described in \Sec{sec:gibbs-noise-model}. Each run was
initialized with different variational parameters
$\{\vtheta,\vp\}$. We then selected the best result among
these five runs for both the noisy and noiseless cases,
respectively.

\begin{figure}
  \includegraphics[width=0.85\linewidth]{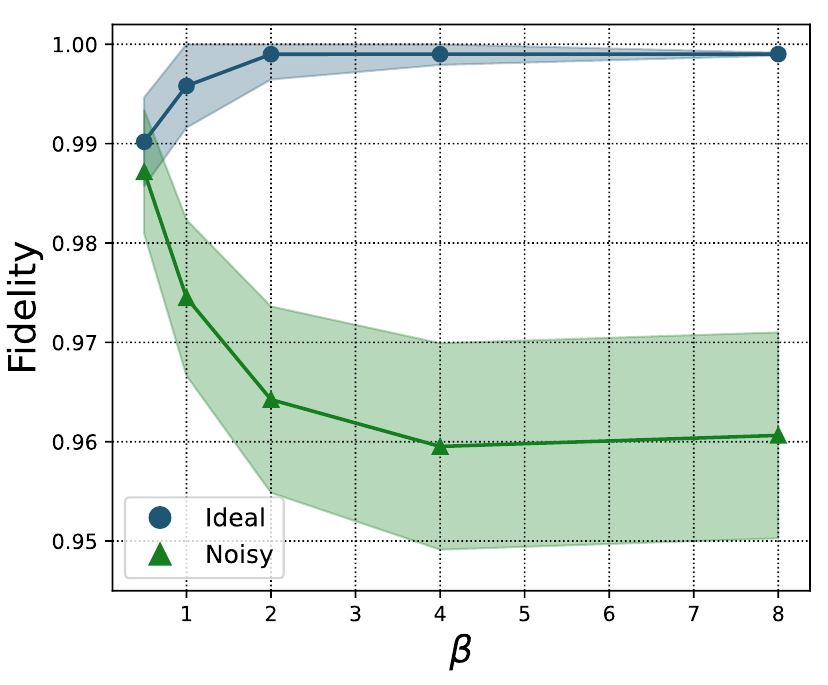}
    \caption{The median fidelity between the estimated and target Gibbs states for 2-local translation-invariant Hamiltonians is shown for both noiseless (blue) and noisy (green) cases. Each data point represents the median fidelity of 200 random Hamiltonians. For each generated Hamiltonian, five optimization runs were performed with the best result taken for each case. The shaded area indicates the uncertainty, represented as one standard deviation.}
    \label{fig:gibbs-random-h-representation}
\end{figure}

In \Fig{fig:gibbs-random-h-representation} we present the median
fidelity between the state estimated by our ansatz and the target
Gibbs state as a function of the inverse temperature $\beta$. For each $\beta$ value the median was calculated over 200 sampled Hamiltonians.

For the noiseless case, the most notable drop in the accuracy occurs
at near room temperature corresponding to $\beta\sim1$. As we
discuss in the \Sec{sec:gibbs-xy-ising-hamiltonians}, when
considering specific Hamiltonians, such dip appears also in other
ansatzes\cc{ref:Selisko2023, ref:Consiglio-XY-Ising2024} and may represent a fundamental limit on the representation
power of the variational ansatzes. Nevertheless, for the ideal case,
the median fidelity in \Fig{fig:gibbs-random-h-representation}
remains above $99\%$ with uncertainty of at most $\sim0.5\%$.

For the noisy case, the median fidelity gradually decreases towards
$96\%$ as $\beta$ increases. Such behavior is expected, as for higher $\beta$ values representing
lower temperatures, the Gibbs state becomes more and more pure,
and in light of \Sec{sec:dissipative-gates}, our dissipative circuit
becomes less resilient to noise.

\subsection{Ising and XY models}
\label{sec:gibbs-xy-ising-hamiltonians}

In this part we focused on two specific, well-studied models: the
transverse field Ising (TFI) and the XY models. As in previous
sections, we considered Gibbs states of translation invariant
Hamiltonians arranged on a ring of $n$ qubits.

The Hamiltonian of the transverse field Ising is given by:
\begin{align}
\label{eq:gibbs-ising-hamil}
  H_{TFI} = -\sum_{j=1}^{n}X_jX_{j+1} - h\sum_{j=1}^n Z_j,
\end{align}
where $h$ represents the strength of the external field. In all our
simulations we consider three values of $h=0.5, 1, 1.5$. The
Hamiltonian of the XY model is:
\begin{align}
\label{eq:gibbs-xy-hamil}
  H_{XY} = -\sum_{j=1}^{n} \left[ \frac{1+\gamma}{2} 
    X_jX_{j+1} + \frac{1-\gamma}{2}Y_j Y_{j+1} \right] 
      - h\sum_{j=1}^{n}Z_j,
\end{align}
where $\gamma$ represents the degree of anisotropy with respect to
the $xy$-plane\cc{ref:LIEB1961407}. In our simulations, in this
section, we fixed $h=0.5$ and considered $\gamma=0.1, 0.5, 0.9$.

In all numerical trials that follow, for each data point, the
optimization procedure was ran ten times with different initial
parameter values, and the best result was selected.

\subsubsection{Ising and XY models: system size dependence}
\label{sec:gibbs-xy-ising-hamiltonians-system-size}

\begin{figure}
  \includegraphics[width=\linewidth]{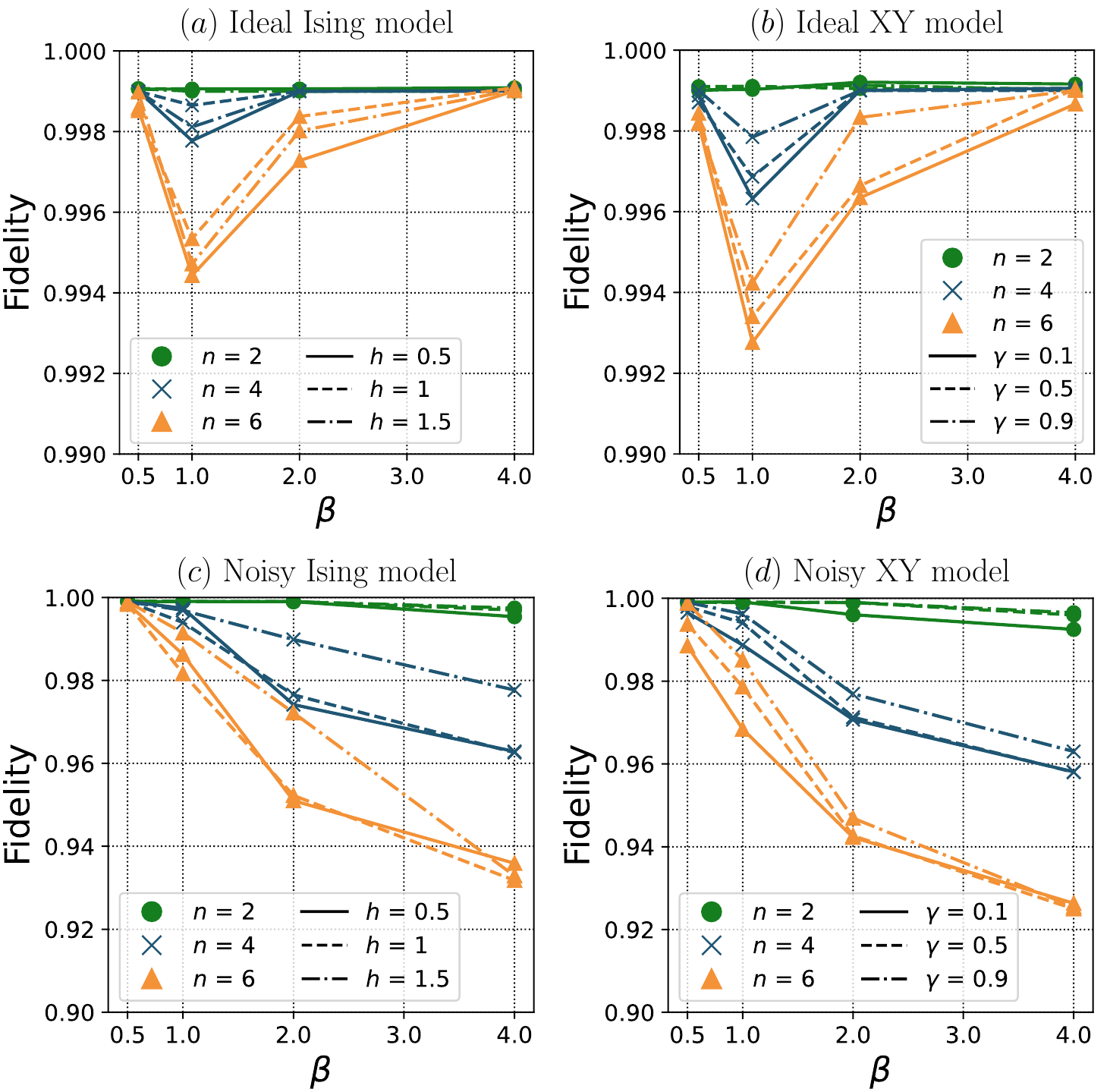}
    \caption{Fidelity between the exact Gibbs state and the state
    produced by our ansatz, versus the inverse temperature $\beta$,
    for several system sizes $n=2,4,6$ and fixed dissipative layer
    depth $D=n$, for each system size. (a, c) Transverse field Ising
    model for ideal and noisy cases, respectively. (b, d) XY model
    for ideal and noise cases, respectively. For each point, ten
    optimization runs were performed with the best result taken.}
    \label{fig:gibbs-system-size}
\end{figure}

In \Fig{fig:gibbs-system-size}, we fixed the number of repetitions
of the dissipative parametrized layer to $D=n$ (see
\Fig{fig:dissipative-gibbs-ansatz}), and show how the precision of
our ansatz varies for system sizes $n=2,4,6$, for the target Gibbs
states of TFI and XY Hamiltonians for several inverse temperature
values and for ideal and noisy cases.

In the noiseless case, \Fig{fig:gibbs-system-size} shows that our
ansatz is capable of preparing the Gibbs state of the target
Hamiltonian with fidelity $>99\%$ for all values of $\beta$
considered, for up to six qubits. Our noiseless ansatz results
closely resemble the numerical results in
\cRef{ref:Consiglio-XY-Ising2024} for the TFI and
XY Hamiltonians, respectively. The most notable drop in the fidelity
for different system sizes occurs at intermediate temperatures where
$\beta\sim1$. Such dip in accuracy was also observed in
\cRef{ref:Consiglio-XY-Ising2024}, where the
authors hinted that this can represent the limit of the ansatz's
representation power. This limitation arises from the higher number
of eigenstates needed to accurately represent the desired Gibbs
state at intermediate temperatures. Naturally, due to this
limitation, the accuracy at intermediate temperatures will continue
decreasing monotonically for larger system sizes.

For the noisy simulations, the noise parameters were set as
described in \Sec{sec:gibbs-noise-model}.
\Fig{fig:gibbs-system-size} illustrates that the fidelity between
the estimated and target Gibbs states gradually decreases with
increasing values of $\beta$, corresponding to lower temperatures,
similar to the trend observed in
\Fig{fig:gibbs-random-h-representation}. As system size increases,
the accuracy of the noisy ansatz significantly drops, particularly
at lower temperatures. This is expected in light of
\Sec{sec:dissipative-gates}, which demonstrates that the noise
resilience of our dissipative ansatz drops as the target state
becomes more and more pure. Such error accumulation for larger
systems and lower temperatures has a more pronounced impact than the
representation limitations observed for intermediate temperatures in
noiseless simulations. Therefore, for large systems with
tens of qubits, the non-coherent noise mitigation capabilities of
our method alone will likely be insufficient to counteract noise
effects, especially at low temperatures. In such cases, our method
should be combined with more advanced error mitigation or
suppression techniques, such as the adaptive error mitigation
presented in \cRef{ref:Henao2023}.

\subsubsection{Ising and XY models: circuit depth}
\label{sec:gibbs-xy-ising-hamiltonians-circuit-depth}

\begin{figure}
  \includegraphics[width=\linewidth]{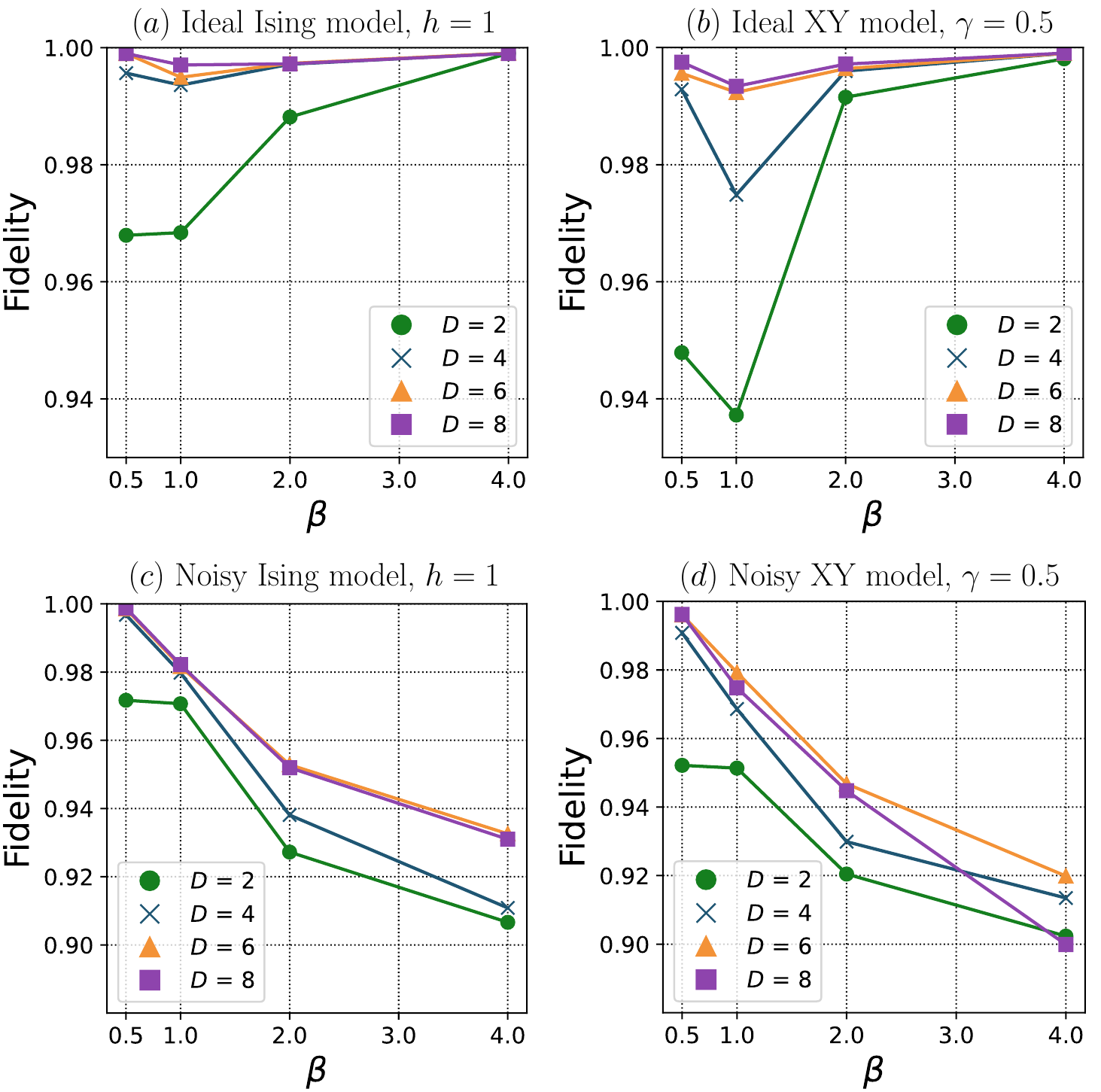}
    \caption{Fidelity between the exact Gibbs state and the state
    produced by our ansatz, versus the inverse temperature $\beta$,
    for different depth values of the dissipative layer $D=2,4,6,8$
    and for fixed system size $n=6$. (a,c) Transverse field Ising
    model for ideal and noisy cases with $h=1$, respectively. (b,d)
    XY model for ideal and noisy cases with $\gamma=0.5$,
    respectively. For each point, ten optimization runs were
    performed with the best result taken.}
    \label{fig:gibbs-circuit-depth}
\end{figure}

In \Fig{fig:gibbs-circuit-depth}, we fixed the system size at $n=6$
and examined how the depth $D$ of the dissipative parametrized layer
affects the accuracy of our ansatz in the noiseless and noisy cases,
for both the TFI Hamiltonian with $h=1$ and the XY Hamiltonian with
$\gamma=0.5$. 

For the noiseless case, shown in \Fig{fig:gibbs-circuit-depth}(a,b),
the accuracy of our ansatz significantly increases as the depth $D$
approaches the system size $n$. For the TFI Hamiltonian, a fidelity
of over $99\%$ is achieved at $D=4$, whereas for the XY Hamiltonian,
this fidelity is reached only at $D=6$. This difference is due to
the higher entanglement in the Gibbs state of the XY model,
requiring deeper circuits to generate the necessary entanglement.
For $D=8$, the fidelity only slightly increases for both models.

For the noisy case, shown in \Fig{fig:gibbs-circuit-depth}(c,d), the
best accuracy is reached at $D=n$. However, unlike the noiseless
case, where increasing the circuit depth from $D=6$ to $D=8$
resulted in a slight increase in accuracy, in the noisy case,
increasing the depth from $D=6$ to $D=8$ may lead to a noticeable
decrease in accuracy, as particularly evident from
\Fig{fig:gibbs-circuit-depth}(d). This behavior is expected in the
noisy case, as adding more layers may lead to an accumulation of
errors that cannot be mitigated by the dissipative circuit, which
might lead to an overall degradation in the fidelity.

\subsubsection{Ising and XY models: comparison with unitary ansatzes}
\label{sec:gibbs-xy-vs-cnot-comparison}

\begin{figure}
  \includegraphics[width=\linewidth]{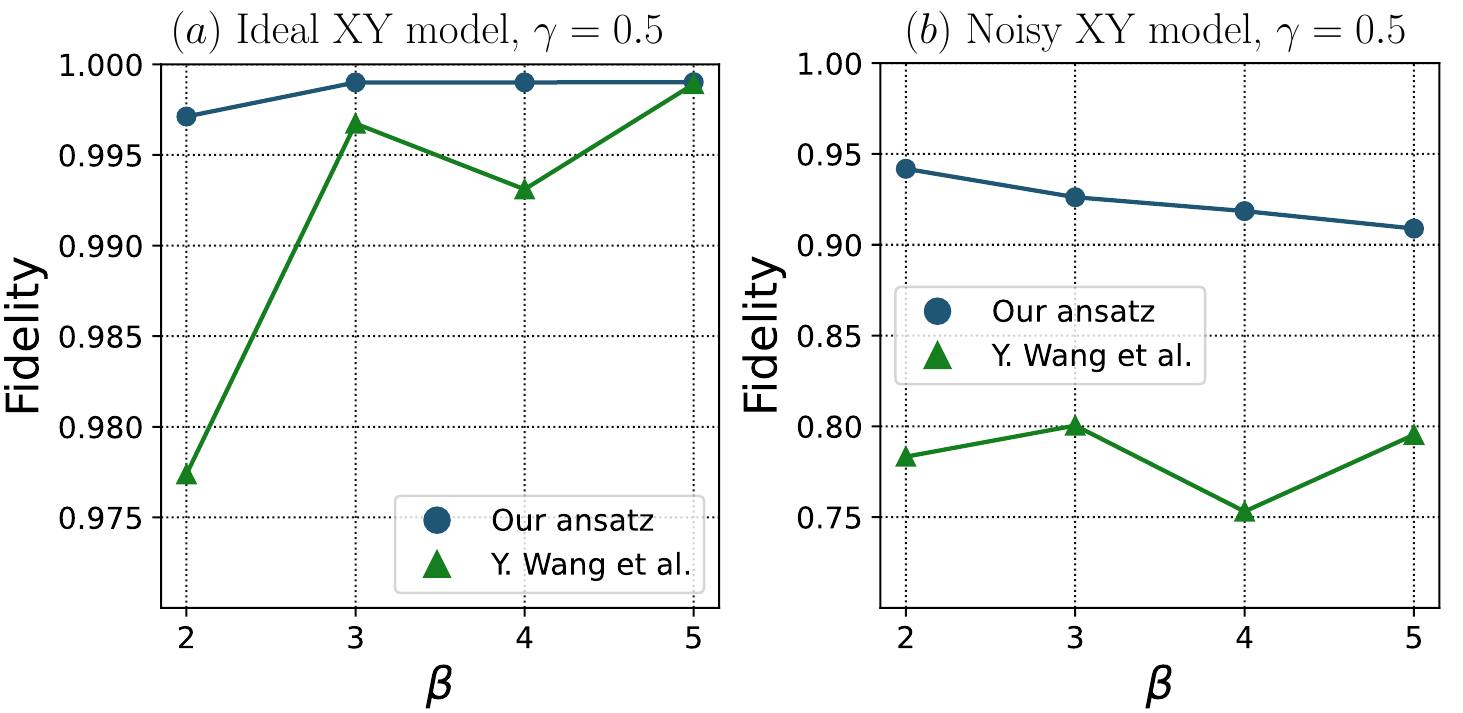}
    \caption{Fidelity comparison between our ansatz and the ansatz
    presented in Fig. 7 in \cRef{ref:Wang2021}, versus the inverse
    temperature $\beta$, for the fixed system size $n=6$. (a) In the
    noiseless case, the ansatzes perform similarly at low
    temperatures. (b) Results for the noisy case, assuming low noise
    regime, the accuracy of our method gradually decreases for lower
    temperatures, while the ansatz presented in Fig. 7 in
    \cRef{ref:Wang2021} reaches $\leq80\%$ fidelity. For each point,
    ten optimization runs were performed with the best result
    taken.} \label{fig:gibbs-xy-vs-cnot}
\end{figure}

In \Fig{fig:gibbs-xy-vs-cnot}, we compare the accuracy of our method
with the full unitary ansatz proposed in \cRef{ref:Wang2021}. The
comparison uses the XY model Hamiltonian defined in
\Eq{eq:gibbs-xy-hamil} on $n=6$ qubits with anisotropic parameter
$\gamma=0.5$ and $h=0.5$.

For our circuits, we set $D=6$, and for the ansatz from
\cRef{ref:Wang2021}, we also chose a depth parameter $d=6$, as shown
in Fig. 7 of \cRef{ref:Wang2021}. The ansatz from
\cRef{ref:Wang2021} uses a single ancilla qubit, requiring a total
of 7 qubits for the Gibbs state preparation on $n=6$ qubits.

We tested both ansatzes under noiseless and noisy conditions,
assuming a low noise level of $\sim 10^{-3}$ for the noisy case.
Both ansatzes were tested at inverse temperatures $\beta=2,3,4,5$
because the ansatz from \cRef{ref:Wang2021} underperformed at lower
$\beta$ values. This underperformance can be attributed to the fact
that the specific ansatz from Fig. 7 in \cRef{ref:Wang2021} was
tested on a smaller system. \Fig{fig:gibbs-xy-vs-cnot} clearly
demonstrates the advantage of using dissipative variational circuits
over unitary circuits in the noisy case, due to their higher noise
resilience.

\begin{table}[h]
    \centering
    \setlength{\tabcolsep}{6pt}
    \begin{tabular}{||c | c | c | c | c||} 
    \hline
     & $\beta=2$ & $\beta=3$ & $\beta=4$ & $\beta=5$ \\ [0.5ex] 
    \hline\hline
    D-VQA & $0.9754$ & $0.9708$ & $0.9662$ & $0.9568$ \\
    \cRef{ref:Consiglio-XY-Ising2024} & $0.8997$ & $0.8502$ & $0.8774$ & $0.825$ \\
    \cRef{ref:Wang2021} & $0.9232$ & $0.9171$ & $0.9015$ & $0.8839$ \\ [1ex]
    \hline
    \end{tabular}
    \caption{Fidelity comparison between the D-VQA ansatz and unitary ansatzes from \cRef{ref:Consiglio-XY-Ising2024} and \cRef{ref:Wang2021}. For each value,
    ten optimization runs were performed with the best result taken.}
    \label{table:ansatz-compare-noisy}
\end{table}

To further illustrate the difference between unitary VQA and our D-VQA, we simulated the XY
model Hamiltonian from \Eq{eq:gibbs-xy-hamil} at inverse
temperatures $\beta=2,3,4,5$ on $n=4$ qubits with anisotropy
parameter $\gamma=0.5$ and field strength $h=0.5$ using three
ansatzes: our D-VQA, the VQA from \cRef{ref:Consiglio-XY-Ising2024}, and the VQA from \cRef{ref:Wang2021}. In ideal conditions, all three ansatzes
consistently achieved fidelities above $99\%$. Table
\ref{table:ansatz-compare-noisy} presents the results from noisy
simulations, where the D-VQA approach consistently maintained
fidelities above $95\%$, while the unitary ansatzes from
\cRef{ref:Consiglio-XY-Ising2024} and \cRef{ref:Wang2021} showed
significantly lower fidelities, particularly at lower temperatures.

Another comparison can be made between our results for the noisy
transverse field Ising (TFI) model presented in
\Fig{fig:gibbs-system-size}(c) and the noisy simulation results
presented in \cRef{ref:Consiglio-XY-Ising2024}, for the case of the
same TFI model. Although the noise model used for the TFI model
simulations in \cRef{ref:Consiglio-XY-Ising2024} is not identical to
the one we used, it is a simplified version derived from the IBMQ
device backend. As demonstrated in \cRef{ref:Ilin2024}, this noise
model primarily consists of dephasing and amplitude damping
processes. Therefore, although a direct comparison between the noisy
TFI results in \cRef{ref:Consiglio-XY-Ising2024} and our noisy TFI
model results in \Fig{fig:gibbs-system-size}(c) is not possible, as
it was in \Fig{fig:gibbs-xy-vs-cnot}, we can still conclude that,
generally, unitary VQAs do not perform well under noisy conditions.

\section{Discussion and outlook}
\label{sec:gibbs-discussion-outlook}

In this article, we introduced the dissipative variational quantum
algorithm (D-VQA) framework, which extends the capabilities of
unitary variational quantum algorithms (VQAs) by incorporating a
dissipative operation as an intrinsic component of the variational
circuit. Specifically, we presented the dissipative $\mathcal{R}$
gate, which re-initializes the single qubit state to a variational
pure state with variational probability $p$. Through analytical
analysis using a single-qubit toy model and various numerical
examples, we demonstrated that incorporating such dissipative
operations enhances the resilience of the variational circuit
against both coherent and non-coherent errors. The resilience to
coherent errors arises naturally from the unitary part of the
variational circuit\cc{ref:Khodjasteh2009, ref:McClean2016,
ref:Sharma2020, ref:Movassagh2021, ref:Berberich2024}, while the
non-coherent noise resilience is directly attributed to the
dissipative component. An additional advantage of our
approach is that it does not require ancillary qubits, making it
more qubit-efficient, especially on NISQ devices, which typically
have limited qubit connectivity.

Our numerical examples showed that a D-VQA-based ansatz can
accurately prepare Gibbs states for a broad range of quantum
many-body Hamiltonians across a wide spectrum of temperatures. We
specifically focused on well-known quantum many-body models, such as
the transverse field Ising model and the XY model. In the ideal
noiseless case, our proposed D-VQA-based ansatz achieved at least
the same degree of accuracy as the latest existing VQA-based
ansatzes\cc{ref:Wang2021, ref:Consiglio-XY-Ising2024, ref:Selisko2023}. In the noisy case, within
the low-noise regime, our ansatz significantly outperformed the
state-of-the-art unitary VQA-based ansatzes from
\cRef{ref:Wang2021} and \cRef{ref:Consiglio-XY-Ising2024}.

Although the low-noise regime is not yet available on publicly
accessible quantum platforms, significant research efforts over the
past years have achieved quantum gate fidelities exceeding $99.9\%$,
suggesting that this noise regime will soon be attainable on NISQ
devices\cc{ref:Zahedinejad2015, ref:Kjaergaard2020, ref:Tianyu2023,
ref:Zhang2020}. Moreover, our framework can be combined with either
error mitigation or error correction techniques --- and will require
a lesser reduction in the noise levels to achieve a certain
fidelity, in comparison with unitary VQAs.

Additionally, using D-VQA-based ansatzes, where the dissipative
operation involves parametrized mid-circuit measurements, may help
classical optimization routines to avoid barren
plateaus\cc{ref:McClean2018, ref:Cerezo2021} by significantly
increasing the variance of the gradients with respect to variational
circuit parameters, as suggested by \cRef{ref:Roeland2023}. In this
context, it would be interesting to investigate the extent to which
D-VQA ansatzes exhibit resilience to the different types of barren
plateaus, specifically the noise-induced barren plateaus identified
in \cRef{ref:Wang2021-noise-barren}.

Our work leaves several open questions. It would be valuable to gain
further theoretical understanding of D-VQAs by exploring more
complex scenarios beyond the single-qubit toy model and the
relatively idealized noise model presented here. Specifically,
understanding how entanglement between qubits influences the D-VQA
circuit's resilience to non-coherent errors would be of great
interest. Further study is also needed to understand how this noise
resilience is impacted by non-Markovian noise in more realistic,
potentially time-varying noise models.

It would also be interesting to implement our D-VQA ansatzes on real
quantum hardware, especially as an added advantage of our approach
is that parameterized mid-circuit measurements allow gradient
computations of the cost function without needing additional
circuits, unlike the circuits used in \cRef{ref:Roeland2023}.
Implementation of the D-VQA on a NISQ device could potentially
demonstrate high-fidelity Gibbs state preparation on a quantum
device for systems larger than just a few qubits. Such
demonstrations, coupled with further analytical investigations of
D-VQAs, could provide more evidence for the importance of
dissipative operations for noise resilience, and potentially
influence future error mitigation and correction approaches.

Another direction of both theoretical and practical significance
could be comparing our work with the probabilistic error correction
(PEC) framework introduced in \cRefs{ref:Temme-PEC2017,
ref:Temme-PEC2023, ref:Temme-PEC2023-2}. It would be interesting to
explore how our D-VQA circuits can be integrated with the PEC
framework to further enhance the noise resilience of the resulting
variational circuits. Although no mitigation method, including PEC,
is perfectly scalable, integrating the PEC with the D-VQA framework
might reduce the exponential cost associated with error mitigation.

\section*{Acknowledgements} 

We thank Amit Kam, Rotem Elimelech, Netanel Lindner and Yoav Sagi
for enlightening discussions. 

This research project was supported by the National Research
Foundation, Singapore and A*STAR under its CQT Bridging Grant. It
was also partially supported by the Helen Diller Quantum Center at
the Technion – Israel Institute of Technology.

\newpage 

\bibliography{main.bib}

\begin{thebibliography}{124}%
\makeatletter
\providecommand \@ifxundefined [1]{%
 \@ifx{#1\undefined}
}%
\providecommand \@ifnum [1]{%
 \ifnum #1\expandafter \@firstoftwo
 \else \expandafter \@secondoftwo
 \fi
}%
\providecommand \@ifx [1]{%
 \ifx #1\expandafter \@firstoftwo
 \else \expandafter \@secondoftwo
 \fi
}%
\providecommand \natexlab [1]{#1}%
\providecommand \enquote  [1]{``#1''}%
\providecommand \bibnamefont  [1]{#1}%
\providecommand \bibfnamefont [1]{#1}%
\providecommand \citenamefont [1]{#1}%
\providecommand \href@noop [0]{\@secondoftwo}%
\providecommand \href [0]{\begingroup \@sanitize@url \@href}%
\providecommand \@href[1]{\@@startlink{#1}\@@href}%
\providecommand \@@href[1]{\endgroup#1\@@endlink}%
\providecommand \@sanitize@url [0]{\catcode `\\12\catcode `\$12\catcode `\&12\catcode `\#12\catcode `\^12\catcode `\_12\catcode `\%12\relax}%
\providecommand \@@startlink[1]{}%
\providecommand \@@endlink[0]{}%
\providecommand \url  [0]{\begingroup\@sanitize@url \@url }%
\providecommand \@url [1]{\endgroup\@href {#1}{\urlprefix }}%
\providecommand \urlprefix  [0]{URL }%
\providecommand \Eprint [0]{\href }%
\providecommand \doibase [0]{https://doi.org/}%
\providecommand \selectlanguage [0]{\@gobble}%
\providecommand \bibinfo  [0]{\@secondoftwo}%
\providecommand \bibfield  [0]{\@secondoftwo}%
\providecommand \translation [1]{[#1]}%
\providecommand \BibitemOpen [0]{}%
\providecommand \bibitemStop [0]{}%
\providecommand \bibitemNoStop [0]{.\EOS\space}%
\providecommand \EOS [0]{\spacefactor3000\relax}%
\providecommand \BibitemShut  [1]{\csname bibitem#1\endcsname}%
\let\auto@bib@innerbib\@empty
\bibitem [{\citenamefont {Peruzzo}\ \emph {et~al.}(2014)\citenamefont {Peruzzo}, \citenamefont {McClean}, \citenamefont {Shadbolt}, \citenamefont {Yung}, \citenamefont {Zhou}, \citenamefont {Love}, \citenamefont {Aspuru-Guzik},\ and\ \citenamefont {O’Brien}}]{ref:Peruzzo2014}%
  \BibitemOpen
  \bibfield  {author} {\bibinfo {author} {\bibfnamefont {A.}~\bibnamefont {Peruzzo}}, \bibinfo {author} {\bibfnamefont {J.}~\bibnamefont {McClean}}, \bibinfo {author} {\bibfnamefont {P.}~\bibnamefont {Shadbolt}}, \bibinfo {author} {\bibfnamefont {M.-H.}\ \bibnamefont {Yung}}, \bibinfo {author} {\bibfnamefont {X.-Q.}\ \bibnamefont {Zhou}}, \bibinfo {author} {\bibfnamefont {P.~J.}\ \bibnamefont {Love}}, \bibinfo {author} {\bibfnamefont {A.}~\bibnamefont {Aspuru-Guzik}},\ and\ \bibinfo {author} {\bibfnamefont {J.~L.}\ \bibnamefont {O’Brien}},\ }\bibfield  {title} {\bibinfo {title} {A variational eigenvalue solver on a photonic quantum processor},\ }\bibfield  {journal} {\bibinfo  {journal} {Nature Communications}\ }\textbf {\bibinfo {volume} {5}},\ \href {https://doi.org/10.1038/ncomms5213} {10.1038/ncomms5213} (\bibinfo {year} {2014})\BibitemShut {NoStop}%
\bibitem [{\citenamefont {Farhi}\ \emph {et~al.}(2014)\citenamefont {Farhi}, \citenamefont {Goldstone},\ and\ \citenamefont {Gutmann}}]{ref:Farhi2014}%
  \BibitemOpen
  \bibfield  {author} {\bibinfo {author} {\bibfnamefont {E.}~\bibnamefont {Farhi}}, \bibinfo {author} {\bibfnamefont {J.}~\bibnamefont {Goldstone}},\ and\ \bibinfo {author} {\bibfnamefont {S.}~\bibnamefont {Gutmann}},\ }\href {https://doi.org/10.48550/ARXIV.1411.4028} {\bibinfo {title} {A quantum approximate optimization algorithm}} (\bibinfo {year} {2014})\BibitemShut {NoStop}%
\bibitem [{\citenamefont {Wecker}\ \emph {et~al.}(2015)\citenamefont {Wecker}, \citenamefont {Hastings},\ and\ \citenamefont {Troyer}}]{ref:Troyer2015}%
  \BibitemOpen
  \bibfield  {author} {\bibinfo {author} {\bibfnamefont {D.}~\bibnamefont {Wecker}}, \bibinfo {author} {\bibfnamefont {M.~B.}\ \bibnamefont {Hastings}},\ and\ \bibinfo {author} {\bibfnamefont {M.}~\bibnamefont {Troyer}},\ }\bibfield  {title} {\bibinfo {title} {Progress towards practical quantum variational algorithms},\ }\href {https://doi.org/10.1103/PhysRevA.92.042303} {\bibfield  {journal} {\bibinfo  {journal} {Phys. Rev. A}\ }\textbf {\bibinfo {volume} {92}},\ \bibinfo {pages} {042303} (\bibinfo {year} {2015})}\BibitemShut {NoStop}%
\bibitem [{\citenamefont {Endo}\ \emph {et~al.}(2020)\citenamefont {Endo}, \citenamefont {Sun}, \citenamefont {Li}, \citenamefont {Benjamin},\ and\ \citenamefont {Yuan}}]{ref:Endo2020}%
  \BibitemOpen
  \bibfield  {author} {\bibinfo {author} {\bibfnamefont {S.}~\bibnamefont {Endo}}, \bibinfo {author} {\bibfnamefont {J.}~\bibnamefont {Sun}}, \bibinfo {author} {\bibfnamefont {Y.}~\bibnamefont {Li}}, \bibinfo {author} {\bibfnamefont {S.~C.}\ \bibnamefont {Benjamin}},\ and\ \bibinfo {author} {\bibfnamefont {X.}~\bibnamefont {Yuan}},\ }\bibfield  {title} {\bibinfo {title} {Variational quantum simulation of general processes},\ }\href {https://doi.org/10.1103/PhysRevLett.125.010501} {\bibfield  {journal} {\bibinfo  {journal} {Phys. Rev. Lett.}\ }\textbf {\bibinfo {volume} {125}},\ \bibinfo {pages} {010501} (\bibinfo {year} {2020})}\BibitemShut {NoStop}%
\bibitem [{\citenamefont {Bittel}\ and\ \citenamefont {Kliesch}(2021)}]{ref:Bittel2021}%
  \BibitemOpen
  \bibfield  {author} {\bibinfo {author} {\bibfnamefont {L.}~\bibnamefont {Bittel}}\ and\ \bibinfo {author} {\bibfnamefont {M.}~\bibnamefont {Kliesch}},\ }\bibfield  {title} {\bibinfo {title} {Training variational quantum algorithms is np-hard},\ }\href {https://doi.org/10.1103/PhysRevLett.127.120502} {\bibfield  {journal} {\bibinfo  {journal} {Phys. Rev. Lett.}\ }\textbf {\bibinfo {volume} {127}},\ \bibinfo {pages} {120502} (\bibinfo {year} {2021})}\BibitemShut {NoStop}%
\bibitem [{\citenamefont {Benedetti}\ \emph {et~al.}(2021)\citenamefont {Benedetti}, \citenamefont {Fiorentini},\ and\ \citenamefont {Lubasch}}]{ref:Benedetti2021}%
  \BibitemOpen
  \bibfield  {author} {\bibinfo {author} {\bibfnamefont {M.}~\bibnamefont {Benedetti}}, \bibinfo {author} {\bibfnamefont {M.}~\bibnamefont {Fiorentini}},\ and\ \bibinfo {author} {\bibfnamefont {M.}~\bibnamefont {Lubasch}},\ }\bibfield  {title} {\bibinfo {title} {Hardware-efficient variational quantum algorithms for time evolution},\ }\href {https://doi.org/10.1103/PhysRevResearch.3.033083} {\bibfield  {journal} {\bibinfo  {journal} {Phys. Rev. Res.}\ }\textbf {\bibinfo {volume} {3}},\ \bibinfo {pages} {033083} (\bibinfo {year} {2021})}\BibitemShut {NoStop}%
\bibitem [{\citenamefont {Bonet-Monroig}\ \emph {et~al.}(2023)\citenamefont {Bonet-Monroig}, \citenamefont {Wang}, \citenamefont {Vermetten}, \citenamefont {Senjean}, \citenamefont {Moussa}, \citenamefont {B\"ack}, \citenamefont {Dunjko},\ and\ \citenamefont {O'Brien}}]{ref:Bonet-Monroig2023}%
  \BibitemOpen
  \bibfield  {author} {\bibinfo {author} {\bibfnamefont {X.}~\bibnamefont {Bonet-Monroig}}, \bibinfo {author} {\bibfnamefont {H.}~\bibnamefont {Wang}}, \bibinfo {author} {\bibfnamefont {D.}~\bibnamefont {Vermetten}}, \bibinfo {author} {\bibfnamefont {B.}~\bibnamefont {Senjean}}, \bibinfo {author} {\bibfnamefont {C.}~\bibnamefont {Moussa}}, \bibinfo {author} {\bibfnamefont {T.}~\bibnamefont {B\"ack}}, \bibinfo {author} {\bibfnamefont {V.}~\bibnamefont {Dunjko}},\ and\ \bibinfo {author} {\bibfnamefont {T.~E.}\ \bibnamefont {O'Brien}},\ }\bibfield  {title} {\bibinfo {title} {Performance comparison of optimization methods on variational quantum algorithms},\ }\href {https://doi.org/10.1103/PhysRevA.107.032407} {\bibfield  {journal} {\bibinfo  {journal} {Phys. Rev. A}\ }\textbf {\bibinfo {volume} {107}},\ \bibinfo {pages} {032407} (\bibinfo {year} {2023})}\BibitemShut {NoStop}%
\bibitem [{\citenamefont {Yuan}\ \emph {et~al.}(2019)\citenamefont {Yuan}, \citenamefont {Endo}, \citenamefont {Zhao}, \citenamefont {Li},\ and\ \citenamefont {Benjamin}}]{ref:Yuan2019}%
  \BibitemOpen
  \bibfield  {author} {\bibinfo {author} {\bibfnamefont {X.}~\bibnamefont {Yuan}}, \bibinfo {author} {\bibfnamefont {S.}~\bibnamefont {Endo}}, \bibinfo {author} {\bibfnamefont {Q.}~\bibnamefont {Zhao}}, \bibinfo {author} {\bibfnamefont {Y.}~\bibnamefont {Li}},\ and\ \bibinfo {author} {\bibfnamefont {S.~C.}\ \bibnamefont {Benjamin}},\ }\bibfield  {title} {\bibinfo {title} {Theory of variational quantum simulation},\ }\href {https://doi.org/10.22331/q-2019-10-07-191} {\bibfield  {journal} {\bibinfo  {journal} {Quantum}\ }\textbf {\bibinfo {volume} {3}},\ \bibinfo {pages} {191} (\bibinfo {year} {2019})}\BibitemShut {NoStop}%
\bibitem [{\citenamefont {Hastings}\ \emph {et~al.}(2014)\citenamefont {Hastings}, \citenamefont {Wecker}, \citenamefont {Bauer},\ and\ \citenamefont {Troyer}}]{ref:Hastings2014}%
  \BibitemOpen
  \bibfield  {author} {\bibinfo {author} {\bibfnamefont {M.~B.}\ \bibnamefont {Hastings}}, \bibinfo {author} {\bibfnamefont {D.}~\bibnamefont {Wecker}}, \bibinfo {author} {\bibfnamefont {B.}~\bibnamefont {Bauer}},\ and\ \bibinfo {author} {\bibfnamefont {M.}~\bibnamefont {Troyer}},\ }\href {https://doi.org/10.48550/ARXIV.1403.1539} {\bibinfo {title} {Improving quantum algorithms for quantum chemistry}} (\bibinfo {year} {2014})\BibitemShut {NoStop}%
\bibitem [{\citenamefont {Kandala}\ \emph {et~al.}(2017)\citenamefont {Kandala}, \citenamefont {Mezzacapo}, \citenamefont {Temme}, \citenamefont {Takita}, \citenamefont {Brink}, \citenamefont {Chow},\ and\ \citenamefont {Gambetta}}]{ref:Kandala2017}%
  \BibitemOpen
  \bibfield  {author} {\bibinfo {author} {\bibfnamefont {A.}~\bibnamefont {Kandala}}, \bibinfo {author} {\bibfnamefont {A.}~\bibnamefont {Mezzacapo}}, \bibinfo {author} {\bibfnamefont {K.}~\bibnamefont {Temme}}, \bibinfo {author} {\bibfnamefont {M.}~\bibnamefont {Takita}}, \bibinfo {author} {\bibfnamefont {M.}~\bibnamefont {Brink}}, \bibinfo {author} {\bibfnamefont {J.~M.}\ \bibnamefont {Chow}},\ and\ \bibinfo {author} {\bibfnamefont {J.~M.}\ \bibnamefont {Gambetta}},\ }\bibfield  {title} {\bibinfo {title} {Hardware-efficient variational quantum eigensolver for small molecules and quantum magnets},\ }\href {https://doi.org/10.1038/nature23879} {\bibfield  {journal} {\bibinfo  {journal} {Nature}\ }\textbf {\bibinfo {volume} {549}},\ \bibinfo {pages} {242–246} (\bibinfo {year} {2017})}\BibitemShut {NoStop}%
\bibitem [{\citenamefont {Li}\ and\ \citenamefont {Benjamin}(2017)}]{ref:Ying2017}%
  \BibitemOpen
  \bibfield  {author} {\bibinfo {author} {\bibfnamefont {Y.}~\bibnamefont {Li}}\ and\ \bibinfo {author} {\bibfnamefont {S.~C.}\ \bibnamefont {Benjamin}},\ }\bibfield  {title} {\bibinfo {title} {Efficient variational quantum simulator incorporating active error minimization},\ }\href {https://doi.org/10.1103/PhysRevX.7.021050} {\bibfield  {journal} {\bibinfo  {journal} {Phys. Rev. X}\ }\textbf {\bibinfo {volume} {7}},\ \bibinfo {pages} {021050} (\bibinfo {year} {2017})}\BibitemShut {NoStop}%
\bibitem [{\citenamefont {Hempel}\ \emph {et~al.}(2018)\citenamefont {Hempel}, \citenamefont {Maier}, \citenamefont {Romero}, \citenamefont {McClean}, \citenamefont {Monz}, \citenamefont {Shen}, \citenamefont {Jurcevic}, \citenamefont {Lanyon}, \citenamefont {Love}, \citenamefont {Babbush}, \citenamefont {Aspuru-Guzik}, \citenamefont {Blatt},\ and\ \citenamefont {Roos}}]{ref:Hempel2018}%
  \BibitemOpen
  \bibfield  {author} {\bibinfo {author} {\bibfnamefont {C.}~\bibnamefont {Hempel}}, \bibinfo {author} {\bibfnamefont {C.}~\bibnamefont {Maier}}, \bibinfo {author} {\bibfnamefont {J.}~\bibnamefont {Romero}}, \bibinfo {author} {\bibfnamefont {J.}~\bibnamefont {McClean}}, \bibinfo {author} {\bibfnamefont {T.}~\bibnamefont {Monz}}, \bibinfo {author} {\bibfnamefont {H.}~\bibnamefont {Shen}}, \bibinfo {author} {\bibfnamefont {P.}~\bibnamefont {Jurcevic}}, \bibinfo {author} {\bibfnamefont {B.~P.}\ \bibnamefont {Lanyon}}, \bibinfo {author} {\bibfnamefont {P.}~\bibnamefont {Love}}, \bibinfo {author} {\bibfnamefont {R.}~\bibnamefont {Babbush}}, \bibinfo {author} {\bibfnamefont {A.}~\bibnamefont {Aspuru-Guzik}}, \bibinfo {author} {\bibfnamefont {R.}~\bibnamefont {Blatt}},\ and\ \bibinfo {author} {\bibfnamefont {C.~F.}\ \bibnamefont {Roos}},\ }\bibfield  {title} {\bibinfo {title} {Quantum chemistry calculations on a trapped-ion quantum simulator},\ }\href {https://doi.org/10.1103/PhysRevX.8.031022} {\bibfield  {journal}
  {\bibinfo  {journal} {Phys. Rev. X}\ }\textbf {\bibinfo {volume} {8}},\ \bibinfo {pages} {031022} (\bibinfo {year} {2018})}\BibitemShut {NoStop}%
\bibitem [{\citenamefont {Romero}\ \emph {et~al.}(2018)\citenamefont {Romero}, \citenamefont {Babbush}, \citenamefont {McClean}, \citenamefont {Hempel}, \citenamefont {Love},\ and\ \citenamefont {Aspuru-Guzik}}]{ref:Romero2018}%
  \BibitemOpen
  \bibfield  {author} {\bibinfo {author} {\bibfnamefont {J.}~\bibnamefont {Romero}}, \bibinfo {author} {\bibfnamefont {R.}~\bibnamefont {Babbush}}, \bibinfo {author} {\bibfnamefont {J.~R.}\ \bibnamefont {McClean}}, \bibinfo {author} {\bibfnamefont {C.}~\bibnamefont {Hempel}}, \bibinfo {author} {\bibfnamefont {P.~J.}\ \bibnamefont {Love}},\ and\ \bibinfo {author} {\bibfnamefont {A.}~\bibnamefont {Aspuru-Guzik}},\ }\bibfield  {title} {\bibinfo {title} {Strategies for quantum computing molecular energies using the unitary coupled cluster ansatz},\ }\href {https://doi.org/10.1088/2058-9565/aad3e4} {\bibfield  {journal} {\bibinfo  {journal} {Quantum Science and Technology}\ }\textbf {\bibinfo {volume} {4}},\ \bibinfo {pages} {014008} (\bibinfo {year} {2018})}\BibitemShut {NoStop}%
\bibitem [{\citenamefont {Xu}\ \emph {et~al.}(2021)\citenamefont {Xu}, \citenamefont {Sun}, \citenamefont {Endo}, \citenamefont {Li}, \citenamefont {Benjamin},\ and\ \citenamefont {Yuan}}]{ref:Xu2021}%
  \BibitemOpen
  \bibfield  {author} {\bibinfo {author} {\bibfnamefont {X.}~\bibnamefont {Xu}}, \bibinfo {author} {\bibfnamefont {J.}~\bibnamefont {Sun}}, \bibinfo {author} {\bibfnamefont {S.}~\bibnamefont {Endo}}, \bibinfo {author} {\bibfnamefont {Y.}~\bibnamefont {Li}}, \bibinfo {author} {\bibfnamefont {S.~C.}\ \bibnamefont {Benjamin}},\ and\ \bibinfo {author} {\bibfnamefont {X.}~\bibnamefont {Yuan}},\ }\bibfield  {title} {\bibinfo {title} {Variational algorithms for linear algebra},\ }\href {https://doi.org/10.1016/j.scib.2021.06.023} {\bibfield  {journal} {\bibinfo  {journal} {Science Bulletin}\ }\textbf {\bibinfo {volume} {66}},\ \bibinfo {pages} {2181–2188} (\bibinfo {year} {2021})}\BibitemShut {NoStop}%
\bibitem [{\citenamefont {Bravo-Prieto}\ \emph {et~al.}(2023)\citenamefont {Bravo-Prieto}, \citenamefont {LaRose}, \citenamefont {Cerezo}, \citenamefont {Subasi}, \citenamefont {Cincio},\ and\ \citenamefont {Coles}}]{ref:BravoPrieto2023}%
  \BibitemOpen
  \bibfield  {author} {\bibinfo {author} {\bibfnamefont {C.}~\bibnamefont {Bravo-Prieto}}, \bibinfo {author} {\bibfnamefont {R.}~\bibnamefont {LaRose}}, \bibinfo {author} {\bibfnamefont {M.}~\bibnamefont {Cerezo}}, \bibinfo {author} {\bibfnamefont {Y.}~\bibnamefont {Subasi}}, \bibinfo {author} {\bibfnamefont {L.}~\bibnamefont {Cincio}},\ and\ \bibinfo {author} {\bibfnamefont {P.~J.}\ \bibnamefont {Coles}},\ }\bibfield  {title} {\bibinfo {title} {Variational quantum linear solver},\ }\href {https://doi.org/10.22331/q-2023-11-22-1188} {\bibfield  {journal} {\bibinfo  {journal} {Quantum}\ }\textbf {\bibinfo {volume} {7}},\ \bibinfo {pages} {1188} (\bibinfo {year} {2023})}\BibitemShut {NoStop}%
\bibitem [{\citenamefont {Huang}\ \emph {et~al.}(2019)\citenamefont {Huang}, \citenamefont {Bharti},\ and\ \citenamefont {Rebentrost}}]{ref:Huang2019}%
  \BibitemOpen
  \bibfield  {author} {\bibinfo {author} {\bibfnamefont {H.-Y.}\ \bibnamefont {Huang}}, \bibinfo {author} {\bibfnamefont {K.}~\bibnamefont {Bharti}},\ and\ \bibinfo {author} {\bibfnamefont {P.}~\bibnamefont {Rebentrost}},\ }\href {https://doi.org/10.48550/ARXIV.1909.07344} {\bibinfo {title} {Near-term quantum algorithms for linear systems of equations}} (\bibinfo {year} {2019})\BibitemShut {NoStop}%
\bibitem [{\citenamefont {Nakanishi}\ \emph {et~al.}(2019)\citenamefont {Nakanishi}, \citenamefont {Mitarai},\ and\ \citenamefont {Fujii}}]{ref:Nakanishi2019}%
  \BibitemOpen
  \bibfield  {author} {\bibinfo {author} {\bibfnamefont {K.~M.}\ \bibnamefont {Nakanishi}}, \bibinfo {author} {\bibfnamefont {K.}~\bibnamefont {Mitarai}},\ and\ \bibinfo {author} {\bibfnamefont {K.}~\bibnamefont {Fujii}},\ }\bibfield  {title} {\bibinfo {title} {Subspace-search variational quantum eigensolver for excited states},\ }\href {https://doi.org/10.1103/PhysRevResearch.1.033062} {\bibfield  {journal} {\bibinfo  {journal} {Phys. Rev. Res.}\ }\textbf {\bibinfo {volume} {1}},\ \bibinfo {pages} {033062} (\bibinfo {year} {2019})}\BibitemShut {NoStop}%
\bibitem [{\citenamefont {Cerezo}\ \emph {et~al.}(2022)\citenamefont {Cerezo}, \citenamefont {Sharma}, \citenamefont {Arrasmith},\ and\ \citenamefont {Coles}}]{ref:Cerezo2022}%
  \BibitemOpen
  \bibfield  {author} {\bibinfo {author} {\bibfnamefont {M.}~\bibnamefont {Cerezo}}, \bibinfo {author} {\bibfnamefont {K.}~\bibnamefont {Sharma}}, \bibinfo {author} {\bibfnamefont {A.}~\bibnamefont {Arrasmith}},\ and\ \bibinfo {author} {\bibfnamefont {P.~J.}\ \bibnamefont {Coles}},\ }\bibfield  {title} {\bibinfo {title} {Variational quantum state eigensolver},\ }\bibfield  {journal} {\bibinfo  {journal} {npj Quantum Information}\ }\textbf {\bibinfo {volume} {8}},\ \href {https://doi.org/10.1038/s41534-022-00611-6} {10.1038/s41534-022-00611-6} (\bibinfo {year} {2022})\BibitemShut {NoStop}%
\bibitem [{\citenamefont {Wang}\ \emph {et~al.}(2021{\natexlab{a}})\citenamefont {Wang}, \citenamefont {Song},\ and\ \citenamefont {Wang}}]{ref:Wang2021-SVD}%
  \BibitemOpen
  \bibfield  {author} {\bibinfo {author} {\bibfnamefont {X.}~\bibnamefont {Wang}}, \bibinfo {author} {\bibfnamefont {Z.}~\bibnamefont {Song}},\ and\ \bibinfo {author} {\bibfnamefont {Y.}~\bibnamefont {Wang}},\ }\bibfield  {title} {\bibinfo {title} {Variational quantum singular value decomposition},\ }\href {https://doi.org/10.22331/q-2021-06-29-483} {\bibfield  {journal} {\bibinfo  {journal} {Quantum}\ }\textbf {\bibinfo {volume} {5}},\ \bibinfo {pages} {483} (\bibinfo {year} {2021}{\natexlab{a}})}\BibitemShut {NoStop}%
\bibitem [{\citenamefont {Chen}\ \emph {et~al.}(2021)\citenamefont {Chen}, \citenamefont {Song}, \citenamefont {Zhao},\ and\ \citenamefont {Wang}}]{ref:Chen2021}%
  \BibitemOpen
  \bibfield  {author} {\bibinfo {author} {\bibfnamefont {R.}~\bibnamefont {Chen}}, \bibinfo {author} {\bibfnamefont {Z.}~\bibnamefont {Song}}, \bibinfo {author} {\bibfnamefont {X.}~\bibnamefont {Zhao}},\ and\ \bibinfo {author} {\bibfnamefont {X.}~\bibnamefont {Wang}},\ }\bibfield  {title} {\bibinfo {title} {Variational quantum algorithms for trace distance and fidelity estimation},\ }\href {https://doi.org/10.1088/2058-9565/ac38ba} {\bibfield  {journal} {\bibinfo  {journal} {Quantum Science and Technology}\ }\textbf {\bibinfo {volume} {7}},\ \bibinfo {pages} {015019} (\bibinfo {year} {2021})}\BibitemShut {NoStop}%
\bibitem [{\citenamefont {McArdle}\ \emph {et~al.}(2019)\citenamefont {McArdle}, \citenamefont {Jones}, \citenamefont {Endo}, \citenamefont {Li}, \citenamefont {Benjamin},\ and\ \citenamefont {Yuan}}]{ref:McArdle2019}%
  \BibitemOpen
  \bibfield  {author} {\bibinfo {author} {\bibfnamefont {S.}~\bibnamefont {McArdle}}, \bibinfo {author} {\bibfnamefont {T.}~\bibnamefont {Jones}}, \bibinfo {author} {\bibfnamefont {S.}~\bibnamefont {Endo}}, \bibinfo {author} {\bibfnamefont {Y.}~\bibnamefont {Li}}, \bibinfo {author} {\bibfnamefont {S.~C.}\ \bibnamefont {Benjamin}},\ and\ \bibinfo {author} {\bibfnamefont {X.}~\bibnamefont {Yuan}},\ }\bibfield  {title} {\bibinfo {title} {Variational ansatz-based quantum simulation of imaginary time evolution},\ }\bibfield  {journal} {\bibinfo  {journal} {npj Quantum Information}\ }\textbf {\bibinfo {volume} {5}},\ \href {https://doi.org/10.1038/s41534-019-0187-2} {10.1038/s41534-019-0187-2} (\bibinfo {year} {2019})\BibitemShut {NoStop}%
\bibitem [{\citenamefont {McClean}\ \emph {et~al.}(2016)\citenamefont {McClean}, \citenamefont {Romero}, \citenamefont {Babbush},\ and\ \citenamefont {Aspuru-Guzik}}]{ref:McClean2016}%
  \BibitemOpen
  \bibfield  {author} {\bibinfo {author} {\bibfnamefont {J.~R.}\ \bibnamefont {McClean}}, \bibinfo {author} {\bibfnamefont {J.}~\bibnamefont {Romero}}, \bibinfo {author} {\bibfnamefont {R.}~\bibnamefont {Babbush}},\ and\ \bibinfo {author} {\bibfnamefont {A.}~\bibnamefont {Aspuru-Guzik}},\ }\bibfield  {title} {\bibinfo {title} {The theory of variational hybrid quantum-classical algorithms},\ }\href {https://doi.org/10.1088/1367-2630/18/2/023023} {\bibfield  {journal} {\bibinfo  {journal} {New Journal of Physics}\ }\textbf {\bibinfo {volume} {18}},\ \bibinfo {pages} {023023} (\bibinfo {year} {2016})}\BibitemShut {NoStop}%
\bibitem [{\citenamefont {Grimsley}\ \emph {et~al.}(2019)\citenamefont {Grimsley}, \citenamefont {Economou}, \citenamefont {Barnes},\ and\ \citenamefont {Mayhall}}]{ref:Grimsley2019}%
  \BibitemOpen
  \bibfield  {author} {\bibinfo {author} {\bibfnamefont {H.~R.}\ \bibnamefont {Grimsley}}, \bibinfo {author} {\bibfnamefont {S.~E.}\ \bibnamefont {Economou}}, \bibinfo {author} {\bibfnamefont {E.}~\bibnamefont {Barnes}},\ and\ \bibinfo {author} {\bibfnamefont {N.~J.}\ \bibnamefont {Mayhall}},\ }\bibfield  {title} {\bibinfo {title} {An adaptive variational algorithm for exact molecular simulations on a quantum computer},\ }\bibfield  {journal} {\bibinfo  {journal} {Nature Communications}\ }\textbf {\bibinfo {volume} {10}},\ \href {https://doi.org/10.1038/s41467-019-10988-2} {10.1038/s41467-019-10988-2} (\bibinfo {year} {2019})\BibitemShut {NoStop}%
\bibitem [{\citenamefont {Khodjasteh}\ and\ \citenamefont {Viola}(2009)}]{ref:Khodjasteh2009}%
  \BibitemOpen
  \bibfield  {author} {\bibinfo {author} {\bibfnamefont {K.}~\bibnamefont {Khodjasteh}}\ and\ \bibinfo {author} {\bibfnamefont {L.}~\bibnamefont {Viola}},\ }\bibfield  {title} {\bibinfo {title} {Dynamically error-corrected gates for universal quantum computation},\ }\href {https://doi.org/10.1103/PhysRevLett.102.080501} {\bibfield  {journal} {\bibinfo  {journal} {Phys. Rev. Lett.}\ }\textbf {\bibinfo {volume} {102}},\ \bibinfo {pages} {080501} (\bibinfo {year} {2009})}\BibitemShut {NoStop}%
\bibitem [{\citenamefont {Sharma}\ \emph {et~al.}(2020)\citenamefont {Sharma}, \citenamefont {Khatri}, \citenamefont {Cerezo},\ and\ \citenamefont {Coles}}]{ref:Sharma2020}%
  \BibitemOpen
  \bibfield  {author} {\bibinfo {author} {\bibfnamefont {K.}~\bibnamefont {Sharma}}, \bibinfo {author} {\bibfnamefont {S.}~\bibnamefont {Khatri}}, \bibinfo {author} {\bibfnamefont {M.}~\bibnamefont {Cerezo}},\ and\ \bibinfo {author} {\bibfnamefont {P.~J.}\ \bibnamefont {Coles}},\ }\bibfield  {title} {\bibinfo {title} {Noise resilience of variational quantum compiling},\ }\href {https://doi.org/10.1088/1367-2630/ab784c} {\bibfield  {journal} {\bibinfo  {journal} {New Journal of Physics}\ }\textbf {\bibinfo {volume} {22}},\ \bibinfo {pages} {043006} (\bibinfo {year} {2020})}\BibitemShut {NoStop}%
\bibitem [{\citenamefont {Shtanko}\ and\ \citenamefont {Movassagh}(2021)}]{ref:Movassagh2021}%
  \BibitemOpen
  \bibfield  {author} {\bibinfo {author} {\bibfnamefont {O.}~\bibnamefont {Shtanko}}\ and\ \bibinfo {author} {\bibfnamefont {R.}~\bibnamefont {Movassagh}},\ }\href {https://doi.org/10.48550/ARXIV.2112.14688} {\bibinfo {title} {Preparing thermal states on noiseless and noisy programmable quantum processors}} (\bibinfo {year} {2021})\BibitemShut {NoStop}%
\bibitem [{\citenamefont {Berberich}\ \emph {et~al.}(2024)\citenamefont {Berberich}, \citenamefont {Fink},\ and\ \citenamefont {Holm}}]{ref:Berberich2024}%
  \BibitemOpen
  \bibfield  {author} {\bibinfo {author} {\bibfnamefont {J.}~\bibnamefont {Berberich}}, \bibinfo {author} {\bibfnamefont {D.}~\bibnamefont {Fink}},\ and\ \bibinfo {author} {\bibfnamefont {C.}~\bibnamefont {Holm}},\ }\bibfield  {title} {\bibinfo {title} {Robustness of quantum algorithms against coherent control errors},\ }\href {https://doi.org/10.1103/PhysRevA.109.012417} {\bibfield  {journal} {\bibinfo  {journal} {Phys. Rev. A}\ }\textbf {\bibinfo {volume} {109}},\ \bibinfo {pages} {012417} (\bibinfo {year} {2024})}\BibitemShut {NoStop}%
\bibitem [{\citenamefont {Uvarov}\ and\ \citenamefont {Biamonte}(2021)}]{ref:Uvarov2021}%
  \BibitemOpen
  \bibfield  {author} {\bibinfo {author} {\bibfnamefont {A.~V.}\ \bibnamefont {Uvarov}}\ and\ \bibinfo {author} {\bibfnamefont {J.~D.}\ \bibnamefont {Biamonte}},\ }\bibfield  {title} {\bibinfo {title} {On barren plateaus and cost function locality in variational quantum algorithms},\ }\href {https://doi.org/10.1088/1751-8121/abfac7} {\bibfield  {journal} {\bibinfo  {journal} {Journal of Physics A: Mathematical and Theoretical}\ }\textbf {\bibinfo {volume} {54}},\ \bibinfo {pages} {245301} (\bibinfo {year} {2021})}\BibitemShut {NoStop}%
\bibitem [{\citenamefont {Larocca}\ \emph {et~al.}(2024)\citenamefont {Larocca}, \citenamefont {Thanasilp}, \citenamefont {Wang}, \citenamefont {Sharma}, \citenamefont {Biamonte}, \citenamefont {Coles}, \citenamefont {Cincio}, \citenamefont {McClean}, \citenamefont {Holmes},\ and\ \citenamefont {Cerezo}}]{ref:Larocca2024}%
  \BibitemOpen
  \bibfield  {author} {\bibinfo {author} {\bibfnamefont {M.}~\bibnamefont {Larocca}}, \bibinfo {author} {\bibfnamefont {S.}~\bibnamefont {Thanasilp}}, \bibinfo {author} {\bibfnamefont {S.}~\bibnamefont {Wang}}, \bibinfo {author} {\bibfnamefont {K.}~\bibnamefont {Sharma}}, \bibinfo {author} {\bibfnamefont {J.}~\bibnamefont {Biamonte}}, \bibinfo {author} {\bibfnamefont {P.~J.}\ \bibnamefont {Coles}}, \bibinfo {author} {\bibfnamefont {L.}~\bibnamefont {Cincio}}, \bibinfo {author} {\bibfnamefont {J.~R.}\ \bibnamefont {McClean}}, \bibinfo {author} {\bibfnamefont {Z.}~\bibnamefont {Holmes}},\ and\ \bibinfo {author} {\bibfnamefont {M.}~\bibnamefont {Cerezo}},\ }\href {https://doi.org/10.48550/ARXIV.2405.00781} {\bibinfo {title} {A review of barren plateaus in variational quantum computing}} (\bibinfo {year} {2024})\BibitemShut {NoStop}%
\bibitem [{\citenamefont {Gonthier}\ \emph {et~al.}(2022)\citenamefont {Gonthier}, \citenamefont {Radin}, \citenamefont {Buda}, \citenamefont {Doskocil}, \citenamefont {Abuan},\ and\ \citenamefont {Romero}}]{ref:Gonthier2022}%
  \BibitemOpen
  \bibfield  {author} {\bibinfo {author} {\bibfnamefont {J.~F.}\ \bibnamefont {Gonthier}}, \bibinfo {author} {\bibfnamefont {M.~D.}\ \bibnamefont {Radin}}, \bibinfo {author} {\bibfnamefont {C.}~\bibnamefont {Buda}}, \bibinfo {author} {\bibfnamefont {E.~J.}\ \bibnamefont {Doskocil}}, \bibinfo {author} {\bibfnamefont {C.~M.}\ \bibnamefont {Abuan}},\ and\ \bibinfo {author} {\bibfnamefont {J.}~\bibnamefont {Romero}},\ }\bibfield  {title} {\bibinfo {title} {Measurements as a roadblock to near-term practical quantum advantage in chemistry: Resource analysis},\ }\href {https://doi.org/10.1103/PhysRevResearch.4.033154} {\bibfield  {journal} {\bibinfo  {journal} {Phys. Rev. Res.}\ }\textbf {\bibinfo {volume} {4}},\ \bibinfo {pages} {033154} (\bibinfo {year} {2022})}\BibitemShut {NoStop}%
\bibitem [{\citenamefont {Anschuetz}\ and\ \citenamefont {Kiani}(2022)}]{ref:Anschuetz2022}%
  \BibitemOpen
  \bibfield  {author} {\bibinfo {author} {\bibfnamefont {E.~R.}\ \bibnamefont {Anschuetz}}\ and\ \bibinfo {author} {\bibfnamefont {B.~T.}\ \bibnamefont {Kiani}},\ }\bibfield  {title} {\bibinfo {title} {Quantum variational algorithms are swamped with traps},\ }\bibfield  {journal} {\bibinfo  {journal} {Nature Communications}\ }\textbf {\bibinfo {volume} {13}},\ \href {https://doi.org/10.1038/s41467-022-35364-5} {10.1038/s41467-022-35364-5} (\bibinfo {year} {2022})\BibitemShut {NoStop}%
\bibitem [{\citenamefont {Lee}\ \emph {et~al.}(2023)\citenamefont {Lee}, \citenamefont {Lee}, \citenamefont {Zhai}, \citenamefont {Tong}, \citenamefont {Dalzell}, \citenamefont {Kumar}, \citenamefont {Helms}, \citenamefont {Gray}, \citenamefont {Cui}, \citenamefont {Liu}, \citenamefont {Kastoryano}, \citenamefont {Babbush}, \citenamefont {Preskill}, \citenamefont {Reichman}, \citenamefont {Campbell}, \citenamefont {Valeev}, \citenamefont {Lin},\ and\ \citenamefont {Chan}}]{ref:Lee2023-noexpVQE}%
  \BibitemOpen
  \bibfield  {author} {\bibinfo {author} {\bibfnamefont {S.}~\bibnamefont {Lee}}, \bibinfo {author} {\bibfnamefont {J.}~\bibnamefont {Lee}}, \bibinfo {author} {\bibfnamefont {H.}~\bibnamefont {Zhai}}, \bibinfo {author} {\bibfnamefont {Y.}~\bibnamefont {Tong}}, \bibinfo {author} {\bibfnamefont {A.~M.}\ \bibnamefont {Dalzell}}, \bibinfo {author} {\bibfnamefont {A.}~\bibnamefont {Kumar}}, \bibinfo {author} {\bibfnamefont {P.}~\bibnamefont {Helms}}, \bibinfo {author} {\bibfnamefont {J.}~\bibnamefont {Gray}}, \bibinfo {author} {\bibfnamefont {Z.-H.}\ \bibnamefont {Cui}}, \bibinfo {author} {\bibfnamefont {W.}~\bibnamefont {Liu}}, \bibinfo {author} {\bibfnamefont {M.}~\bibnamefont {Kastoryano}}, \bibinfo {author} {\bibfnamefont {R.}~\bibnamefont {Babbush}}, \bibinfo {author} {\bibfnamefont {J.}~\bibnamefont {Preskill}}, \bibinfo {author} {\bibfnamefont {D.~R.}\ \bibnamefont {Reichman}}, \bibinfo {author} {\bibfnamefont {E.~T.}\ \bibnamefont {Campbell}}, \bibinfo {author} {\bibfnamefont {E.~F.}\ \bibnamefont
  {Valeev}}, \bibinfo {author} {\bibfnamefont {L.}~\bibnamefont {Lin}},\ and\ \bibinfo {author} {\bibfnamefont {G.~K.-L.}\ \bibnamefont {Chan}},\ }\bibfield  {title} {\bibinfo {title} {Evaluating the evidence for exponential quantum advantage in ground-state quantum chemistry},\ }\href {https://doi.org/10.1038/s41467-023-37587-6} {\bibfield  {journal} {\bibinfo  {journal} {Nature Communications}\ }\textbf {\bibinfo {volume} {14}},\ \bibinfo {pages} {1952} (\bibinfo {year} {2023})}\BibitemShut {NoStop}%
\bibitem [{\citenamefont {van~den Berg}\ \emph {et~al.}(2023)\citenamefont {van~den Berg}, \citenamefont {Minev}, \citenamefont {Kandala},\ and\ \citenamefont {Temme}}]{ref:Temme-PEC2023}%
  \BibitemOpen
  \bibfield  {author} {\bibinfo {author} {\bibfnamefont {E.}~\bibnamefont {van~den Berg}}, \bibinfo {author} {\bibfnamefont {Z.~K.}\ \bibnamefont {Minev}}, \bibinfo {author} {\bibfnamefont {A.}~\bibnamefont {Kandala}},\ and\ \bibinfo {author} {\bibfnamefont {K.}~\bibnamefont {Temme}},\ }\bibfield  {title} {\bibinfo {title} {Probabilistic error cancellation with sparse pauli–lindblad models on noisy quantum processors},\ }\href {https://doi.org/10.1038/s41567-023-02042-2} {\bibfield  {journal} {\bibinfo  {journal} {Nature Physics}\ }\textbf {\bibinfo {volume} {19}},\ \bibinfo {pages} {1116–1121} (\bibinfo {year} {2023})}\BibitemShut {NoStop}%
\bibitem [{\citenamefont {Seif}\ \emph {et~al.}(2023)\citenamefont {Seif}, \citenamefont {Cian}, \citenamefont {Zhou}, \citenamefont {Chen},\ and\ \citenamefont {Jiang}}]{ref:Seif2023}%
  \BibitemOpen
  \bibfield  {author} {\bibinfo {author} {\bibfnamefont {A.}~\bibnamefont {Seif}}, \bibinfo {author} {\bibfnamefont {Z.-P.}\ \bibnamefont {Cian}}, \bibinfo {author} {\bibfnamefont {S.}~\bibnamefont {Zhou}}, \bibinfo {author} {\bibfnamefont {S.}~\bibnamefont {Chen}},\ and\ \bibinfo {author} {\bibfnamefont {L.}~\bibnamefont {Jiang}},\ }\bibfield  {title} {\bibinfo {title} {Shadow distillation: Quantum error mitigation with classical shadows for near-term quantum processors},\ }\href {https://doi.org/10.1103/PRXQuantum.4.010303} {\bibfield  {journal} {\bibinfo  {journal} {PRX Quantum}\ }\textbf {\bibinfo {volume} {4}},\ \bibinfo {pages} {010303} (\bibinfo {year} {2023})}\BibitemShut {NoStop}%
\bibitem [{\citenamefont {Strikis}\ \emph {et~al.}(2021)\citenamefont {Strikis}, \citenamefont {Qin}, \citenamefont {Chen}, \citenamefont {Benjamin},\ and\ \citenamefont {Li}}]{ref:Strikis2021}%
  \BibitemOpen
  \bibfield  {author} {\bibinfo {author} {\bibfnamefont {A.}~\bibnamefont {Strikis}}, \bibinfo {author} {\bibfnamefont {D.}~\bibnamefont {Qin}}, \bibinfo {author} {\bibfnamefont {Y.}~\bibnamefont {Chen}}, \bibinfo {author} {\bibfnamefont {S.~C.}\ \bibnamefont {Benjamin}},\ and\ \bibinfo {author} {\bibfnamefont {Y.}~\bibnamefont {Li}},\ }\bibfield  {title} {\bibinfo {title} {Learning-based quantum error mitigation},\ }\href {https://doi.org/10.1103/PRXQuantum.2.040330} {\bibfield  {journal} {\bibinfo  {journal} {PRX Quantum}\ }\textbf {\bibinfo {volume} {2}},\ \bibinfo {pages} {040330} (\bibinfo {year} {2021})}\BibitemShut {NoStop}%
\bibitem [{\citenamefont {Cai}(2021)}]{ref:Cai2021}%
  \BibitemOpen
  \bibfield  {author} {\bibinfo {author} {\bibfnamefont {Z.}~\bibnamefont {Cai}},\ }\bibfield  {title} {\bibinfo {title} {Quantum error mitigation using symmetry expansion},\ }\href {https://doi.org/10.22331/q-2021-09-21-548} {\bibfield  {journal} {\bibinfo  {journal} {Quantum}\ }\textbf {\bibinfo {volume} {5}},\ \bibinfo {pages} {548} (\bibinfo {year} {2021})}\BibitemShut {NoStop}%
\bibitem [{\citenamefont {Huggins}\ \emph {et~al.}(2021)\citenamefont {Huggins}, \citenamefont {McArdle}, \citenamefont {O'Brien}, \citenamefont {Lee}, \citenamefont {Rubin}, \citenamefont {Boixo}, \citenamefont {Whaley}, \citenamefont {Babbush},\ and\ \citenamefont {McClean}}]{ref:Huggins2021}%
  \BibitemOpen
  \bibfield  {author} {\bibinfo {author} {\bibfnamefont {W.~J.}\ \bibnamefont {Huggins}}, \bibinfo {author} {\bibfnamefont {S.}~\bibnamefont {McArdle}}, \bibinfo {author} {\bibfnamefont {T.~E.}\ \bibnamefont {O'Brien}}, \bibinfo {author} {\bibfnamefont {J.}~\bibnamefont {Lee}}, \bibinfo {author} {\bibfnamefont {N.~C.}\ \bibnamefont {Rubin}}, \bibinfo {author} {\bibfnamefont {S.}~\bibnamefont {Boixo}}, \bibinfo {author} {\bibfnamefont {K.~B.}\ \bibnamefont {Whaley}}, \bibinfo {author} {\bibfnamefont {R.}~\bibnamefont {Babbush}},\ and\ \bibinfo {author} {\bibfnamefont {J.~R.}\ \bibnamefont {McClean}},\ }\bibfield  {title} {\bibinfo {title} {Virtual distillation for quantum error mitigation},\ }\href {https://doi.org/10.1103/PhysRevX.11.041036} {\bibfield  {journal} {\bibinfo  {journal} {Phys. Rev. X}\ }\textbf {\bibinfo {volume} {11}},\ \bibinfo {pages} {041036} (\bibinfo {year} {2021})}\BibitemShut {NoStop}%
\bibitem [{\citenamefont {Ravi}\ \emph {et~al.}(2022)\citenamefont {Ravi}, \citenamefont {Smith}, \citenamefont {Gokhale}, \citenamefont {Mari}, \citenamefont {Earnest}, \citenamefont {Javadi-Abhari},\ and\ \citenamefont {Chong}}]{ref:Ravi2022}%
  \BibitemOpen
  \bibfield  {author} {\bibinfo {author} {\bibfnamefont {G.~S.}\ \bibnamefont {Ravi}}, \bibinfo {author} {\bibfnamefont {K.~N.}\ \bibnamefont {Smith}}, \bibinfo {author} {\bibfnamefont {P.}~\bibnamefont {Gokhale}}, \bibinfo {author} {\bibfnamefont {A.}~\bibnamefont {Mari}}, \bibinfo {author} {\bibfnamefont {N.}~\bibnamefont {Earnest}}, \bibinfo {author} {\bibfnamefont {A.}~\bibnamefont {Javadi-Abhari}},\ and\ \bibinfo {author} {\bibfnamefont {F.~T.}\ \bibnamefont {Chong}},\ }\bibfield  {title} {\bibinfo {title} {Vaqem: A variational approach to quantum error mitigation},\ }in\ \href {https://doi.org/10.1109/hpca53966.2022.00029} {\emph {\bibinfo {booktitle} {2022 IEEE International Symposium on High-Performance Computer Architecture (HPCA)}}}\ (\bibinfo  {publisher} {IEEE},\ \bibinfo {year} {2022})\BibitemShut {NoStop}%
\bibitem [{\citenamefont {Ilin}\ and\ \citenamefont {Arad}(2024)}]{ref:Ilin2024}%
  \BibitemOpen
  \bibfield  {author} {\bibinfo {author} {\bibfnamefont {Y.}~\bibnamefont {Ilin}}\ and\ \bibinfo {author} {\bibfnamefont {I.}~\bibnamefont {Arad}},\ }\bibfield  {title} {\bibinfo {title} {Learning a quantum channel from its steady-state},\ }\href {https://doi.org/10.1088/1367-2630/ad5464} {\bibfield  {journal} {\bibinfo  {journal} {New Journal of Physics}\ }\textbf {\bibinfo {volume} {26}},\ \bibinfo {pages} {073003} (\bibinfo {year} {2024})}\BibitemShut {NoStop}%
\bibitem [{\citenamefont {Wiersema}\ \emph {et~al.}(2023)\citenamefont {Wiersema}, \citenamefont {Zhou}, \citenamefont {Carrasquilla},\ and\ \citenamefont {Kim}}]{ref:Roeland2023}%
  \BibitemOpen
  \bibfield  {author} {\bibinfo {author} {\bibfnamefont {R.}~\bibnamefont {Wiersema}}, \bibinfo {author} {\bibfnamefont {C.}~\bibnamefont {Zhou}}, \bibinfo {author} {\bibfnamefont {J.~F.}\ \bibnamefont {Carrasquilla}},\ and\ \bibinfo {author} {\bibfnamefont {Y.~B.}\ \bibnamefont {Kim}},\ }\bibfield  {title} {\bibinfo {title} {{Measurement-induced entanglement phase transitions in variational quantum circuits}},\ }\href {https://doi.org/10.21468/SciPostPhys.14.6.147} {\bibfield  {journal} {\bibinfo  {journal} {SciPost Phys.}\ }\textbf {\bibinfo {volume} {14}},\ \bibinfo {pages} {147} (\bibinfo {year} {2023})}\BibitemShut {NoStop}%
\bibitem [{\citenamefont {Botelho}\ \emph {et~al.}(2022)\citenamefont {Botelho}, \citenamefont {Glos}, \citenamefont {Kundu}, \citenamefont {Miszczak}, \citenamefont {Salehi},\ and\ \citenamefont {Zimbor\'as}}]{ref:Botelho2022}%
  \BibitemOpen
  \bibfield  {author} {\bibinfo {author} {\bibfnamefont {L.}~\bibnamefont {Botelho}}, \bibinfo {author} {\bibfnamefont {A.}~\bibnamefont {Glos}}, \bibinfo {author} {\bibfnamefont {A.}~\bibnamefont {Kundu}}, \bibinfo {author} {\bibfnamefont {J.~A.}\ \bibnamefont {Miszczak}}, \bibinfo {author} {\bibfnamefont {O.}~\bibnamefont {Salehi}},\ and\ \bibinfo {author} {\bibfnamefont {Z.}~\bibnamefont {Zimbor\'as}},\ }\bibfield  {title} {\bibinfo {title} {Error mitigation for variational quantum algorithms through mid-circuit measurements},\ }\href {https://doi.org/10.1103/PhysRevA.105.022441} {\bibfield  {journal} {\bibinfo  {journal} {Phys. Rev. A}\ }\textbf {\bibinfo {volume} {105}},\ \bibinfo {pages} {022441} (\bibinfo {year} {2022})}\BibitemShut {NoStop}%
\bibitem [{\citenamefont {Volya}\ and\ \citenamefont {Mishra}(2024)}]{ref:Volya2024}%
  \BibitemOpen
  \bibfield  {author} {\bibinfo {author} {\bibfnamefont {D.}~\bibnamefont {Volya}}\ and\ \bibinfo {author} {\bibfnamefont {P.}~\bibnamefont {Mishra}},\ }\bibfield  {title} {\bibinfo {title} {State preparation on quantum computers via quantum steering},\ }\href {https://doi.org/10.1109/tqe.2024.3358193} {\bibfield  {journal} {\bibinfo  {journal} {IEEE Transactions on Quantum Engineering}\ }\textbf {\bibinfo {volume} {5}},\ \bibinfo {pages} {1–14} (\bibinfo {year} {2024})}\BibitemShut {NoStop}%
\bibitem [{\citenamefont {Chertkov}\ \emph {et~al.}(2023)\citenamefont {Chertkov}, \citenamefont {Cheng}, \citenamefont {Potter}, \citenamefont {Gopalakrishnan}, \citenamefont {Gatterman}, \citenamefont {Gerber}, \citenamefont {Gilmore}, \citenamefont {Gresh}, \citenamefont {Hall}, \citenamefont {Hankin}, \citenamefont {Matheny}, \citenamefont {Mengle}, \citenamefont {Hayes}, \citenamefont {Neyenhuis}, \citenamefont {Stutz},\ and\ \citenamefont {Foss-Feig}}]{ref:Chertkov2023}%
  \BibitemOpen
  \bibfield  {author} {\bibinfo {author} {\bibfnamefont {E.}~\bibnamefont {Chertkov}}, \bibinfo {author} {\bibfnamefont {Z.}~\bibnamefont {Cheng}}, \bibinfo {author} {\bibfnamefont {A.~C.}\ \bibnamefont {Potter}}, \bibinfo {author} {\bibfnamefont {S.}~\bibnamefont {Gopalakrishnan}}, \bibinfo {author} {\bibfnamefont {T.~M.}\ \bibnamefont {Gatterman}}, \bibinfo {author} {\bibfnamefont {J.~A.}\ \bibnamefont {Gerber}}, \bibinfo {author} {\bibfnamefont {K.}~\bibnamefont {Gilmore}}, \bibinfo {author} {\bibfnamefont {D.}~\bibnamefont {Gresh}}, \bibinfo {author} {\bibfnamefont {A.}~\bibnamefont {Hall}}, \bibinfo {author} {\bibfnamefont {A.}~\bibnamefont {Hankin}}, \bibinfo {author} {\bibfnamefont {M.}~\bibnamefont {Matheny}}, \bibinfo {author} {\bibfnamefont {T.}~\bibnamefont {Mengle}}, \bibinfo {author} {\bibfnamefont {D.}~\bibnamefont {Hayes}}, \bibinfo {author} {\bibfnamefont {B.}~\bibnamefont {Neyenhuis}}, \bibinfo {author} {\bibfnamefont {R.}~\bibnamefont {Stutz}},\ and\ \bibinfo {author} {\bibfnamefont
  {M.}~\bibnamefont {Foss-Feig}},\ }\bibfield  {title} {\bibinfo {title} {Characterizing a non-equilibrium phase transition on a quantum computer},\ }\href {https://doi.org/10.1038/s41567-023-02199-w} {\bibfield  {journal} {\bibinfo  {journal} {Nature Physics}\ }\textbf {\bibinfo {volume} {19}},\ \bibinfo {pages} {1799–1804} (\bibinfo {year} {2023})}\BibitemShut {NoStop}%
\bibitem [{\citenamefont {DeCross}\ \emph {et~al.}(2023)\citenamefont {DeCross}, \citenamefont {Chertkov}, \citenamefont {Kohagen},\ and\ \citenamefont {Foss-Feig}}]{ref:DeCross2023}%
  \BibitemOpen
  \bibfield  {author} {\bibinfo {author} {\bibfnamefont {M.}~\bibnamefont {DeCross}}, \bibinfo {author} {\bibfnamefont {E.}~\bibnamefont {Chertkov}}, \bibinfo {author} {\bibfnamefont {M.}~\bibnamefont {Kohagen}},\ and\ \bibinfo {author} {\bibfnamefont {M.}~\bibnamefont {Foss-Feig}},\ }\bibfield  {title} {\bibinfo {title} {Qubit-reuse compilation with mid-circuit measurement and reset},\ }\href {https://doi.org/10.1103/PhysRevX.13.041057} {\bibfield  {journal} {\bibinfo  {journal} {Phys. Rev. X}\ }\textbf {\bibinfo {volume} {13}},\ \bibinfo {pages} {041057} (\bibinfo {year} {2023})}\BibitemShut {NoStop}%
\bibitem [{\citenamefont {Hua}\ \emph {et~al.}(2022)\citenamefont {Hua}, \citenamefont {Jin}, \citenamefont {Chen}, \citenamefont {Vittal}, \citenamefont {Krsulich}, \citenamefont {Bishop}, \citenamefont {Lapeyre}, \citenamefont {Javadi-Abhari},\ and\ \citenamefont {Zhang}}]{ref:Hua2022}%
  \BibitemOpen
  \bibfield  {author} {\bibinfo {author} {\bibfnamefont {F.}~\bibnamefont {Hua}}, \bibinfo {author} {\bibfnamefont {Y.}~\bibnamefont {Jin}}, \bibinfo {author} {\bibfnamefont {Y.}~\bibnamefont {Chen}}, \bibinfo {author} {\bibfnamefont {S.}~\bibnamefont {Vittal}}, \bibinfo {author} {\bibfnamefont {K.}~\bibnamefont {Krsulich}}, \bibinfo {author} {\bibfnamefont {L.~S.}\ \bibnamefont {Bishop}}, \bibinfo {author} {\bibfnamefont {J.}~\bibnamefont {Lapeyre}}, \bibinfo {author} {\bibfnamefont {A.}~\bibnamefont {Javadi-Abhari}},\ and\ \bibinfo {author} {\bibfnamefont {E.~Z.}\ \bibnamefont {Zhang}},\ }\href {https://doi.org/10.48550/ARXIV.2211.01925} {\bibinfo {title} {Exploiting qubit reuse through mid-circuit measurement and reset}} (\bibinfo {year} {2022})\BibitemShut {NoStop}%
\bibitem [{\citenamefont {Brandhofer}\ \emph {et~al.}(2023)\citenamefont {Brandhofer}, \citenamefont {Polian},\ and\ \citenamefont {Krsulich}}]{ref:Brandhofer2023}%
  \BibitemOpen
  \bibfield  {author} {\bibinfo {author} {\bibfnamefont {S.}~\bibnamefont {Brandhofer}}, \bibinfo {author} {\bibfnamefont {I.}~\bibnamefont {Polian}},\ and\ \bibinfo {author} {\bibfnamefont {K.}~\bibnamefont {Krsulich}},\ }\bibfield  {title} {\bibinfo {title} {Optimal qubit reuse for near-term quantum computers},\ }in\ \href {https://doi.org/10.1109/qce57702.2023.00100} {\emph {\bibinfo {booktitle} {2023 IEEE International Conference on Quantum Computing and Engineering (QCE)}}}\ (\bibinfo  {publisher} {IEEE},\ \bibinfo {year} {2023})\BibitemShut {NoStop}%
\bibitem [{\citenamefont {Biamonte}\ \emph {et~al.}(2017)\citenamefont {Biamonte}, \citenamefont {Wittek}, \citenamefont {Pancotti}, \citenamefont {Rebentrost}, \citenamefont {Wiebe},\ and\ \citenamefont {Lloyd}}]{ref:Biamonte2017}%
  \BibitemOpen
  \bibfield  {author} {\bibinfo {author} {\bibfnamefont {J.}~\bibnamefont {Biamonte}}, \bibinfo {author} {\bibfnamefont {P.}~\bibnamefont {Wittek}}, \bibinfo {author} {\bibfnamefont {N.}~\bibnamefont {Pancotti}}, \bibinfo {author} {\bibfnamefont {P.}~\bibnamefont {Rebentrost}}, \bibinfo {author} {\bibfnamefont {N.}~\bibnamefont {Wiebe}},\ and\ \bibinfo {author} {\bibfnamefont {S.}~\bibnamefont {Lloyd}},\ }\bibfield  {title} {\bibinfo {title} {Quantum machine learning},\ }\href {https://doi.org/10.1038/nature23474} {\bibfield  {journal} {\bibinfo  {journal} {Nature}\ }\textbf {\bibinfo {volume} {549}},\ \bibinfo {pages} {195–202} (\bibinfo {year} {2017})}\BibitemShut {NoStop}%
\bibitem [{\citenamefont {Lifshitz}\ \emph {et~al.}(2021)\citenamefont {Lifshitz}, \citenamefont {Bairey}, \citenamefont {Arbel}, \citenamefont {Aleksandrowicz}, \citenamefont {Landa},\ and\ \citenamefont {Arad}}]{ref:Lifshitz2021-GibbsLearn}%
  \BibitemOpen
  \bibfield  {author} {\bibinfo {author} {\bibfnamefont {Y.~Y.}\ \bibnamefont {Lifshitz}}, \bibinfo {author} {\bibfnamefont {E.}~\bibnamefont {Bairey}}, \bibinfo {author} {\bibfnamefont {E.}~\bibnamefont {Arbel}}, \bibinfo {author} {\bibfnamefont {G.}~\bibnamefont {Aleksandrowicz}}, \bibinfo {author} {\bibfnamefont {H.}~\bibnamefont {Landa}},\ and\ \bibinfo {author} {\bibfnamefont {I.}~\bibnamefont {Arad}},\ }\href {https://doi.org/10.48550/ARXIV.2112.10418} {\bibinfo {title} {Practical quantum state tomography for gibbs states}} (\bibinfo {year} {2021})\BibitemShut {NoStop}%
\bibitem [{\citenamefont {Zoufal}\ \emph {et~al.}(2021)\citenamefont {Zoufal}, \citenamefont {Lucchi},\ and\ \citenamefont {Woerner}}]{ref:Zoufal2021-1}%
  \BibitemOpen
  \bibfield  {author} {\bibinfo {author} {\bibfnamefont {C.}~\bibnamefont {Zoufal}}, \bibinfo {author} {\bibfnamefont {A.}~\bibnamefont {Lucchi}},\ and\ \bibinfo {author} {\bibfnamefont {S.}~\bibnamefont {Woerner}},\ }\bibfield  {title} {\bibinfo {title} {Variational quantum boltzmann machines},\ }\bibfield  {journal} {\bibinfo  {journal} {Quantum Machine Intelligence}\ }\textbf {\bibinfo {volume} {3}},\ \href {https://doi.org/10.1007/s42484-020-00033-7} {10.1007/s42484-020-00033-7} (\bibinfo {year} {2021})\BibitemShut {NoStop}%
\bibitem [{\citenamefont {Zoufal}(2021)}]{ref:Zoufal2021-2}%
  \BibitemOpen
  \bibfield  {author} {\bibinfo {author} {\bibfnamefont {C.}~\bibnamefont {Zoufal}},\ }\href {https://doi.org/10.48550/ARXIV.2111.12738} {\bibinfo {title} {Generative quantum machine learning}} (\bibinfo {year} {2021})\BibitemShut {NoStop}%
\bibitem [{\citenamefont {Torlai}\ and\ \citenamefont {Melko}(2020)}]{ref:Torlai2020}%
  \BibitemOpen
  \bibfield  {author} {\bibinfo {author} {\bibfnamefont {G.}~\bibnamefont {Torlai}}\ and\ \bibinfo {author} {\bibfnamefont {R.~G.}\ \bibnamefont {Melko}},\ }\bibfield  {title} {\bibinfo {title} {Machine-learning quantum states in the nisq era},\ }\href {https://doi.org/10.1146/annurev-conmatphys-031119-050651} {\bibfield  {journal} {\bibinfo  {journal} {Annual Review of Condensed Matter Physics}\ }\textbf {\bibinfo {volume} {11}},\ \bibinfo {pages} {325–344} (\bibinfo {year} {2020})}\BibitemShut {NoStop}%
\bibitem [{\citenamefont {Childs}\ \emph {et~al.}(2018)\citenamefont {Childs}, \citenamefont {Maslov}, \citenamefont {Nam}, \citenamefont {Ross},\ and\ \citenamefont {Su}}]{ref:Childs2018}%
  \BibitemOpen
  \bibfield  {author} {\bibinfo {author} {\bibfnamefont {A.~M.}\ \bibnamefont {Childs}}, \bibinfo {author} {\bibfnamefont {D.}~\bibnamefont {Maslov}}, \bibinfo {author} {\bibfnamefont {Y.}~\bibnamefont {Nam}}, \bibinfo {author} {\bibfnamefont {N.~J.}\ \bibnamefont {Ross}},\ and\ \bibinfo {author} {\bibfnamefont {Y.}~\bibnamefont {Su}},\ }\bibfield  {title} {\bibinfo {title} {Toward the first quantum simulation with quantum speedup},\ }\href {https://doi.org/10.1073/pnas.1801723115} {\bibfield  {journal} {\bibinfo  {journal} {Proceedings of the National Academy of Sciences}\ }\textbf {\bibinfo {volume} {115}},\ \bibinfo {pages} {9456–9461} (\bibinfo {year} {2018})}\BibitemShut {NoStop}%
\bibitem [{\citenamefont {Vernier}\ \emph {et~al.}(2023)\citenamefont {Vernier}, \citenamefont {Bertini}, \citenamefont {Giudici},\ and\ \citenamefont {Piroli}}]{ref:Vernier2023}%
  \BibitemOpen
  \bibfield  {author} {\bibinfo {author} {\bibfnamefont {E.}~\bibnamefont {Vernier}}, \bibinfo {author} {\bibfnamefont {B.}~\bibnamefont {Bertini}}, \bibinfo {author} {\bibfnamefont {G.}~\bibnamefont {Giudici}},\ and\ \bibinfo {author} {\bibfnamefont {L.}~\bibnamefont {Piroli}},\ }\bibfield  {title} {\bibinfo {title} {Integrable digital quantum simulation: Generalized gibbs ensembles and trotter transitions},\ }\href {https://doi.org/10.1103/PhysRevLett.130.260401} {\bibfield  {journal} {\bibinfo  {journal} {Phys. Rev. Lett.}\ }\textbf {\bibinfo {volume} {130}},\ \bibinfo {pages} {260401} (\bibinfo {year} {2023})}\BibitemShut {NoStop}%
\bibitem [{\citenamefont {Somma}\ \emph {et~al.}(2008)\citenamefont {Somma}, \citenamefont {Boixo}, \citenamefont {Barnum},\ and\ \citenamefont {Knill}}]{ref:Somma2008}%
  \BibitemOpen
  \bibfield  {author} {\bibinfo {author} {\bibfnamefont {R.~D.}\ \bibnamefont {Somma}}, \bibinfo {author} {\bibfnamefont {S.}~\bibnamefont {Boixo}}, \bibinfo {author} {\bibfnamefont {H.}~\bibnamefont {Barnum}},\ and\ \bibinfo {author} {\bibfnamefont {E.}~\bibnamefont {Knill}},\ }\bibfield  {title} {\bibinfo {title} {Quantum simulations of classical annealing processes},\ }\href {https://doi.org/10.1103/PhysRevLett.101.130504} {\bibfield  {journal} {\bibinfo  {journal} {Phys. Rev. Lett.}\ }\textbf {\bibinfo {volume} {101}},\ \bibinfo {pages} {130504} (\bibinfo {year} {2008})}\BibitemShut {NoStop}%
\bibitem [{\citenamefont {Mohseni}\ \emph {et~al.}(2022)\citenamefont {Mohseni}, \citenamefont {McMahon},\ and\ \citenamefont {Byrnes}}]{ref:Mohseni2022}%
  \BibitemOpen
  \bibfield  {author} {\bibinfo {author} {\bibfnamefont {N.}~\bibnamefont {Mohseni}}, \bibinfo {author} {\bibfnamefont {P.~L.}\ \bibnamefont {McMahon}},\ and\ \bibinfo {author} {\bibfnamefont {T.}~\bibnamefont {Byrnes}},\ }\bibfield  {title} {\bibinfo {title} {Ising machines as hardware solvers of combinatorial optimization problems},\ }\href {https://doi.org/10.1038/s42254-022-00440-8} {\bibfield  {journal} {\bibinfo  {journal} {Nature Reviews Physics}\ }\textbf {\bibinfo {volume} {4}},\ \bibinfo {pages} {363–379} (\bibinfo {year} {2022})}\BibitemShut {NoStop}%
\bibitem [{\citenamefont {Wiebe}\ \emph {et~al.}(2014)\citenamefont {Wiebe}, \citenamefont {Kapoor},\ and\ \citenamefont {Svore}}]{ref:Wiebe2014}%
  \BibitemOpen
  \bibfield  {author} {\bibinfo {author} {\bibfnamefont {N.}~\bibnamefont {Wiebe}}, \bibinfo {author} {\bibfnamefont {A.}~\bibnamefont {Kapoor}},\ and\ \bibinfo {author} {\bibfnamefont {K.~M.}\ \bibnamefont {Svore}},\ }\href {https://doi.org/10.48550/ARXIV.1412.3489} {\bibinfo {title} {Quantum deep learning}} (\bibinfo {year} {2014})\BibitemShut {NoStop}%
\bibitem [{\citenamefont {Kieferov\'a}\ and\ \citenamefont {Wiebe}(2017)}]{ref:Kieferova2017}%
  \BibitemOpen
  \bibfield  {author} {\bibinfo {author} {\bibfnamefont {M.}~\bibnamefont {Kieferov\'a}}\ and\ \bibinfo {author} {\bibfnamefont {N.}~\bibnamefont {Wiebe}},\ }\bibfield  {title} {\bibinfo {title} {Tomography and generative training with quantum boltzmann machines},\ }\href {https://doi.org/10.1103/PhysRevA.96.062327} {\bibfield  {journal} {\bibinfo  {journal} {Phys. Rev. A}\ }\textbf {\bibinfo {volume} {96}},\ \bibinfo {pages} {062327} (\bibinfo {year} {2017})}\BibitemShut {NoStop}%
\bibitem [{\citenamefont {Amin}\ \emph {et~al.}(2018)\citenamefont {Amin}, \citenamefont {Andriyash}, \citenamefont {Rolfe}, \citenamefont {Kulchytskyy},\ and\ \citenamefont {Melko}}]{ref:Amin2018}%
  \BibitemOpen
  \bibfield  {author} {\bibinfo {author} {\bibfnamefont {M.~H.}\ \bibnamefont {Amin}}, \bibinfo {author} {\bibfnamefont {E.}~\bibnamefont {Andriyash}}, \bibinfo {author} {\bibfnamefont {J.}~\bibnamefont {Rolfe}}, \bibinfo {author} {\bibfnamefont {B.}~\bibnamefont {Kulchytskyy}},\ and\ \bibinfo {author} {\bibfnamefont {R.}~\bibnamefont {Melko}},\ }\bibfield  {title} {\bibinfo {title} {Quantum boltzmann machine},\ }\href {https://doi.org/10.1103/PhysRevX.8.021050} {\bibfield  {journal} {\bibinfo  {journal} {Phys. Rev. X}\ }\textbf {\bibinfo {volume} {8}},\ \bibinfo {pages} {021050} (\bibinfo {year} {2018})}\BibitemShut {NoStop}%
\bibitem [{\citenamefont {Poulin}\ and\ \citenamefont {Wocjan}(2009)}]{ref:Poulin2009}%
  \BibitemOpen
  \bibfield  {author} {\bibinfo {author} {\bibfnamefont {D.}~\bibnamefont {Poulin}}\ and\ \bibinfo {author} {\bibfnamefont {P.}~\bibnamefont {Wocjan}},\ }\bibfield  {title} {\bibinfo {title} {Sampling from the thermal quantum gibbs state and evaluating partition functions with a quantum computer},\ }\href {https://doi.org/10.1103/PhysRevLett.103.220502} {\bibfield  {journal} {\bibinfo  {journal} {Phys. Rev. Lett.}\ }\textbf {\bibinfo {volume} {103}},\ \bibinfo {pages} {220502} (\bibinfo {year} {2009})}\BibitemShut {NoStop}%
\bibitem [{\citenamefont {Bilgin}\ and\ \citenamefont {Boixo}(2010)}]{ref:Boixo2010}%
  \BibitemOpen
  \bibfield  {author} {\bibinfo {author} {\bibfnamefont {E.}~\bibnamefont {Bilgin}}\ and\ \bibinfo {author} {\bibfnamefont {S.}~\bibnamefont {Boixo}},\ }\bibfield  {title} {\bibinfo {title} {Preparing thermal states of quantum systems by dimension reduction},\ }\href {https://doi.org/10.1103/PhysRevLett.105.170405} {\bibfield  {journal} {\bibinfo  {journal} {Phys. Rev. Lett.}\ }\textbf {\bibinfo {volume} {105}},\ \bibinfo {pages} {170405} (\bibinfo {year} {2010})}\BibitemShut {NoStop}%
\bibitem [{\citenamefont {Riera}\ \emph {et~al.}(2012)\citenamefont {Riera}, \citenamefont {Gogolin},\ and\ \citenamefont {Eisert}}]{ref:Riera-Eisert2012}%
  \BibitemOpen
  \bibfield  {author} {\bibinfo {author} {\bibfnamefont {A.}~\bibnamefont {Riera}}, \bibinfo {author} {\bibfnamefont {C.}~\bibnamefont {Gogolin}},\ and\ \bibinfo {author} {\bibfnamefont {J.}~\bibnamefont {Eisert}},\ }\bibfield  {title} {\bibinfo {title} {Thermalization in nature and on a quantum computer},\ }\href {https://doi.org/10.1103/PhysRevLett.108.080402} {\bibfield  {journal} {\bibinfo  {journal} {Phys. Rev. Lett.}\ }\textbf {\bibinfo {volume} {108}},\ \bibinfo {pages} {080402} (\bibinfo {year} {2012})}\BibitemShut {NoStop}%
\bibitem [{\citenamefont {Su}\ and\ \citenamefont {Li}(2020)}]{ref:Su2020}%
  \BibitemOpen
  \bibfield  {author} {\bibinfo {author} {\bibfnamefont {H.-Y.}\ \bibnamefont {Su}}\ and\ \bibinfo {author} {\bibfnamefont {Y.}~\bibnamefont {Li}},\ }\bibfield  {title} {\bibinfo {title} {Quantum algorithm for the simulation of open-system dynamics and thermalization},\ }\href {https://doi.org/10.1103/PhysRevA.101.012328} {\bibfield  {journal} {\bibinfo  {journal} {Phys. Rev. A}\ }\textbf {\bibinfo {volume} {101}},\ \bibinfo {pages} {012328} (\bibinfo {year} {2020})}\BibitemShut {NoStop}%
\bibitem [{\citenamefont {Holmes}\ \emph {et~al.}(2022)\citenamefont {Holmes}, \citenamefont {Muraleedharan}, \citenamefont {Somma}, \citenamefont {Subasi},\ and\ \citenamefont {Şahinoğlu}}]{ref:Holmes2022-theoretical}%
  \BibitemOpen
  \bibfield  {author} {\bibinfo {author} {\bibfnamefont {Z.}~\bibnamefont {Holmes}}, \bibinfo {author} {\bibfnamefont {G.}~\bibnamefont {Muraleedharan}}, \bibinfo {author} {\bibfnamefont {R.~D.}\ \bibnamefont {Somma}}, \bibinfo {author} {\bibfnamefont {Y.}~\bibnamefont {Subasi}},\ and\ \bibinfo {author} {\bibfnamefont {B.}~\bibnamefont {Şahinoğlu}},\ }\bibfield  {title} {\bibinfo {title} {Quantum algorithms from fluctuation theorems: Thermal-state preparation},\ }\href {https://doi.org/10.22331/q-2022-10-06-825} {\bibfield  {journal} {\bibinfo  {journal} {Quantum}\ }\textbf {\bibinfo {volume} {6}},\ \bibinfo {pages} {825} (\bibinfo {year} {2022})}\BibitemShut {NoStop}%
\bibitem [{\citenamefont {Temme}\ \emph {et~al.}(2011)\citenamefont {Temme}, \citenamefont {Osborne}, \citenamefont {Vollbrecht}, \citenamefont {Poulin},\ and\ \citenamefont {Verstraete}}]{ref:Temme2011}%
  \BibitemOpen
  \bibfield  {author} {\bibinfo {author} {\bibfnamefont {K.}~\bibnamefont {Temme}}, \bibinfo {author} {\bibfnamefont {T.~J.}\ \bibnamefont {Osborne}}, \bibinfo {author} {\bibfnamefont {K.~G.}\ \bibnamefont {Vollbrecht}}, \bibinfo {author} {\bibfnamefont {D.}~\bibnamefont {Poulin}},\ and\ \bibinfo {author} {\bibfnamefont {F.}~\bibnamefont {Verstraete}},\ }\bibfield  {title} {\bibinfo {title} {Quantum metropolis sampling},\ }\href {https://doi.org/10.1038/nature09770} {\bibfield  {journal} {\bibinfo  {journal} {Nature}\ }\textbf {\bibinfo {volume} {471}},\ \bibinfo {pages} {87–90} (\bibinfo {year} {2011})}\BibitemShut {NoStop}%
\bibitem [{\citenamefont {Yung}\ and\ \citenamefont {Aspuru-Guzik}(2012)}]{ref:Yung2012}%
  \BibitemOpen
  \bibfield  {author} {\bibinfo {author} {\bibfnamefont {M.-H.}\ \bibnamefont {Yung}}\ and\ \bibinfo {author} {\bibfnamefont {A.}~\bibnamefont {Aspuru-Guzik}},\ }\bibfield  {title} {\bibinfo {title} {A quantum–quantum metropolis algorithm},\ }\href {https://doi.org/10.1073/pnas.1111758109} {\bibfield  {journal} {\bibinfo  {journal} {Proceedings of the National Academy of Sciences}\ }\textbf {\bibinfo {volume} {109}},\ \bibinfo {pages} {754–759} (\bibinfo {year} {2012})}\BibitemShut {NoStop}%
\bibitem [{\citenamefont {Metcalf}\ \emph {et~al.}(2020)\citenamefont {Metcalf}, \citenamefont {Moussa}, \citenamefont {de~Jong},\ and\ \citenamefont {Sarovar}}]{ref:Metcalf2020}%
  \BibitemOpen
  \bibfield  {author} {\bibinfo {author} {\bibfnamefont {M.}~\bibnamefont {Metcalf}}, \bibinfo {author} {\bibfnamefont {J.~E.}\ \bibnamefont {Moussa}}, \bibinfo {author} {\bibfnamefont {W.~A.}\ \bibnamefont {de~Jong}},\ and\ \bibinfo {author} {\bibfnamefont {M.}~\bibnamefont {Sarovar}},\ }\bibfield  {title} {\bibinfo {title} {Engineered thermalization and cooling of quantum many-body systems},\ }\href {https://doi.org/10.1103/PhysRevResearch.2.023214} {\bibfield  {journal} {\bibinfo  {journal} {Phys. Rev. Res.}\ }\textbf {\bibinfo {volume} {2}},\ \bibinfo {pages} {023214} (\bibinfo {year} {2020})}\BibitemShut {NoStop}%
\bibitem [{\citenamefont {Verstraete}\ \emph {et~al.}(2004)\citenamefont {Verstraete}, \citenamefont {Garc\'{\i}a-Ripoll},\ and\ \citenamefont {Cirac}}]{ref:Cirac2004}%
  \BibitemOpen
  \bibfield  {author} {\bibinfo {author} {\bibfnamefont {F.}~\bibnamefont {Verstraete}}, \bibinfo {author} {\bibfnamefont {J.~J.}\ \bibnamefont {Garc\'{\i}a-Ripoll}},\ and\ \bibinfo {author} {\bibfnamefont {J.~I.}\ \bibnamefont {Cirac}},\ }\bibfield  {title} {\bibinfo {title} {Matrix product density operators: Simulation of finite-temperature and dissipative systems},\ }\href {https://doi.org/10.1103/PhysRevLett.93.207204} {\bibfield  {journal} {\bibinfo  {journal} {Phys. Rev. Lett.}\ }\textbf {\bibinfo {volume} {93}},\ \bibinfo {pages} {207204} (\bibinfo {year} {2004})}\BibitemShut {NoStop}%
\bibitem [{\citenamefont {Motta}\ \emph {et~al.}(2019)\citenamefont {Motta}, \citenamefont {Sun}, \citenamefont {Tan}, \citenamefont {O’Rourke}, \citenamefont {Ye}, \citenamefont {Minnich}, \citenamefont {Brandão},\ and\ \citenamefont {Chan}}]{ref:Motta2019}%
  \BibitemOpen
  \bibfield  {author} {\bibinfo {author} {\bibfnamefont {M.}~\bibnamefont {Motta}}, \bibinfo {author} {\bibfnamefont {C.}~\bibnamefont {Sun}}, \bibinfo {author} {\bibfnamefont {A.~T.~K.}\ \bibnamefont {Tan}}, \bibinfo {author} {\bibfnamefont {M.~J.}\ \bibnamefont {O’Rourke}}, \bibinfo {author} {\bibfnamefont {E.}~\bibnamefont {Ye}}, \bibinfo {author} {\bibfnamefont {A.~J.}\ \bibnamefont {Minnich}}, \bibinfo {author} {\bibfnamefont {F.~G. S.~L.}\ \bibnamefont {Brandão}},\ and\ \bibinfo {author} {\bibfnamefont {G.~K.-L.}\ \bibnamefont {Chan}},\ }\bibfield  {title} {\bibinfo {title} {Determining eigenstates and thermal states on a quantum computer using quantum imaginary time evolution},\ }\href {https://doi.org/10.1038/s41567-019-0704-4} {\bibfield  {journal} {\bibinfo  {journal} {Nature Physics}\ }\textbf {\bibinfo {volume} {16}},\ \bibinfo {pages} {205–210} (\bibinfo {year} {2019})}\BibitemShut {NoStop}%
\bibitem [{\citenamefont {Chen}\ \emph {et~al.}(2023)\citenamefont {Chen}, \citenamefont {Kastoryano}, \citenamefont {Brandão},\ and\ \citenamefont {Gilyén}}]{ref:Gilyen2023}%
  \BibitemOpen
  \bibfield  {author} {\bibinfo {author} {\bibfnamefont {C.-F.}\ \bibnamefont {Chen}}, \bibinfo {author} {\bibfnamefont {M.~J.}\ \bibnamefont {Kastoryano}}, \bibinfo {author} {\bibfnamefont {F.~G. S.~L.}\ \bibnamefont {Brandão}},\ and\ \bibinfo {author} {\bibfnamefont {A.}~\bibnamefont {Gilyén}},\ }\href {https://doi.org/10.48550/ARXIV.2303.18224} {\bibinfo {title} {Quantum thermal state preparation}} (\bibinfo {year} {2023})\BibitemShut {NoStop}%
\bibitem [{\citenamefont {Layden}\ \emph {et~al.}(2023)\citenamefont {Layden}, \citenamefont {Mazzola}, \citenamefont {Mishmash}, \citenamefont {Motta}, \citenamefont {Wocjan}, \citenamefont {Kim},\ and\ \citenamefont {Sheldon}}]{ref:Layden2023}%
  \BibitemOpen
  \bibfield  {author} {\bibinfo {author} {\bibfnamefont {D.}~\bibnamefont {Layden}}, \bibinfo {author} {\bibfnamefont {G.}~\bibnamefont {Mazzola}}, \bibinfo {author} {\bibfnamefont {R.~V.}\ \bibnamefont {Mishmash}}, \bibinfo {author} {\bibfnamefont {M.}~\bibnamefont {Motta}}, \bibinfo {author} {\bibfnamefont {P.}~\bibnamefont {Wocjan}}, \bibinfo {author} {\bibfnamefont {J.-S.}\ \bibnamefont {Kim}},\ and\ \bibinfo {author} {\bibfnamefont {S.}~\bibnamefont {Sheldon}},\ }\bibfield  {title} {\bibinfo {title} {Quantum-enhanced markov chain monte carlo},\ }\href {https://doi.org/10.1038/s41586-023-06095-4} {\bibfield  {journal} {\bibinfo  {journal} {Nature}\ }\textbf {\bibinfo {volume} {619}},\ \bibinfo {pages} {282–287} (\bibinfo {year} {2023})}\BibitemShut {NoStop}%
\bibitem [{\citenamefont {Coopmans}\ \emph {et~al.}(2023)\citenamefont {Coopmans}, \citenamefont {Kikuchi},\ and\ \citenamefont {Benedetti}}]{ref:Benedetti2023}%
  \BibitemOpen
  \bibfield  {author} {\bibinfo {author} {\bibfnamefont {L.}~\bibnamefont {Coopmans}}, \bibinfo {author} {\bibfnamefont {Y.}~\bibnamefont {Kikuchi}},\ and\ \bibinfo {author} {\bibfnamefont {M.}~\bibnamefont {Benedetti}},\ }\bibfield  {title} {\bibinfo {title} {Predicting gibbs-state expectation values with pure thermal shadows},\ }\href {https://doi.org/10.1103/PRXQuantum.4.010305} {\bibfield  {journal} {\bibinfo  {journal} {PRX Quantum}\ }\textbf {\bibinfo {volume} {4}},\ \bibinfo {pages} {010305} (\bibinfo {year} {2023})}\BibitemShut {NoStop}%
\bibitem [{\citenamefont {Eassa}\ \emph {et~al.}(2023)\citenamefont {Eassa}, \citenamefont {Moustafa}, \citenamefont {Banerjee},\ and\ \citenamefont {Cohn}}]{ref:Cohn2023}%
  \BibitemOpen
  \bibfield  {author} {\bibinfo {author} {\bibfnamefont {N.~M.}\ \bibnamefont {Eassa}}, \bibinfo {author} {\bibfnamefont {M.~M.}\ \bibnamefont {Moustafa}}, \bibinfo {author} {\bibfnamefont {A.}~\bibnamefont {Banerjee}},\ and\ \bibinfo {author} {\bibfnamefont {J.}~\bibnamefont {Cohn}},\ }\href {https://doi.org/10.48550/ARXIV.2310.20129} {\bibinfo {title} {Gibbs state sampling via cluster expansions}} (\bibinfo {year} {2023})\BibitemShut {NoStop}%
\bibitem [{\citenamefont {Arad}\ \emph {et~al.}(2024)\citenamefont {Arad}, \citenamefont {Firanko},\ and\ \citenamefont {Gurevich}}]{ref:Gurevich2024}%
  \BibitemOpen
  \bibfield  {author} {\bibinfo {author} {\bibfnamefont {I.}~\bibnamefont {Arad}}, \bibinfo {author} {\bibfnamefont {R.}~\bibnamefont {Firanko}},\ and\ \bibinfo {author} {\bibfnamefont {O.}~\bibnamefont {Gurevich}},\ }\href {https://doi.org/10.48550/ARXIV.2408.08672} {\bibinfo {title} {Local quantum channels giving rise to quasi-local gibbs states}} (\bibinfo {year} {2024})\BibitemShut {NoStop}%
\bibitem [{\citenamefont {Zhang}\ \emph {et~al.}(2023)\citenamefont {Zhang}, \citenamefont {Bosse},\ and\ \citenamefont {Cubitt}}]{ref:Cubitt2023}%
  \BibitemOpen
  \bibfield  {author} {\bibinfo {author} {\bibfnamefont {D.}~\bibnamefont {Zhang}}, \bibinfo {author} {\bibfnamefont {J.~L.}\ \bibnamefont {Bosse}},\ and\ \bibinfo {author} {\bibfnamefont {T.}~\bibnamefont {Cubitt}},\ }\href {https://doi.org/10.48550/ARXIV.2304.04526} {\bibinfo {title} {Dissipative quantum gibbs sampling}} (\bibinfo {year} {2023})\BibitemShut {NoStop}%
\bibitem [{\citenamefont {Brandão}\ and\ \citenamefont {Kastoryano}(2018)}]{ref:Brando2018}%
  \BibitemOpen
  \bibfield  {author} {\bibinfo {author} {\bibfnamefont {F.~G. S.~L.}\ \bibnamefont {Brandão}}\ and\ \bibinfo {author} {\bibfnamefont {M.~J.}\ \bibnamefont {Kastoryano}},\ }\bibfield  {title} {\bibinfo {title} {Finite correlation length implies efficient preparation of quantum thermal states},\ }\href {https://doi.org/10.1007/s00220-018-3150-8} {\bibfield  {journal} {\bibinfo  {journal} {Communications in Mathematical Physics}\ }\textbf {\bibinfo {volume} {365}},\ \bibinfo {pages} {1–16} (\bibinfo {year} {2018})}\BibitemShut {NoStop}%
\bibitem [{\citenamefont {Chowdhury}\ and\ \citenamefont {Somma}(2016)}]{ref:Chowdhury2016}%
  \BibitemOpen
  \bibfield  {author} {\bibinfo {author} {\bibfnamefont {A.~N.}\ \bibnamefont {Chowdhury}}\ and\ \bibinfo {author} {\bibfnamefont {R.~D.}\ \bibnamefont {Somma}},\ }\href@noop {} {\bibinfo {title} {Quantum algorithms for gibbs sampling and hitting-time estimation}} (\bibinfo {year} {2016}),\ \Eprint {https://arxiv.org/abs/1603.02940} {arXiv:1603.02940 [quant-ph]} \BibitemShut {NoStop}%
\bibitem [{\citenamefont {Gily\'en}\ \emph {et~al.}(2022)\citenamefont {Gily\'en}, \citenamefont {Lloyd}, \citenamefont {Marvian}, \citenamefont {Quek},\ and\ \citenamefont {Wilde}}]{ref:Gilyen2022}%
  \BibitemOpen
  \bibfield  {author} {\bibinfo {author} {\bibfnamefont {A.}~\bibnamefont {Gily\'en}}, \bibinfo {author} {\bibfnamefont {S.}~\bibnamefont {Lloyd}}, \bibinfo {author} {\bibfnamefont {I.}~\bibnamefont {Marvian}}, \bibinfo {author} {\bibfnamefont {Y.}~\bibnamefont {Quek}},\ and\ \bibinfo {author} {\bibfnamefont {M.~M.}\ \bibnamefont {Wilde}},\ }\bibfield  {title} {\bibinfo {title} {Quantum algorithm for petz recovery channels and pretty good measurements},\ }\href {https://doi.org/10.1103/PhysRevLett.128.220502} {\bibfield  {journal} {\bibinfo  {journal} {Phys. Rev. Lett.}\ }\textbf {\bibinfo {volume} {128}},\ \bibinfo {pages} {220502} (\bibinfo {year} {2022})}\BibitemShut {NoStop}%
\bibitem [{\citenamefont {Lu}\ \emph {et~al.}(2021)\citenamefont {Lu}, \citenamefont {Ba\~nuls},\ and\ \citenamefont {Cirac}}]{ref:Lu2021}%
  \BibitemOpen
  \bibfield  {author} {\bibinfo {author} {\bibfnamefont {S.}~\bibnamefont {Lu}}, \bibinfo {author} {\bibfnamefont {M.~C.}\ \bibnamefont {Ba\~nuls}},\ and\ \bibinfo {author} {\bibfnamefont {J.~I.}\ \bibnamefont {Cirac}},\ }\bibfield  {title} {\bibinfo {title} {Algorithms for quantum simulation at finite energies},\ }\href {https://doi.org/10.1103/PRXQuantum.2.020321} {\bibfield  {journal} {\bibinfo  {journal} {PRX Quantum}\ }\textbf {\bibinfo {volume} {2}},\ \bibinfo {pages} {020321} (\bibinfo {year} {2021})}\BibitemShut {NoStop}%
\bibitem [{\citenamefont {Alhambra}(2023)}]{ref:Alhambra2023}%
  \BibitemOpen
  \bibfield  {author} {\bibinfo {author} {\bibfnamefont {A.~M.}\ \bibnamefont {Alhambra}},\ }\bibfield  {title} {\bibinfo {title} {Quantum many-body systems in thermal equilibrium},\ }\href {https://doi.org/10.1103/PRXQuantum.4.040201} {\bibfield  {journal} {\bibinfo  {journal} {PRX Quantum}\ }\textbf {\bibinfo {volume} {4}},\ \bibinfo {pages} {040201} (\bibinfo {year} {2023})}\BibitemShut {NoStop}%
\bibitem [{\citenamefont {Tilly}\ \emph {et~al.}(2022)\citenamefont {Tilly}, \citenamefont {Chen}, \citenamefont {Cao}, \citenamefont {Picozzi}, \citenamefont {Setia}, \citenamefont {Li}, \citenamefont {Grant}, \citenamefont {Wossnig}, \citenamefont {Rungger}, \citenamefont {Booth},\ and\ \citenamefont {Tennyson}}]{ref:Tilly2022}%
  \BibitemOpen
  \bibfield  {author} {\bibinfo {author} {\bibfnamefont {J.}~\bibnamefont {Tilly}}, \bibinfo {author} {\bibfnamefont {H.}~\bibnamefont {Chen}}, \bibinfo {author} {\bibfnamefont {S.}~\bibnamefont {Cao}}, \bibinfo {author} {\bibfnamefont {D.}~\bibnamefont {Picozzi}}, \bibinfo {author} {\bibfnamefont {K.}~\bibnamefont {Setia}}, \bibinfo {author} {\bibfnamefont {Y.}~\bibnamefont {Li}}, \bibinfo {author} {\bibfnamefont {E.}~\bibnamefont {Grant}}, \bibinfo {author} {\bibfnamefont {L.}~\bibnamefont {Wossnig}}, \bibinfo {author} {\bibfnamefont {I.}~\bibnamefont {Rungger}}, \bibinfo {author} {\bibfnamefont {G.~H.}\ \bibnamefont {Booth}},\ and\ \bibinfo {author} {\bibfnamefont {J.}~\bibnamefont {Tennyson}},\ }\bibfield  {title} {\bibinfo {title} {The variational quantum eigensolver: A review of methods and best practices},\ }\href {https://doi.org/10.1016/j.physrep.2022.08.003} {\bibfield  {journal} {\bibinfo  {journal} {Physics Reports}\ }\textbf {\bibinfo {volume} {986}},\ \bibinfo {pages} {1–128} (\bibinfo {year}
  {2022})}\BibitemShut {NoStop}%
\bibitem [{\citenamefont {Cerezo}\ \emph {et~al.}(2021)\citenamefont {Cerezo}, \citenamefont {Arrasmith}, \citenamefont {Babbush}, \citenamefont {Benjamin}, \citenamefont {Endo}, \citenamefont {Fujii}, \citenamefont {McClean}, \citenamefont {Mitarai}, \citenamefont {Yuan}, \citenamefont {Cincio},\ and\ \citenamefont {Coles}}]{ref:Cerezo2021}%
  \BibitemOpen
  \bibfield  {author} {\bibinfo {author} {\bibfnamefont {M.}~\bibnamefont {Cerezo}}, \bibinfo {author} {\bibfnamefont {A.}~\bibnamefont {Arrasmith}}, \bibinfo {author} {\bibfnamefont {R.}~\bibnamefont {Babbush}}, \bibinfo {author} {\bibfnamefont {S.~C.}\ \bibnamefont {Benjamin}}, \bibinfo {author} {\bibfnamefont {S.}~\bibnamefont {Endo}}, \bibinfo {author} {\bibfnamefont {K.}~\bibnamefont {Fujii}}, \bibinfo {author} {\bibfnamefont {J.~R.}\ \bibnamefont {McClean}}, \bibinfo {author} {\bibfnamefont {K.}~\bibnamefont {Mitarai}}, \bibinfo {author} {\bibfnamefont {X.}~\bibnamefont {Yuan}}, \bibinfo {author} {\bibfnamefont {L.}~\bibnamefont {Cincio}},\ and\ \bibinfo {author} {\bibfnamefont {P.~J.}\ \bibnamefont {Coles}},\ }\bibfield  {title} {\bibinfo {title} {Variational quantum algorithms},\ }\href {https://doi.org/10.1038/s42254-021-00348-9} {\bibfield  {journal} {\bibinfo  {journal} {Nature Reviews Physics}\ }\textbf {\bibinfo {volume} {3}},\ \bibinfo {pages} {625–644} (\bibinfo {year} {2021})}\BibitemShut
  {NoStop}%
\bibitem [{\citenamefont {Sagastizabal}\ \emph {et~al.}(2021)\citenamefont {Sagastizabal}, \citenamefont {Premaratne}, \citenamefont {Klaver}, \citenamefont {Rol}, \citenamefont {Negîrneac}, \citenamefont {Moreira}, \citenamefont {Zou}, \citenamefont {Johri}, \citenamefont {Muthusubramanian}, \citenamefont {Beekman}, \citenamefont {Zachariadis}, \citenamefont {Ostroukh}, \citenamefont {Haider}, \citenamefont {Bruno}, \citenamefont {Matsuura},\ and\ \citenamefont {DiCarlo}}]{ref:Sagastizabal2021}%
  \BibitemOpen
  \bibfield  {author} {\bibinfo {author} {\bibfnamefont {R.}~\bibnamefont {Sagastizabal}}, \bibinfo {author} {\bibfnamefont {S.~P.}\ \bibnamefont {Premaratne}}, \bibinfo {author} {\bibfnamefont {B.~A.}\ \bibnamefont {Klaver}}, \bibinfo {author} {\bibfnamefont {M.~A.}\ \bibnamefont {Rol}}, \bibinfo {author} {\bibfnamefont {V.}~\bibnamefont {Negîrneac}}, \bibinfo {author} {\bibfnamefont {M.~S.}\ \bibnamefont {Moreira}}, \bibinfo {author} {\bibfnamefont {X.}~\bibnamefont {Zou}}, \bibinfo {author} {\bibfnamefont {S.}~\bibnamefont {Johri}}, \bibinfo {author} {\bibfnamefont {N.}~\bibnamefont {Muthusubramanian}}, \bibinfo {author} {\bibfnamefont {M.}~\bibnamefont {Beekman}}, \bibinfo {author} {\bibfnamefont {C.}~\bibnamefont {Zachariadis}}, \bibinfo {author} {\bibfnamefont {V.~P.}\ \bibnamefont {Ostroukh}}, \bibinfo {author} {\bibfnamefont {N.}~\bibnamefont {Haider}}, \bibinfo {author} {\bibfnamefont {A.}~\bibnamefont {Bruno}}, \bibinfo {author} {\bibfnamefont {A.~Y.}\ \bibnamefont {Matsuura}},\ and\ \bibinfo
  {author} {\bibfnamefont {L.}~\bibnamefont {DiCarlo}},\ }\bibfield  {title} {\bibinfo {title} {Variational preparation of finite-temperature states on a quantum computer},\ }\bibfield  {journal} {\bibinfo  {journal} {npj Quantum Information}\ }\textbf {\bibinfo {volume} {7}},\ \href {https://doi.org/10.1038/s41534-021-00468-1} {10.1038/s41534-021-00468-1} (\bibinfo {year} {2021})\BibitemShut {NoStop}%
\bibitem [{\citenamefont {Wang}\ \emph {et~al.}(2023)\citenamefont {Wang}, \citenamefont {Feng}, \citenamefont {Hartung}, \citenamefont {Jansen},\ and\ \citenamefont {Stornati}}]{ref:Wang2023}%
  \BibitemOpen
  \bibfield  {author} {\bibinfo {author} {\bibfnamefont {X.}~\bibnamefont {Wang}}, \bibinfo {author} {\bibfnamefont {X.}~\bibnamefont {Feng}}, \bibinfo {author} {\bibfnamefont {T.}~\bibnamefont {Hartung}}, \bibinfo {author} {\bibfnamefont {K.}~\bibnamefont {Jansen}},\ and\ \bibinfo {author} {\bibfnamefont {P.}~\bibnamefont {Stornati}},\ }\bibfield  {title} {\bibinfo {title} {Critical behavior of the ising model by preparing the thermal state on a quantum computer},\ }\bibfield  {journal} {\bibinfo  {journal} {Physical Review A}\ }\textbf {\bibinfo {volume} {108}},\ \href {https://doi.org/10.1103/physreva.108.022612} {10.1103/physreva.108.022612} (\bibinfo {year} {2023})\BibitemShut {NoStop}%
\bibitem [{\citenamefont {Patti}\ \emph {et~al.}(2022)\citenamefont {Patti}, \citenamefont {Shehab}, \citenamefont {Najafi},\ and\ \citenamefont {Yelin}}]{ref:Patti2022}%
  \BibitemOpen
  \bibfield  {author} {\bibinfo {author} {\bibfnamefont {T.~L.}\ \bibnamefont {Patti}}, \bibinfo {author} {\bibfnamefont {O.}~\bibnamefont {Shehab}}, \bibinfo {author} {\bibfnamefont {K.}~\bibnamefont {Najafi}},\ and\ \bibinfo {author} {\bibfnamefont {S.~F.}\ \bibnamefont {Yelin}},\ }\bibfield  {title} {\bibinfo {title} {Markov chain monte carlo enhanced variational quantum algorithms},\ }\href {https://doi.org/10.1088/2058-9565/aca821} {\bibfield  {journal} {\bibinfo  {journal} {Quantum Science and Technology}\ }\textbf {\bibinfo {volume} {8}},\ \bibinfo {pages} {015019} (\bibinfo {year} {2022})}\BibitemShut {NoStop}%
\bibitem [{\citenamefont {Cohn}\ \emph {et~al.}(2020)\citenamefont {Cohn}, \citenamefont {Yang}, \citenamefont {Najafi}, \citenamefont {Jones},\ and\ \citenamefont {Freericks}}]{ref:Cohn2020}%
  \BibitemOpen
  \bibfield  {author} {\bibinfo {author} {\bibfnamefont {J.}~\bibnamefont {Cohn}}, \bibinfo {author} {\bibfnamefont {F.}~\bibnamefont {Yang}}, \bibinfo {author} {\bibfnamefont {K.}~\bibnamefont {Najafi}}, \bibinfo {author} {\bibfnamefont {B.}~\bibnamefont {Jones}},\ and\ \bibinfo {author} {\bibfnamefont {J.~K.}\ \bibnamefont {Freericks}},\ }\bibfield  {title} {\bibinfo {title} {Minimal effective gibbs ansatz: A simple protocol for extracting an accurate thermal representation for quantum simulation},\ }\href {https://doi.org/10.1103/PhysRevA.102.022622} {\bibfield  {journal} {\bibinfo  {journal} {Phys. Rev. A}\ }\textbf {\bibinfo {volume} {102}},\ \bibinfo {pages} {022622} (\bibinfo {year} {2020})}\BibitemShut {NoStop}%
\bibitem [{\citenamefont {Verdon}\ \emph {et~al.}(2017)\citenamefont {Verdon}, \citenamefont {Broughton},\ and\ \citenamefont {Biamonte}}]{ref:Verdon2017}%
  \BibitemOpen
  \bibfield  {author} {\bibinfo {author} {\bibfnamefont {G.}~\bibnamefont {Verdon}}, \bibinfo {author} {\bibfnamefont {M.}~\bibnamefont {Broughton}},\ and\ \bibinfo {author} {\bibfnamefont {J.}~\bibnamefont {Biamonte}},\ }\href {https://doi.org/10.48550/ARXIV.1712.05304} {\bibinfo {title} {A quantum algorithm to train neural networks using low-depth circuits}} (\bibinfo {year} {2017})\BibitemShut {NoStop}%
\bibitem [{\citenamefont {Liu}\ \emph {et~al.}(2021)\citenamefont {Liu}, \citenamefont {Mao}, \citenamefont {Zhang},\ and\ \citenamefont {Wang}}]{ref:Liu2021}%
  \BibitemOpen
  \bibfield  {author} {\bibinfo {author} {\bibfnamefont {J.-G.}\ \bibnamefont {Liu}}, \bibinfo {author} {\bibfnamefont {L.}~\bibnamefont {Mao}}, \bibinfo {author} {\bibfnamefont {P.}~\bibnamefont {Zhang}},\ and\ \bibinfo {author} {\bibfnamefont {L.}~\bibnamefont {Wang}},\ }\bibfield  {title} {\bibinfo {title} {Solving quantum statistical mechanics with variational autoregressive networks and quantum circuits},\ }\href {https://doi.org/10.1088/2632-2153/aba19d} {\bibfield  {journal} {\bibinfo  {journal} {Machine Learning: Science and Technology}\ }\textbf {\bibinfo {volume} {2}},\ \bibinfo {pages} {025011} (\bibinfo {year} {2021})}\BibitemShut {NoStop}%
\bibitem [{\citenamefont {Zhang}\ \emph {et~al.}(2024)\citenamefont {Zhang}, \citenamefont {Miao},\ and\ \citenamefont {Hsieh}}]{ref:Guveina2024}%
  \BibitemOpen
  \bibfield  {author} {\bibinfo {author} {\bibfnamefont {S.-X.}\ \bibnamefont {Zhang}}, \bibinfo {author} {\bibfnamefont {J.}~\bibnamefont {Miao}},\ and\ \bibinfo {author} {\bibfnamefont {C.-Y.}\ \bibnamefont {Hsieh}},\ }\href {https://doi.org/10.48550/ARXIV.2402.07605} {\bibinfo {title} {Variational post-selection for ground states and thermal states simulation}} (\bibinfo {year} {2024})\BibitemShut {NoStop}%
\bibitem [{\citenamefont {Martyn}\ and\ \citenamefont {Swingle}(2019)}]{ref:Swingle2019}%
  \BibitemOpen
  \bibfield  {author} {\bibinfo {author} {\bibfnamefont {J.}~\bibnamefont {Martyn}}\ and\ \bibinfo {author} {\bibfnamefont {B.}~\bibnamefont {Swingle}},\ }\bibfield  {title} {\bibinfo {title} {Product spectrum ansatz and the simplicity of thermal states},\ }\href {https://doi.org/10.1103/PhysRevA.100.032107} {\bibfield  {journal} {\bibinfo  {journal} {Phys. Rev. A}\ }\textbf {\bibinfo {volume} {100}},\ \bibinfo {pages} {032107} (\bibinfo {year} {2019})}\BibitemShut {NoStop}%
\bibitem [{\citenamefont {Verdon}\ \emph {et~al.}(2019)\citenamefont {Verdon}, \citenamefont {Marks}, \citenamefont {Nanda}, \citenamefont {Leichenauer},\ and\ \citenamefont {Hidary}}]{ref:Hidary2019}%
  \BibitemOpen
  \bibfield  {author} {\bibinfo {author} {\bibfnamefont {G.}~\bibnamefont {Verdon}}, \bibinfo {author} {\bibfnamefont {J.}~\bibnamefont {Marks}}, \bibinfo {author} {\bibfnamefont {S.}~\bibnamefont {Nanda}}, \bibinfo {author} {\bibfnamefont {S.}~\bibnamefont {Leichenauer}},\ and\ \bibinfo {author} {\bibfnamefont {J.}~\bibnamefont {Hidary}},\ }\href {https://doi.org/10.48550/ARXIV.1910.02071} {\bibinfo {title} {Quantum hamiltonian-based models and the variational quantum thermalizer algorithm}} (\bibinfo {year} {2019})\BibitemShut {NoStop}%
\bibitem [{\citenamefont {Zhu}\ \emph {et~al.}(2020)\citenamefont {Zhu}, \citenamefont {Johri}, \citenamefont {Linke}, \citenamefont {Landsman}, \citenamefont {Huerta~Alderete}, \citenamefont {Nguyen}, \citenamefont {Matsuura}, \citenamefont {Hsieh},\ and\ \citenamefont {Monroe}}]{ref:Zhu2020}%
  \BibitemOpen
  \bibfield  {author} {\bibinfo {author} {\bibfnamefont {D.}~\bibnamefont {Zhu}}, \bibinfo {author} {\bibfnamefont {S.}~\bibnamefont {Johri}}, \bibinfo {author} {\bibfnamefont {N.~M.}\ \bibnamefont {Linke}}, \bibinfo {author} {\bibfnamefont {K.~A.}\ \bibnamefont {Landsman}}, \bibinfo {author} {\bibfnamefont {C.}~\bibnamefont {Huerta~Alderete}}, \bibinfo {author} {\bibfnamefont {N.~H.}\ \bibnamefont {Nguyen}}, \bibinfo {author} {\bibfnamefont {A.~Y.}\ \bibnamefont {Matsuura}}, \bibinfo {author} {\bibfnamefont {T.~H.}\ \bibnamefont {Hsieh}},\ and\ \bibinfo {author} {\bibfnamefont {C.}~\bibnamefont {Monroe}},\ }\bibfield  {title} {\bibinfo {title} {Generation of thermofield double states and critical ground states with a quantum computer},\ }\href {https://doi.org/10.1073/pnas.2006337117} {\bibfield  {journal} {\bibinfo  {journal} {Proceedings of the National Academy of Sciences}\ }\textbf {\bibinfo {volume} {117}},\ \bibinfo {pages} {25402–25406} (\bibinfo {year} {2020})}\BibitemShut {NoStop}%
\bibitem [{\citenamefont {Wang}\ \emph {et~al.}(2021{\natexlab{b}})\citenamefont {Wang}, \citenamefont {Li},\ and\ \citenamefont {Wang}}]{ref:Wang2021}%
  \BibitemOpen
  \bibfield  {author} {\bibinfo {author} {\bibfnamefont {Y.}~\bibnamefont {Wang}}, \bibinfo {author} {\bibfnamefont {G.}~\bibnamefont {Li}},\ and\ \bibinfo {author} {\bibfnamefont {X.}~\bibnamefont {Wang}},\ }\bibfield  {title} {\bibinfo {title} {Variational quantum gibbs state preparation with a truncated taylor series},\ }\href {https://doi.org/10.1103/PhysRevApplied.16.054035} {\bibfield  {journal} {\bibinfo  {journal} {Phys. Rev. Appl.}\ }\textbf {\bibinfo {volume} {16}},\ \bibinfo {pages} {054035} (\bibinfo {year} {2021}{\natexlab{b}})}\BibitemShut {NoStop}%
\bibitem [{\citenamefont {Foldager}\ \emph {et~al.}(2022)\citenamefont {Foldager}, \citenamefont {Pesah},\ and\ \citenamefont {Hansen}}]{ref:Foldager2022}%
  \BibitemOpen
  \bibfield  {author} {\bibinfo {author} {\bibfnamefont {J.}~\bibnamefont {Foldager}}, \bibinfo {author} {\bibfnamefont {A.}~\bibnamefont {Pesah}},\ and\ \bibinfo {author} {\bibfnamefont {L.~K.}\ \bibnamefont {Hansen}},\ }\bibfield  {title} {\bibinfo {title} {Noise-assisted variational quantum thermalization},\ }\bibfield  {journal} {\bibinfo  {journal} {Scientific Reports}\ }\textbf {\bibinfo {volume} {12}},\ \href {https://doi.org/10.1038/s41598-022-07296-z} {10.1038/s41598-022-07296-z} (\bibinfo {year} {2022})\BibitemShut {NoStop}%
\bibitem [{\citenamefont {Consiglio}\ \emph {et~al.}(2024)\citenamefont {Consiglio}, \citenamefont {Settino}, \citenamefont {Giordano}, \citenamefont {Mastroianni}, \citenamefont {Plastina}, \citenamefont {Lorenzo}, \citenamefont {Maniscalco}, \citenamefont {Goold},\ and\ \citenamefont {Apollaro}}]{ref:Consiglio-XY-Ising2024}%
  \BibitemOpen
  \bibfield  {author} {\bibinfo {author} {\bibfnamefont {M.}~\bibnamefont {Consiglio}}, \bibinfo {author} {\bibfnamefont {J.}~\bibnamefont {Settino}}, \bibinfo {author} {\bibfnamefont {A.}~\bibnamefont {Giordano}}, \bibinfo {author} {\bibfnamefont {C.}~\bibnamefont {Mastroianni}}, \bibinfo {author} {\bibfnamefont {F.}~\bibnamefont {Plastina}}, \bibinfo {author} {\bibfnamefont {S.}~\bibnamefont {Lorenzo}}, \bibinfo {author} {\bibfnamefont {S.}~\bibnamefont {Maniscalco}}, \bibinfo {author} {\bibfnamefont {J.}~\bibnamefont {Goold}},\ and\ \bibinfo {author} {\bibfnamefont {T.~J.~G.}\ \bibnamefont {Apollaro}},\ }\bibfield  {title} {\bibinfo {title} {Variational gibbs state preparation on noisy intermediate-scale quantum devices},\ }\href {https://doi.org/10.1103/PhysRevA.110.012445} {\bibfield  {journal} {\bibinfo  {journal} {Phys. Rev. A}\ }\textbf {\bibinfo {volume} {110}},\ \bibinfo {pages} {012445} (\bibinfo {year} {2024})}\BibitemShut {NoStop}%
\bibitem [{\citenamefont {Selisko}\ \emph {et~al.}(2023)\citenamefont {Selisko}, \citenamefont {Amsler}, \citenamefont {Hammerschmidt}, \citenamefont {Drautz},\ and\ \citenamefont {Eckl}}]{ref:Selisko2023}%
  \BibitemOpen
  \bibfield  {author} {\bibinfo {author} {\bibfnamefont {J.}~\bibnamefont {Selisko}}, \bibinfo {author} {\bibfnamefont {M.}~\bibnamefont {Amsler}}, \bibinfo {author} {\bibfnamefont {T.}~\bibnamefont {Hammerschmidt}}, \bibinfo {author} {\bibfnamefont {R.}~\bibnamefont {Drautz}},\ and\ \bibinfo {author} {\bibfnamefont {T.}~\bibnamefont {Eckl}},\ }\bibfield  {title} {\bibinfo {title} {Extending the variational quantum eigensolver to finite temperatures},\ }\href {https://doi.org/10.1088/2058-9565/ad1340} {\bibfield  {journal} {\bibinfo  {journal} {Quantum Science and Technology}\ }\textbf {\bibinfo {volume} {9}},\ \bibinfo {pages} {015026} (\bibinfo {year} {2023})}\BibitemShut {NoStop}%
\bibitem [{\citenamefont {Reif}(1965)}]{ref:Reif1965-StatMech}%
  \BibitemOpen
  \bibfield  {author} {\bibinfo {author} {\bibfnamefont {F.}~\bibnamefont {Reif}},\ }\href@noop {} {\emph {\bibinfo {title} {{Fundamentals of Statistical and Thermal Physics}}}}\ (\bibinfo  {publisher} {McGraw-Hill},\ \bibinfo {address} {New York},\ \bibinfo {year} {1965})\BibitemShut {NoStop}%
\bibitem [{\citenamefont {Wu}\ and\ \citenamefont {Hsieh}(2019)}]{ref:Hsieh2019}%
  \BibitemOpen
  \bibfield  {author} {\bibinfo {author} {\bibfnamefont {J.}~\bibnamefont {Wu}}\ and\ \bibinfo {author} {\bibfnamefont {T.~H.}\ \bibnamefont {Hsieh}},\ }\bibfield  {title} {\bibinfo {title} {Variational thermal quantum simulation via thermofield double states},\ }\href {https://doi.org/10.1103/PhysRevLett.123.220502} {\bibfield  {journal} {\bibinfo  {journal} {Phys. Rev. Lett.}\ }\textbf {\bibinfo {volume} {123}},\ \bibinfo {pages} {220502} (\bibinfo {year} {2019})}\BibitemShut {NoStop}%
\bibitem [{\citenamefont {Sewell}\ \emph {et~al.}(2022)\citenamefont {Sewell}, \citenamefont {White},\ and\ \citenamefont {Swingle}}]{ref:Swingle2022}%
  \BibitemOpen
  \bibfield  {author} {\bibinfo {author} {\bibfnamefont {T.~J.}\ \bibnamefont {Sewell}}, \bibinfo {author} {\bibfnamefont {C.~D.}\ \bibnamefont {White}},\ and\ \bibinfo {author} {\bibfnamefont {B.}~\bibnamefont {Swingle}},\ }\href {https://doi.org/10.48550/ARXIV.2210.16419} {\bibinfo {title} {Thermal multi-scale entanglement renormalization ansatz for variational gibbs state preparation}} (\bibinfo {year} {2022})\BibitemShut {NoStop}%
\bibitem [{\citenamefont {Chowdhury}\ \emph {et~al.}(2020)\citenamefont {Chowdhury}, \citenamefont {Low},\ and\ \citenamefont {Wiebe}}]{ref:Chowdhury2020}%
  \BibitemOpen
  \bibfield  {author} {\bibinfo {author} {\bibfnamefont {A.~N.}\ \bibnamefont {Chowdhury}}, \bibinfo {author} {\bibfnamefont {G.~H.}\ \bibnamefont {Low}},\ and\ \bibinfo {author} {\bibfnamefont {N.}~\bibnamefont {Wiebe}},\ }\href {https://doi.org/10.48550/ARXIV.2002.00055} {\bibinfo {title} {A variational quantum algorithm for preparing quantum gibbs states}} (\bibinfo {year} {2020})\BibitemShut {NoStop}%
\bibitem [{\citenamefont {Araz}\ \emph {et~al.}(2024)\citenamefont {Araz}, \citenamefont {Jha}, \citenamefont {Ringer},\ and\ \citenamefont {Sambasivam}}]{ref:Araz2024}%
  \BibitemOpen
  \bibfield  {author} {\bibinfo {author} {\bibfnamefont {J.~Y.}\ \bibnamefont {Araz}}, \bibinfo {author} {\bibfnamefont {R.~G.}\ \bibnamefont {Jha}}, \bibinfo {author} {\bibfnamefont {F.}~\bibnamefont {Ringer}},\ and\ \bibinfo {author} {\bibfnamefont {B.}~\bibnamefont {Sambasivam}},\ }\href {https://doi.org/10.48550/ARXIV.2406.15545} {\bibinfo {title} {Thermal state preparation of the syk model using a variational quantum algorithm}} (\bibinfo {year} {2024})\BibitemShut {NoStop}%
\bibitem [{\citenamefont {Warren}\ \emph {et~al.}(2022)\citenamefont {Warren}, \citenamefont {Zhu}, \citenamefont {Barnes},\ and\ \citenamefont {Economou}}]{ref:Warren2022adaptive}%
  \BibitemOpen
  \bibfield  {author} {\bibinfo {author} {\bibfnamefont {A.}~\bibnamefont {Warren}}, \bibinfo {author} {\bibfnamefont {L.}~\bibnamefont {Zhu}}, \bibinfo {author} {\bibfnamefont {E.}~\bibnamefont {Barnes}},\ and\ \bibinfo {author} {\bibfnamefont {S.}~\bibnamefont {Economou}},\ }\bibfield  {title} {\bibinfo {title} {Adaptive variational algorithms for quantum gibbs state preparation},\ }\href@noop {} {\bibfield  {journal} {\bibinfo  {journal} {Bulletin of the American Physical Society}\ }\textbf {\bibinfo {volume} {67}} (\bibinfo {year} {2022})}\BibitemShut {NoStop}%
\bibitem [{\citenamefont {Temme}\ \emph {et~al.}(2017)\citenamefont {Temme}, \citenamefont {Bravyi},\ and\ \citenamefont {Gambetta}}]{ref:Temme-PEC2017}%
  \BibitemOpen
  \bibfield  {author} {\bibinfo {author} {\bibfnamefont {K.}~\bibnamefont {Temme}}, \bibinfo {author} {\bibfnamefont {S.}~\bibnamefont {Bravyi}},\ and\ \bibinfo {author} {\bibfnamefont {J.~M.}\ \bibnamefont {Gambetta}},\ }\bibfield  {title} {\bibinfo {title} {Error mitigation for short-depth quantum circuits},\ }\href {https://doi.org/10.1103/PhysRevLett.119.180509} {\bibfield  {journal} {\bibinfo  {journal} {Phys. Rev. Lett.}\ }\textbf {\bibinfo {volume} {119}},\ \bibinfo {pages} {180509} (\bibinfo {year} {2017})}\BibitemShut {NoStop}%
\bibitem [{\citenamefont {Gupta}\ \emph {et~al.}(2023)\citenamefont {Gupta}, \citenamefont {Berg}, \citenamefont {Takita}, \citenamefont {Riste}, \citenamefont {Temme},\ and\ \citenamefont {Kandala}}]{ref:Temme-PEC2023-2}%
  \BibitemOpen
  \bibfield  {author} {\bibinfo {author} {\bibfnamefont {R.~S.}\ \bibnamefont {Gupta}}, \bibinfo {author} {\bibfnamefont {E.~v.~d.}\ \bibnamefont {Berg}}, \bibinfo {author} {\bibfnamefont {M.}~\bibnamefont {Takita}}, \bibinfo {author} {\bibfnamefont {D.}~\bibnamefont {Riste}}, \bibinfo {author} {\bibfnamefont {K.}~\bibnamefont {Temme}},\ and\ \bibinfo {author} {\bibfnamefont {A.}~\bibnamefont {Kandala}},\ }\href {https://doi.org/10.48550/ARXIV.2310.07825} {\bibinfo {title} {Probabilistic error cancellation for dynamic quantum circuits}} (\bibinfo {year} {2023})\BibitemShut {NoStop}%
\bibitem [{\citenamefont {McClean}\ \emph {et~al.}(2018)\citenamefont {McClean}, \citenamefont {Boixo}, \citenamefont {Smelyanskiy}, \citenamefont {Babbush},\ and\ \citenamefont {Neven}}]{ref:McClean2018}%
  \BibitemOpen
  \bibfield  {author} {\bibinfo {author} {\bibfnamefont {J.~R.}\ \bibnamefont {McClean}}, \bibinfo {author} {\bibfnamefont {S.}~\bibnamefont {Boixo}}, \bibinfo {author} {\bibfnamefont {V.~N.}\ \bibnamefont {Smelyanskiy}}, \bibinfo {author} {\bibfnamefont {R.}~\bibnamefont {Babbush}},\ and\ \bibinfo {author} {\bibfnamefont {H.}~\bibnamefont {Neven}},\ }\bibfield  {title} {\bibinfo {title} {Barren plateaus in quantum neural network training landscapes},\ }\bibfield  {journal} {\bibinfo  {journal} {Nature Communications}\ }\textbf {\bibinfo {volume} {9}},\ \href {https://doi.org/10.1038/s41467-018-07090-4} {10.1038/s41467-018-07090-4} (\bibinfo {year} {2018})\BibitemShut {NoStop}%
\bibitem [{\citenamefont {Vatan}\ and\ \citenamefont {Williams}(2004)}]{ref:Farrokh2004}%
  \BibitemOpen
  \bibfield  {author} {\bibinfo {author} {\bibfnamefont {F.}~\bibnamefont {Vatan}}\ and\ \bibinfo {author} {\bibfnamefont {C.}~\bibnamefont {Williams}},\ }\bibfield  {title} {\bibinfo {title} {Optimal quantum circuits for general two-qubit gates},\ }\href {https://doi.org/10.1103/PhysRevA.69.032315} {\bibfield  {journal} {\bibinfo  {journal} {Phys. Rev. A}\ }\textbf {\bibinfo {volume} {69}},\ \bibinfo {pages} {032315} (\bibinfo {year} {2004})}\BibitemShut {NoStop}%
\bibitem [{\citenamefont {Schuld}\ \emph {et~al.}(2019)\citenamefont {Schuld}, \citenamefont {Bergholm}, \citenamefont {Gogolin}, \citenamefont {Izaac},\ and\ \citenamefont {Killoran}}]{ref:Schuld2019}%
  \BibitemOpen
  \bibfield  {author} {\bibinfo {author} {\bibfnamefont {M.}~\bibnamefont {Schuld}}, \bibinfo {author} {\bibfnamefont {V.}~\bibnamefont {Bergholm}}, \bibinfo {author} {\bibfnamefont {C.}~\bibnamefont {Gogolin}}, \bibinfo {author} {\bibfnamefont {J.}~\bibnamefont {Izaac}},\ and\ \bibinfo {author} {\bibfnamefont {N.}~\bibnamefont {Killoran}},\ }\bibfield  {title} {\bibinfo {title} {Evaluating analytic gradients on quantum hardware},\ }\href {https://doi.org/10.1103/PhysRevA.99.032331} {\bibfield  {journal} {\bibinfo  {journal} {Phys. Rev. A}\ }\textbf {\bibinfo {volume} {99}},\ \bibinfo {pages} {032331} (\bibinfo {year} {2019})}\BibitemShut {NoStop}%
\bibitem [{\citenamefont {Crooks}(2019)}]{ref:crooks2019gradients}%
  \BibitemOpen
  \bibfield  {author} {\bibinfo {author} {\bibfnamefont {G.~E.}\ \bibnamefont {Crooks}},\ }\href {https://doi.org/10.48550/ARXIV.1905.13311} {\bibinfo {title} {Gradients of parameterized quantum gates using the parameter-shift rule and gate decomposition}} (\bibinfo {year} {2019})\BibitemShut {NoStop}%
\bibitem [{ref(2022{\natexlab{a}})}]{ref:IBMQ}%
  \BibitemOpen
  \href@noop {} {\bibinfo {title} {Ibm quantum. https://quantum-computing.ibm.com/, 2022}} (\bibinfo {year} {2022}{\natexlab{a}})\BibitemShut {NoStop}%
\bibitem [{ref(2022{\natexlab{b}})}]{ref:Qiskit}%
  \BibitemOpen
  \href {https://doi.org/10.5281/zenodo.2573505} {\bibinfo {title} {Qiskit: An open-source framework for quantum computing}} (\bibinfo {year} {2022}{\natexlab{b}})\BibitemShut {NoStop}%
\bibitem [{\citenamefont {Zeng}\ \emph {et~al.}(2021)\citenamefont {Zeng}, \citenamefont {Wu}, \citenamefont {Cao}, \citenamefont {Zhang}, \citenamefont {Hou}, \citenamefont {Xu},\ and\ \citenamefont {Zeng}}]{ref:Zeng2021}%
  \BibitemOpen
  \bibfield  {author} {\bibinfo {author} {\bibfnamefont {J.}~\bibnamefont {Zeng}}, \bibinfo {author} {\bibfnamefont {Z.}~\bibnamefont {Wu}}, \bibinfo {author} {\bibfnamefont {C.}~\bibnamefont {Cao}}, \bibinfo {author} {\bibfnamefont {C.}~\bibnamefont {Zhang}}, \bibinfo {author} {\bibfnamefont {S.}~\bibnamefont {Hou}}, \bibinfo {author} {\bibfnamefont {P.}~\bibnamefont {Xu}},\ and\ \bibinfo {author} {\bibfnamefont {B.}~\bibnamefont {Zeng}},\ }\bibfield  {title} {\bibinfo {title} {Simulating noisy variational quantum eigensolver with local noise models},\ }\bibfield  {journal} {\bibinfo  {journal} {Quantum Engineering}\ }\textbf {\bibinfo {volume} {3}},\ \href {https://doi.org/10.1002/que2.77} {10.1002/que2.77} (\bibinfo {year} {2021})\BibitemShut {NoStop}%
\bibitem [{\citenamefont {Fontana}\ \emph {et~al.}(2021)\citenamefont {Fontana}, \citenamefont {Fitzpatrick}, \citenamefont {Ramo}, \citenamefont {Duncan},\ and\ \citenamefont {Rungger}}]{ref:Fontana2021}%
  \BibitemOpen
  \bibfield  {author} {\bibinfo {author} {\bibfnamefont {E.}~\bibnamefont {Fontana}}, \bibinfo {author} {\bibfnamefont {N.}~\bibnamefont {Fitzpatrick}}, \bibinfo {author} {\bibfnamefont {D.~M.~n.}\ \bibnamefont {Ramo}}, \bibinfo {author} {\bibfnamefont {R.}~\bibnamefont {Duncan}},\ and\ \bibinfo {author} {\bibfnamefont {I.}~\bibnamefont {Rungger}},\ }\bibfield  {title} {\bibinfo {title} {Evaluating the noise resilience of variational quantum algorithms},\ }\href {https://doi.org/10.1103/PhysRevA.104.022403} {\bibfield  {journal} {\bibinfo  {journal} {Phys. Rev. A}\ }\textbf {\bibinfo {volume} {104}},\ \bibinfo {pages} {022403} (\bibinfo {year} {2021})}\BibitemShut {NoStop}%
\bibitem [{\citenamefont {Nielsen}\ and\ \citenamefont {Chuang}(2010)}]{ref:nielsen2010quantum}%
  \BibitemOpen
  \bibfield  {author} {\bibinfo {author} {\bibfnamefont {M.~A.}\ \bibnamefont {Nielsen}}\ and\ \bibinfo {author} {\bibfnamefont {I.~L.}\ \bibnamefont {Chuang}},\ }\href@noop {} {\emph {\bibinfo {title} {Quantum computation and quantum information}}}\ (\bibinfo  {publisher} {Cambridge university press},\ \bibinfo {year} {2010})\BibitemShut {NoStop}%
\bibitem [{\citenamefont {Kjaergaard}\ \emph {et~al.}(2020)\citenamefont {Kjaergaard}, \citenamefont {Schwartz}, \citenamefont {Braumuller}, \citenamefont {Krantz}, \citenamefont {Wang}, \citenamefont {Gustavsson},\ and\ \citenamefont {Oliver}}]{ref:Kjaergaard2020}%
  \BibitemOpen
  \bibfield  {author} {\bibinfo {author} {\bibfnamefont {M.}~\bibnamefont {Kjaergaard}}, \bibinfo {author} {\bibfnamefont {M.~E.}\ \bibnamefont {Schwartz}}, \bibinfo {author} {\bibfnamefont {J.}~\bibnamefont {Braumuller}}, \bibinfo {author} {\bibfnamefont {P.}~\bibnamefont {Krantz}}, \bibinfo {author} {\bibfnamefont {J.~I.-J.}\ \bibnamefont {Wang}}, \bibinfo {author} {\bibfnamefont {S.}~\bibnamefont {Gustavsson}},\ and\ \bibinfo {author} {\bibfnamefont {W.~D.}\ \bibnamefont {Oliver}},\ }\bibfield  {title} {\bibinfo {title} {Superconducting qubits: Current state of play},\ }\href {https://doi.org/10.1146/annurev-conmatphys-031119-050605} {\bibfield  {journal} {\bibinfo  {journal} {Annual Review of Condensed Matter Physics}\ }\textbf {\bibinfo {volume} {11}},\ \bibinfo {pages} {369} (\bibinfo {year} {2020})},\ \Eprint {https://arxiv.org/abs/https://doi.org/10.1146/annurev-conmatphys-031119-050605} {https://doi.org/10.1146/annurev-conmatphys-031119-050605} \BibitemShut {NoStop}%
\bibitem [{\citenamefont {Jurcevic}\ \emph {et~al.}(2021)\citenamefont {Jurcevic}, \citenamefont {Javadi-Abhari}, \citenamefont {Bishop}, \citenamefont {Lauer}, \citenamefont {Bogorin}, \citenamefont {Brink}, \citenamefont {Capelluto}, \citenamefont {Günlük}, \citenamefont {Itoko},\ and\ \citenamefont {Kanazawa}}]{ref:Jurcevic_2021}%
  \BibitemOpen
  \bibfield  {author} {\bibinfo {author} {\bibfnamefont {P.}~\bibnamefont {Jurcevic}}, \bibinfo {author} {\bibfnamefont {A.}~\bibnamefont {Javadi-Abhari}}, \bibinfo {author} {\bibfnamefont {L.~S.}\ \bibnamefont {Bishop}}, \bibinfo {author} {\bibfnamefont {I.}~\bibnamefont {Lauer}}, \bibinfo {author} {\bibfnamefont {D.~F.}\ \bibnamefont {Bogorin}}, \bibinfo {author} {\bibfnamefont {M.}~\bibnamefont {Brink}}, \bibinfo {author} {\bibfnamefont {L.}~\bibnamefont {Capelluto}}, \bibinfo {author} {\bibfnamefont {O.}~\bibnamefont {Günlük}}, \bibinfo {author} {\bibfnamefont {T.}~\bibnamefont {Itoko}},\ and\ \bibinfo {author} {\bibfnamefont {N.}~\bibnamefont {Kanazawa}},\ }\bibfield  {title} {\bibinfo {title} {Demonstration of quantum volume 64 on a superconducting quantum computing system},\ }\href {https://doi.org/10.1088/2058-9565/abe519} {\bibfield  {journal} {\bibinfo  {journal} {Quantum Science and Technology}\ }\textbf {\bibinfo {volume} {6}},\ \bibinfo {pages} {025020} (\bibinfo {year} {2021})}\BibitemShut
  {NoStop}%
\bibitem [{\citenamefont {{Rost}}\ \emph {et~al.}(2021)\citenamefont {{Rost}}, \citenamefont {{Del Re}}, \citenamefont {{Earnest}}, \citenamefont {{Kemper}}, \citenamefont {{Jones}},\ and\ \citenamefont {{Freericks}}}]{ref:Rost2021}%
  \BibitemOpen
  \bibfield  {author} {\bibinfo {author} {\bibfnamefont {B.}~\bibnamefont {{Rost}}}, \bibinfo {author} {\bibfnamefont {L.}~\bibnamefont {{Del Re}}}, \bibinfo {author} {\bibfnamefont {N.}~\bibnamefont {{Earnest}}}, \bibinfo {author} {\bibfnamefont {A.~F.}\ \bibnamefont {{Kemper}}}, \bibinfo {author} {\bibfnamefont {B.}~\bibnamefont {{Jones}}},\ and\ \bibinfo {author} {\bibfnamefont {J.~K.}\ \bibnamefont {{Freericks}}},\ }\bibfield  {title} {\bibinfo {title} {{Demonstrating robust simulation of driven-dissipative problems on near-term quantum computers}},\ }\href {https://doi.org/10.48550/arXiv.2108.01183} {\bibfield  {journal} {\bibinfo  {journal} {arXiv}\ ,\ \bibinfo {eid} {arXiv:2108.01183}} (\bibinfo {year} {2021})}\BibitemShut {NoStop}%
\bibitem [{\citenamefont {Uhlmann}(2010)}]{ref:Uhlmann2010}%
  \BibitemOpen
  \bibfield  {author} {\bibinfo {author} {\bibfnamefont {A.}~\bibnamefont {Uhlmann}},\ }\bibfield  {title} {\bibinfo {title} {Transition probability (fidelity) and its relatives},\ }\href {https://doi.org/10.1007/s10701-009-9381-y} {\bibfield  {journal} {\bibinfo  {journal} {Foundations of Physics}\ }\textbf {\bibinfo {volume} {41}},\ \bibinfo {pages} {288–298} (\bibinfo {year} {2010})}\BibitemShut {NoStop}%
\bibitem [{\citenamefont {Paszke}\ \emph {et~al.}(2019)\citenamefont {Paszke}, \citenamefont {Gross}, \citenamefont {Massa}, \citenamefont {Lerer}, \citenamefont {Bradbury}, \citenamefont {Chanan}, \citenamefont {Killeen}, \citenamefont {Lin}, \citenamefont {Gimelshein}, \citenamefont {Antiga}, \citenamefont {Desmaison}, \citenamefont {Kopf}, \citenamefont {Yang}, \citenamefont {DeVito}, \citenamefont {Raison}, \citenamefont {Tejani}, \citenamefont {Chilamkurthy}, \citenamefont {Steiner}, \citenamefont {Fang}, \citenamefont {Bai},\ and\ \citenamefont {Chintala}}]{ref:NEURIPS2019_9015}%
  \BibitemOpen
  \bibfield  {author} {\bibinfo {author} {\bibfnamefont {A.}~\bibnamefont {Paszke}}, \bibinfo {author} {\bibfnamefont {S.}~\bibnamefont {Gross}}, \bibinfo {author} {\bibfnamefont {F.}~\bibnamefont {Massa}}, \bibinfo {author} {\bibfnamefont {A.}~\bibnamefont {Lerer}}, \bibinfo {author} {\bibfnamefont {J.}~\bibnamefont {Bradbury}}, \bibinfo {author} {\bibfnamefont {G.}~\bibnamefont {Chanan}}, \bibinfo {author} {\bibfnamefont {T.}~\bibnamefont {Killeen}}, \bibinfo {author} {\bibfnamefont {Z.}~\bibnamefont {Lin}}, \bibinfo {author} {\bibfnamefont {N.}~\bibnamefont {Gimelshein}}, \bibinfo {author} {\bibfnamefont {L.}~\bibnamefont {Antiga}}, \bibinfo {author} {\bibfnamefont {A.}~\bibnamefont {Desmaison}}, \bibinfo {author} {\bibfnamefont {A.}~\bibnamefont {Kopf}}, \bibinfo {author} {\bibfnamefont {E.}~\bibnamefont {Yang}}, \bibinfo {author} {\bibfnamefont {Z.}~\bibnamefont {DeVito}}, \bibinfo {author} {\bibfnamefont {M.}~\bibnamefont {Raison}}, \bibinfo {author} {\bibfnamefont {A.}~\bibnamefont {Tejani}}, \bibinfo
  {author} {\bibfnamefont {S.}~\bibnamefont {Chilamkurthy}}, \bibinfo {author} {\bibfnamefont {B.}~\bibnamefont {Steiner}}, \bibinfo {author} {\bibfnamefont {L.}~\bibnamefont {Fang}}, \bibinfo {author} {\bibfnamefont {J.}~\bibnamefont {Bai}},\ and\ \bibinfo {author} {\bibfnamefont {S.}~\bibnamefont {Chintala}},\ }\bibfield  {title} {\bibinfo {title} {Pytorch:an imperative style, high-performance deep learning library},\ }in\ \href {http://papers.neurips.cc/paper/9015-pytorch-an-imperative-style-high-performance-deep-learning-library.pdf} {\emph {\bibinfo {booktitle} {Advances in Neural Information Processing Systems 32}}}\ (\bibinfo  {publisher} {Curran Associates, Inc.},\ \bibinfo {year} {2019})\ pp.\ \bibinfo {pages} {8024--8035}\BibitemShut {NoStop}%
\bibitem [{\citenamefont {Kingma}\ and\ \citenamefont {Ba}(2015)}]{ref:Kingma2015}%
  \BibitemOpen
  \bibfield  {author} {\bibinfo {author} {\bibfnamefont {D.~P.}\ \bibnamefont {Kingma}}\ and\ \bibinfo {author} {\bibfnamefont {J.~L.}\ \bibnamefont {Ba}},\ }\bibfield  {title} {\bibinfo {title} {{Adam: A method for stochastic optimization}},\ }\href@noop {} {\bibfield  {journal} {\bibinfo  {journal} {3rd International Conference on Learning Representations, ICLR 2015 - Conference Track Proceedings}\ ,\ \bibinfo {pages} {1}} (\bibinfo {year} {2015})},\ \Eprint {https://arxiv.org/abs/1412.6980} {arXiv:1412.6980} \BibitemShut {NoStop}%
\bibitem [{\citenamefont {Lieb}\ \emph {et~al.}(1961)\citenamefont {Lieb}, \citenamefont {Schultz},\ and\ \citenamefont {Mattis}}]{ref:LIEB1961407}%
  \BibitemOpen
  \bibfield  {author} {\bibinfo {author} {\bibfnamefont {E.}~\bibnamefont {Lieb}}, \bibinfo {author} {\bibfnamefont {T.}~\bibnamefont {Schultz}},\ and\ \bibinfo {author} {\bibfnamefont {D.}~\bibnamefont {Mattis}},\ }\bibfield  {title} {\bibinfo {title} {Two soluble models of an antiferromagnetic chain},\ }\href {https://doi.org/https://doi.org/10.1016/0003-4916(61)90115-4} {\bibfield  {journal} {\bibinfo  {journal} {Annals of Physics}\ }\textbf {\bibinfo {volume} {16}},\ \bibinfo {pages} {407} (\bibinfo {year} {1961})}\BibitemShut {NoStop}%
\bibitem [{\citenamefont {Henao}\ \emph {et~al.}(2023)\citenamefont {Henao}, \citenamefont {Santos},\ and\ \citenamefont {Uzdin}}]{ref:Henao2023}%
  \BibitemOpen
  \bibfield  {author} {\bibinfo {author} {\bibfnamefont {I.}~\bibnamefont {Henao}}, \bibinfo {author} {\bibfnamefont {J.~P.}\ \bibnamefont {Santos}},\ and\ \bibinfo {author} {\bibfnamefont {R.}~\bibnamefont {Uzdin}},\ }\bibfield  {title} {\bibinfo {title} {Adaptive quantum error mitigation using pulse-based inverse evolutions},\ }\bibfield  {journal} {\bibinfo  {journal} {npj Quantum Information}\ }\textbf {\bibinfo {volume} {9}},\ \href {https://doi.org/10.1038/s41534-023-00785-7} {10.1038/s41534-023-00785-7} (\bibinfo {year} {2023})\BibitemShut {NoStop}%
\bibitem [{\citenamefont {Zahedinejad}\ \emph {et~al.}(2015)\citenamefont {Zahedinejad}, \citenamefont {Ghosh},\ and\ \citenamefont {Sanders}}]{ref:Zahedinejad2015}%
  \BibitemOpen
  \bibfield  {author} {\bibinfo {author} {\bibfnamefont {E.}~\bibnamefont {Zahedinejad}}, \bibinfo {author} {\bibfnamefont {J.}~\bibnamefont {Ghosh}},\ and\ \bibinfo {author} {\bibfnamefont {B.~C.}\ \bibnamefont {Sanders}},\ }\bibfield  {title} {\bibinfo {title} {High-fidelity single-shot toffoli gate via quantum control},\ }\href {https://doi.org/10.1103/PhysRevLett.114.200502} {\bibfield  {journal} {\bibinfo  {journal} {Phys. Rev. Lett.}\ }\textbf {\bibinfo {volume} {114}},\ \bibinfo {pages} {200502} (\bibinfo {year} {2015})}\BibitemShut {NoStop}%
\bibitem [{\citenamefont {Xie}\ \emph {et~al.}(2023)\citenamefont {Xie}, \citenamefont {Zhao}, \citenamefont {Xu}, \citenamefont {Kong}, \citenamefont {Yang}, \citenamefont {Wang}, \citenamefont {Wang}, \citenamefont {Shi},\ and\ \citenamefont {Du}}]{ref:Tianyu2023}%
  \BibitemOpen
  \bibfield  {author} {\bibinfo {author} {\bibfnamefont {T.}~\bibnamefont {Xie}}, \bibinfo {author} {\bibfnamefont {Z.}~\bibnamefont {Zhao}}, \bibinfo {author} {\bibfnamefont {S.}~\bibnamefont {Xu}}, \bibinfo {author} {\bibfnamefont {X.}~\bibnamefont {Kong}}, \bibinfo {author} {\bibfnamefont {Z.}~\bibnamefont {Yang}}, \bibinfo {author} {\bibfnamefont {M.}~\bibnamefont {Wang}}, \bibinfo {author} {\bibfnamefont {Y.}~\bibnamefont {Wang}}, \bibinfo {author} {\bibfnamefont {F.}~\bibnamefont {Shi}},\ and\ \bibinfo {author} {\bibfnamefont {J.}~\bibnamefont {Du}},\ }\bibfield  {title} {\bibinfo {title} {99.92\%-fidelity cnot gates in solids by noise filtering},\ }\href {https://doi.org/10.1103/PhysRevLett.130.030601} {\bibfield  {journal} {\bibinfo  {journal} {Phys. Rev. Lett.}\ }\textbf {\bibinfo {volume} {130}},\ \bibinfo {pages} {030601} (\bibinfo {year} {2023})}\BibitemShut {NoStop}%
\bibitem [{\citenamefont {Zhang}\ \emph {et~al.}(2020)\citenamefont {Zhang}, \citenamefont {Srinivasan}, \citenamefont {Sundaresan}, \citenamefont {Bogorin}, \citenamefont {Martin}, \citenamefont {Hertzberg}, \citenamefont {Timmerwilke}, \citenamefont {Pritchett}, \citenamefont {Yau}, \citenamefont {Wang}, \citenamefont {Landers}, \citenamefont {Lewandowski}, \citenamefont {Narasgond}, \citenamefont {Rosenblatt}, \citenamefont {Keefe}, \citenamefont {Lauer}, \citenamefont {Rothwell}, \citenamefont {McClure}, \citenamefont {Dial}, \citenamefont {Orcutt}, \citenamefont {Brink},\ and\ \citenamefont {Chow}}]{ref:Zhang2020}%
  \BibitemOpen
  \bibfield  {author} {\bibinfo {author} {\bibfnamefont {E.~J.}\ \bibnamefont {Zhang}}, \bibinfo {author} {\bibfnamefont {S.}~\bibnamefont {Srinivasan}}, \bibinfo {author} {\bibfnamefont {N.}~\bibnamefont {Sundaresan}}, \bibinfo {author} {\bibfnamefont {D.~F.}\ \bibnamefont {Bogorin}}, \bibinfo {author} {\bibfnamefont {Y.}~\bibnamefont {Martin}}, \bibinfo {author} {\bibfnamefont {J.~B.}\ \bibnamefont {Hertzberg}}, \bibinfo {author} {\bibfnamefont {J.}~\bibnamefont {Timmerwilke}}, \bibinfo {author} {\bibfnamefont {E.~J.}\ \bibnamefont {Pritchett}}, \bibinfo {author} {\bibfnamefont {J.-B.}\ \bibnamefont {Yau}}, \bibinfo {author} {\bibfnamefont {C.}~\bibnamefont {Wang}}, \bibinfo {author} {\bibfnamefont {W.}~\bibnamefont {Landers}}, \bibinfo {author} {\bibfnamefont {E.~P.}\ \bibnamefont {Lewandowski}}, \bibinfo {author} {\bibfnamefont {A.}~\bibnamefont {Narasgond}}, \bibinfo {author} {\bibfnamefont {S.}~\bibnamefont {Rosenblatt}}, \bibinfo {author} {\bibfnamefont {G.~A.}\ \bibnamefont {Keefe}}, \bibinfo {author}
  {\bibfnamefont {I.}~\bibnamefont {Lauer}}, \bibinfo {author} {\bibfnamefont {M.~B.}\ \bibnamefont {Rothwell}}, \bibinfo {author} {\bibfnamefont {D.~T.}\ \bibnamefont {McClure}}, \bibinfo {author} {\bibfnamefont {O.~E.}\ \bibnamefont {Dial}}, \bibinfo {author} {\bibfnamefont {J.~S.}\ \bibnamefont {Orcutt}}, \bibinfo {author} {\bibfnamefont {M.}~\bibnamefont {Brink}},\ and\ \bibinfo {author} {\bibfnamefont {J.~M.}\ \bibnamefont {Chow}},\ }\href {https://doi.org/10.48550/ARXIV.2012.08475} {\bibinfo {title} {High-fidelity superconducting quantum processors via laser-annealing of transmon qubits}} (\bibinfo {year} {2020})\BibitemShut {NoStop}%
\bibitem [{\citenamefont {Wang}\ \emph {et~al.}(2021{\natexlab{c}})\citenamefont {Wang}, \citenamefont {Fontana}, \citenamefont {Cerezo}, \citenamefont {Sharma}, \citenamefont {Sone}, \citenamefont {Cincio},\ and\ \citenamefont {Coles}}]{ref:Wang2021-noise-barren}%
  \BibitemOpen
  \bibfield  {author} {\bibinfo {author} {\bibfnamefont {S.}~\bibnamefont {Wang}}, \bibinfo {author} {\bibfnamefont {E.}~\bibnamefont {Fontana}}, \bibinfo {author} {\bibfnamefont {M.}~\bibnamefont {Cerezo}}, \bibinfo {author} {\bibfnamefont {K.}~\bibnamefont {Sharma}}, \bibinfo {author} {\bibfnamefont {A.}~\bibnamefont {Sone}}, \bibinfo {author} {\bibfnamefont {L.}~\bibnamefont {Cincio}},\ and\ \bibinfo {author} {\bibfnamefont {P.~J.}\ \bibnamefont {Coles}},\ }\bibfield  {title} {\bibinfo {title} {Noise-induced barren plateaus in variational quantum algorithms},\ }\bibfield  {journal} {\bibinfo  {journal} {Nature Communications}\ }\textbf {\bibinfo {volume} {12}},\ \href {https://doi.org/10.1038/s41467-021-27045-6} {10.1038/s41467-021-27045-6} (\bibinfo {year} {2021}{\natexlab{c}})\BibitemShut {NoStop}%
\end{thebibliography}%

\end{document}